\documentclass[pdflatex,sn-mathphys-ay]{sn-jnl}

\usepackage{graphicx}%
\usepackage{multirow}%
\usepackage{amsmath,amssymb,amsfonts}%
\usepackage{amsthm}%
\usepackage{mathrsfs}%
\usepackage[title]{appendix}%
\usepackage{xcolor}%
\usepackage{textcomp}%
\usepackage{manyfoot}%
\usepackage{booktabs}%
\usepackage{algorithm}%
\usepackage{algorithmicx}%
\usepackage{algpseudocode}%
\usepackage{listings}%
\usepackage{makecell}

\theoremstyle{thmstyleone}%
\theoremstyle{thmstyletwo}%

\theoremstyle{thmstylethree}%

\newcommand{\hi}{H\,\sc{i}}

\newcommand{\lya}{Ly$\alpha$}
\newcommand{\rsun}{R$_\odot$}
\newcommand{\kms}{km s$^{-1}$}

\begin{document}

\title[UV Observations of Solar Phenomena]{Ultraviolet Observations of Solar Active Phenomena}


\author[1,2,3]{\fnm{P. F.} \sur{Chen}}\email{chenpf@nju.edu.cn}
\equalcont{These authors contributed equally to this work.}

\author*[4,11]{\fnm{G.} \sur{Del Zanna}}\email{gd232@cam.ac.uk}
\equalcont{These authors contributed equally to this work.}

\author[5,6]{\fnm{L.} \sur{Feng}}\email{lfeng@pmo.ac.cn}
\equalcont{These authors contributed equally to this work.}

\author[1,2]{\fnm{J. H.} \sur{Guo}}\email{jinhan.guo@nju.edu.cn}
\equalcont{These authors contributed equally to this work.}
 
\author[7,8]{\fnm{Y. J.} \sur{Hou}}\email{yijunhou@nao.cas.cn}
\equalcont{These authors contributed equally to this work.}

\author[9]{\fnm{H.} \sur{Tian}}\email{huitian@pku.edu.cn}
\equalcont{These authors contributed equally to this work.}

\author[10]{\fnm{D.} \sur{Tripathi}}\email{durgesh@iucaa.in}
\equalcont{These authors contributed equally to this work.}



\affil[1]{\orgdiv{School of Astronomy and Space Science}, \orgname{Nanjing University}, \orgaddress{\city{Nanjing} \postcode{210023}, \country{China}}}

\affil[2]{\orgdiv{Key Laboratory for Modern Astronomy and Astrophysics}, \orgname{Nanjing University}, \orgaddress{\street{163 Xianlin Avenue}, \city{Nanjing} \postcode{210023}, \country{China}}}

\affil[3]{\orgdiv{State Key Laboratory of Lunar and Planetary Sciences}, \orgname{Macau University of Science and Technology}, \city{Macau} \postcode{999078}, \country{China}}

\affil[4]{\orgdiv{DAMTP, Centre for Mathematical Sciences}, \orgname{University of Cambridge}, \orgaddress{\street{Wilberforce Road}, \city{Cambridge} \postcode{CB3 0WA}, \country{UK}}}

\affil[5]{\orgdiv{Research Division of Dark Matter and Space Astronomy}, \orgname{Purple Mountain Observatory, Chinese Academy of Sciences}, \city{Nanjing} \postcode{210023}, \country{China}}

\affil[6]{\orgdiv{School of Astronomy and Space Science}, \orgname{University of Science and Technology of China}, \city{Hefei} \postcode{230026}, \country{China}}

\affil[7]{\orgdiv{State Key Laboratory of Solar Activity and Space Weather}, \orgname{National Astronomical Observatories, Chinese Academy of Science}, \city{Beijing} \postcode{100101}, \country{China}}

\affil[8]{\orgdiv{School of Astronomy and Space Science}, \orgname{University of Chinese Academy of Sciences}, \city{Beijing} \postcode{100049}, \country{China}}

\affil[9]{\orgdiv{School of Earth and Space Sciences}, \orgname{Peking University}, \city{Beijing} \postcode{100871}, \country{China}}

\affil[10]{\orgname{Inter-University Centre for Astronomy and Astrophysics}, \state{Post Bag 4, Maharashtra}, \postcode{411007} \city{Pune}, \country{India}}

\affil[11]{\orgdiv{School of Physics \& Astronomy}, \orgname{University of Leicester},
  \orgaddress{\city{Leicester} \postcode{LE
  1 7RH},   \country{UK}}}


\abstract{Ultraviolet observations of the Sun have revolutionized our understanding of the structure and dynamics of the solar atmosphere. From the chromosphere all the way to the corona, there are ubiquitous jets due to magnetic reconnection and associated with type III radio bursts. Sporadic flares occur in the corona, representing an inevitable breakup of magnetic energy accumulation as new magnetic flux continues to pump into the solar atmosphere, which is further shuffled by solar surface motions. Once large-scale eruptions occur, magnetized plasmas are expelled out into the interplanetary space, leaving globally propagating waves and the following dimmings. The waves can be utilized to diagnose the coronal magnetic field which cannot be measured directly, and the dimmings can be used to estimate the ejected mass of eruptions and magnetic connectivity. Extreme ultraviolet spectroscopic observations provide information on the fractionation of different chemical elements, in particular those elements with low first ionization potential. Thanks to the high spatial resolution of solar observations, all these results are believed to shed light on our understanding of the activity on other stars which cannot be imaged.}





 \maketitle

\section{Introduction}\label{sec1}

As a main-sequence star, the Sun is mildly active, which is meaningful for two reasons. First, it is not so active that our Earth is not exposed to drastic bombardment by solar eruptions, so that life can survive and thrive. Second, each year the Sun still produces thousands of large eruptive events such as solar flares and coronal mass ejections, and much more small-scale eruptions such as jets. These solar activities, with rich details, provide a cornerstone for us to understand remote activities on other stars, accretion disks, and even galactic evolution. 

The solar activity can be monitored in various wavelengths, including some visible lines such as H$\alpha$ and Ca H\&K, and many lines in the ultraviolet (UV, 10--400 nm), near UV (NUV, 300--400 nm), extreme UV (EUV, 10--121 nm), and X-ray (0.01--10 nm). Among them, UV emission lines 
provide crucial information on the thermal structure and the dynamics of the solar atmosphere.

The solar atmosphere is composed by the photosphere (with a thickness of $\sim$500 km), the chromosphere (with a thickness of $\sim$1500 km), a thin transition region, and the extended corona. The UV line emission originates mainly from the chromosphere and transition region, with a few 
spectral lines emitted by the corona.

{\bf Chromosphere:} The solar chromosphere can loosely be defined as the region of the atmosphere ranging between the temperature minimum above the photosphere around 4400 K and about 20,000 K where hydrogen is almost fully ionized. Some chromospheric lines, such as Ca {\footnotesize II} K, are mostly emitted by the supergranular internetwork, where magnetic flux concentrates. The size of the cells is about 30 Mm. The distribution of the radiance in both the continuum and the lines was found to be lognormal from Skylab studies \citep{Vernazza1981}. 
The cell and network structures are nearly the same across the solar surface. In coronal lines formed above 1 MK, the emission becomes diffuse and is 1--2 orders of magnitude reduced in coronal holes but is bright in coronal loops. Strong spatial (on scales less than 700 km) and temporal variability is present, especially in the supergranular network. 

{\bf Transition region:} The solar transition region is a very thin region of about 100--200 km where emission lines formed between the upper chromosphere (20,000 K) and the corona (1 MK) are observed. For an older but still relevant review, see \cite{mariska:1992}. Every feature of the transition region is a complete mystery. It has overall the same cell/supergranular network structure across the disk, although the contrast decreases with the formation temperature. Exceptions are bright plage in active regions and the very large `dark halos' regions surrounding them \citep{andretta_delzanna:2014,lezzi+2023}. Spatial and temporal variability is even more pronounced than in chromospheric lines. All lines formed at temperatures up to $5\times 10^5$ K show a predominant pattern of redshifts which have a peak in the middle of the region, where also non-thermal widths are the largest. The redshifts are thought to be the result of radiative cooling of coronal plasma. The non-thermal widths were once thought to be caused by Alfv\'en waves, but are more likely just due to a superposition of flows, as the Interface Region Imaging Spectrograph \citep[IRIS,][]{DePontieu2014} observations indicate \citep[see the review in][]{rao_etal:2022ntw}.

It is still not clear if the observed transition region emission is from plasma physically connecting the chromosphere to the corona, as e.g. in the classical model by \cite{gabriel:1976}.  This is caused by the lack of 
instruments  observing simultaneously the
chromosphere, the transition region and corona with the same high spatial/temporal scales. However, it is likely that a large fraction of the transition region emission is completely unrelated to the coronal one.

By measuring the electron density and the radiances in transition region lines, the volume where the plasma
is emitting has been found to be much smaller (by a factor of 1000--3000)
than the typical size of the transition region features
\citep[the so-called filling factor, see e.g.][]{dere_etal:1987,rao_etal:2022pl}.
For the above reasons, relating the observed transition region emission
to the coronal one or the chromospheric ones with one-dimensional (1D) models is likely unrealistic.

Despite this, there is abundant literature on static 1D emission measure models for the Sun and stars, as all the lines, except some of the strongest ones, are essentially optically thin, which in principle simplifies the modeling. In reality, many physical processes affect the formation of these lines. Most of them have rarely been considered in the case of the Sun, and even less in the case of stars. 

{\bf Corona:} EUV might be the best wavelength to observe the solar corona. As shown below, the EUV observations from the Solar Dynamics Observatory \citep[SDO,][]{Pesnell2012} have revolutionized our understanding of coronal activities. It is noted that, although not many far UV observations have been made, many spatially-resolved off-limb observations of the Sun showed a large number of forbidden lines formed in the corona or during flares.
They offer many plasma diagnostics which are unavailable to stellar observations,
as only a few of the brightest lines are visible against the stellar background emission. 
Notable cases are the coronal forbidden lines, such as Fe {\footnotesize XII} 1242 and 1349~\AA\, plus the Fe \textsc{xviii} 974~\AA, the Fe \textsc{xix} 1118~\AA\ and the Fe \textsc{xxi} 1354~\AA\ in active stars or during flares \citep[cf. the review by][]{linsky:2018}. 

Within the past three decades, several UV/EUV satellites were launched to observe the solar atmosphere, providing a wealth of information. Several results were presented by solar physicists at the 
  ISSI-Beijing workshop entitled ``Space-based Stellar Astrophysics in the Ultraviolet". This paper presents a collection of those results, organized as follows: we consider first solar jets, flares, EUV dimmings, EUV waves, and the first ionization potential effect in Sections \ref{sec2}--\ref{sec6}, respectively; Section \ref{sec7} introduces two recent space satellites from China and India, while Section \ref{sec8} discusses the lessons 
   we have learned within solar physics that are relevant to stellar physics.

\section{Solar Jets}\label{sec2}

Solar jets are transient, collimated plasma ejections found in the solar atmosphere throughout the chromosphere and the corona. They are observed as spicules, macro-spicules, surges, and coronal jets with a width ranging from hundreds of kilometers to hundreds of thousands of kilometers, manifesting the reconfiguration of solar magnetic fields and plasma in diverse solar regions including quiet-Sun regions, coronal holes, sunspots, and active regions \citep[see e.g.,][]{Shibata1992, Yokoyama1995, Asai2001,  MulTDM_2006, ChiYI_2008, KaySMT_2013, ChaGMT_2015, HouYJ2017, Samanta2019, LiT2020, HouZY2021, Tri_2021, HouYJ2022, MulTMD_2023, YangLH2025}. These jetting activities usually occur in association with photospheric magnetic flux emergence and cancellation, type III radio bursts, filament eruptions, flares, and coronal mass ejections (CMEs). As a result, studies of solar jets are fundamental to our understanding of solar activities at different scales and the underlying basic physical processes like magnetic reconnection, magnetohydrodynamic (MHD) waves, and particle acceleration. In addition, by continuously transporting energy and mass into the upper atmosphere, solar jets can also serve as important sources for heating coronal plasma and accelerating  the solar wind \citep[see e.g.,][]{Raouafi2016, ShenYD2021, TriNS_2021, UpenT_2022, MalUT_2024}. This section aims to introduce the fundamental observational characteristics and primary physical models of solar jets, with a special focus on new insights gained from UV observations, as well as the jet phenomena occurring within sunspots, which have been extensively studied in recent years.

\subsection{Observational Features}\label{sec2_{1}}
Early studies of solar jets before the 1990s were mainly based on H$\alpha$ observations from small-aperture ground-based telescopes and white light eclipse photographs, which identified ``surges" and ``spicules" as jet-like structures at the solar limb or on the solar disk \citep{Kleczek1960, Beckers1968}. These observations established the transient and collimated nature of solar jets but were limited by atmospheric seeing, temporal and spatial resolutions, which prevented resolving fine structures of solar jets and their dynamical evolutions. Benefiting from the launch of a series of space telescopes since the 1990s, solar jets have been studied intensively through multi-wavelength observations with high spatiotemporal resolutions.

The X-ray and UV observations in the space era revolutionized solar jet research by opening new observational windows, revealing high-temperature components and dynamic evolution previously inaccessible from the ground. X-ray telescopes such as Yohkoh/SXT and Hinode/XRT detected hot (3--8 MK) jet components with the following distinct features: bright jet bases and collimated, narrow, high-velocity (hundreds of km s$^{-1}$) outflows \citep{Shimojo2000, Savcheva2007}. 
Instruments such as the EUV Imaging Telescope (EIT) onboard the Solar and Heliospheric Observatory \citep[SOHO;][]{Domingo1995} and the Atmospheric Imaging Assembly (AIA) onboard SDO produced observations of chromospheric and transition-region jets. Some studies based on UV and EUV observations have reported that mini-filaments were initially confined by a small arch surrounded by open field lines, between which external reconnection then occurred and produced the hot jet component. Such external reconnection removed the confining fields above the mini-filament, which then erupted, resulting in the appearance of a cool jet component \citep{ShenYD2012}.

Moreover, fine structural features including outward plasmoids (or plasma blobs) and vortex structures were also observed within or at the outer edges of solar jets in UV/EUV observations. They were thought to be caused, respectively, by the tearing-mode instability of the current sheets and the Kelvin-Helmholtz instability at the interface between two fluids with strong velocity shear \citep{ZhangQM2014, NiL2017, LiXH2018}. 
High spatiotemporal-resolution UV/EUV observations showed that solar jets display complex motions such as rotation, lateral expansions with variable velocity, and transverse oscillations, which were explained, respectively, by magnetic reconnection between erupting twisted filaments and their surrounding fields and reconnection between an emerging arch and the ambient open field \citep{Morton2012, Moore2013}.

Depending on the observing wavelengths, solar jets were once classified as white-light jets, H$\alpha$ jets, EUV jets, and X-ray jets. However, recent simultaneous multi-wavelength observations revealed that solar jets are often composed of plasmas covering a wide temperature range, and we should consider an alternative classification method for solar jets according to their physical properties, morphologies, dynamics, and triggering/driving mechanisms. For example, coronal jets can be classified into anemone jets (straight, inverted-Y-shaped with a bright dome-like base) and two-sided-loop jets (paired plasma ejections in opposite directions from the source region) according to their morphology \citep{Shibata2007, ZhengRS2018, ShenYD2019, WeiHY2020, TanS2023}.

\begin{figure}[h]
\centering
\includegraphics[width=0.8\textwidth]{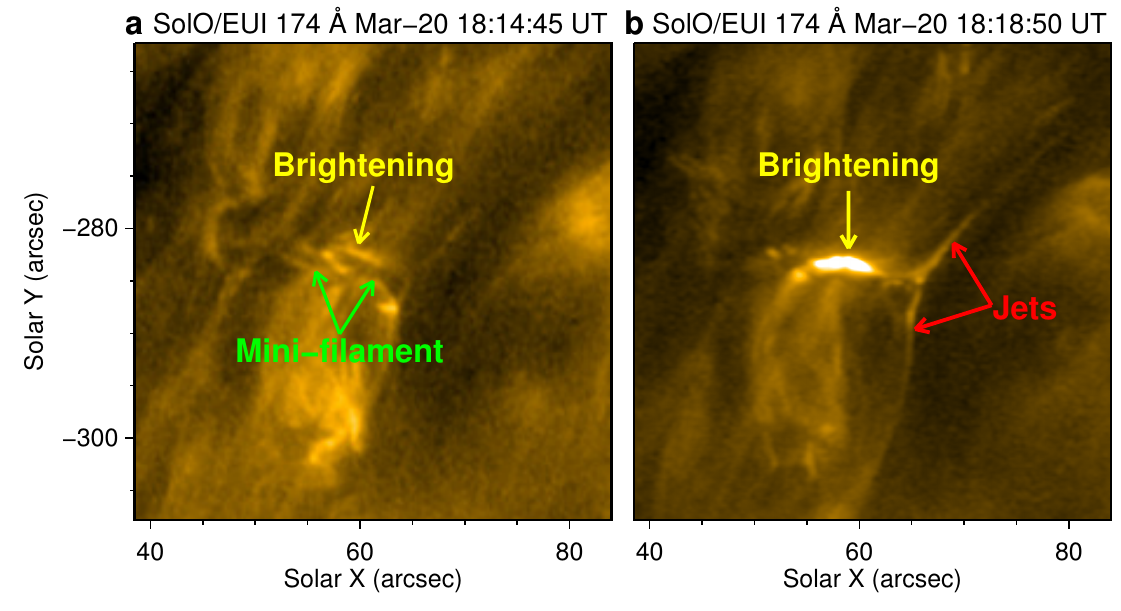}
\caption{EUV 174 {\AA} observations by the Solar Orbiter Extreme Ultraviolet Imager (EUI) of two-sided-loop jets produced by the eruption of a mini-filament in a filament channel.}
\label{fig21}
\end{figure}

For straight anemone jets, a fan--spine magnetic topology has been revealed by recent high-resolution observations combined with 3D coronal magnetic field extrapolations. In the fan--spine topology, a dome-shaped fan corresponds to the closed separatrix surface dividing two domains with different connectivity, and the inner and outer spines meet at a null point \citep{Masson2009, HouYJ2016a, LiT2018, HouYJ2019, ChengX2023, DuanYD2024}. The outer spine can be open or connected to a remote location on the solar surface. The fan--spine topology usually arises when a new dipole emerges into a pre-existing unipolar magnetic field. The ensuing  null-point magnetic reconnection can accelerate plasmas, forming straight anemone jets along the outer spine. On the other hand, two-sided-loop jets are observed to be associated with mini-filament eruptions inside the base arches caused by magnetic flux cancellations (Figure \ref{fig21}), and material is ejected from below into an overlying large filament via magnetic reconnection \citep{WeiHY2020, YanXL2021}. A third type of jets  has also been proposed: blowout jets \citep{Sterling2015}, corresponding to mini-filament eruptions.

\subsection{Physical models}\label{sec2_{2}}
In early studies, an emerging-reconnection model was proposed to explain the formation of straight anemone jets and two-sided-loop jets \citep{Heyvaerts1977, Shibata2007}, as well as the onset of CMEs \citep{ChenPF2000}. In this scenario, hot X-ray jets are first accelerated by the strong tension force of the reconnected magnetic field lines and then by the thermal pressure gradient force after the jets collide with the ambient magnetic field to form a shock. And the simultaneous cool H$\alpha$ surge is formed by the cold chromospheric plasma that is carried up by the emerging flux and then accelerated by slingshot effect of the reconnected field lines \citep{Shibata1992, Yokoyama1995, Yokoyama1996}. On the other hand, an alternative model should be responsible for blowout jets, which are usually accompanied by the eruption of mini-filaments and magnetic flux cancellations \citep{Moore2010, Sterling2015}. 
They were proposed to be mini-filament eruptions, which share common characteristics and thus could be the miniature version of large-scale solar eruptions \citep{Sterling2015}. In this context, with numerical simulations \cite{Wyper2017} found that external reconnection between the confining field above a twisted mini-filament and the ambient open field can remove the confining field, allowing the mini-filament to rise. Subsequent internal reconnection underneath the rising mini-filament eventually leads to the violent eruption of the system and the formation of a rotating jet. This positive-feedback mechanism between filament ejection and reconnection was classified as the traditional  ``magnetic breakout model", which was believed to occur on various scales \citep{ChenY2016, Wyper2018, Kumar2018, HongJC2019}.

\subsection{UV Observations of Jets within Sunspots}\label{sec2_{3}}

UV observations have improved our understanding of solar jets over the past decade. In particular, the community has discovered numerous jet-like activities even in sunspot regions with a strong magnetic field that appeared relatively quiet in the visible waveband. It is crucial to understand the driving mechanisms of solar jets in strongly magnetized environments as well as the evolution of magnetic field and plasma in the solar atmosphere above sunspots.

\begin{figure}[h]
\centering
\includegraphics[width=0.8\textwidth]{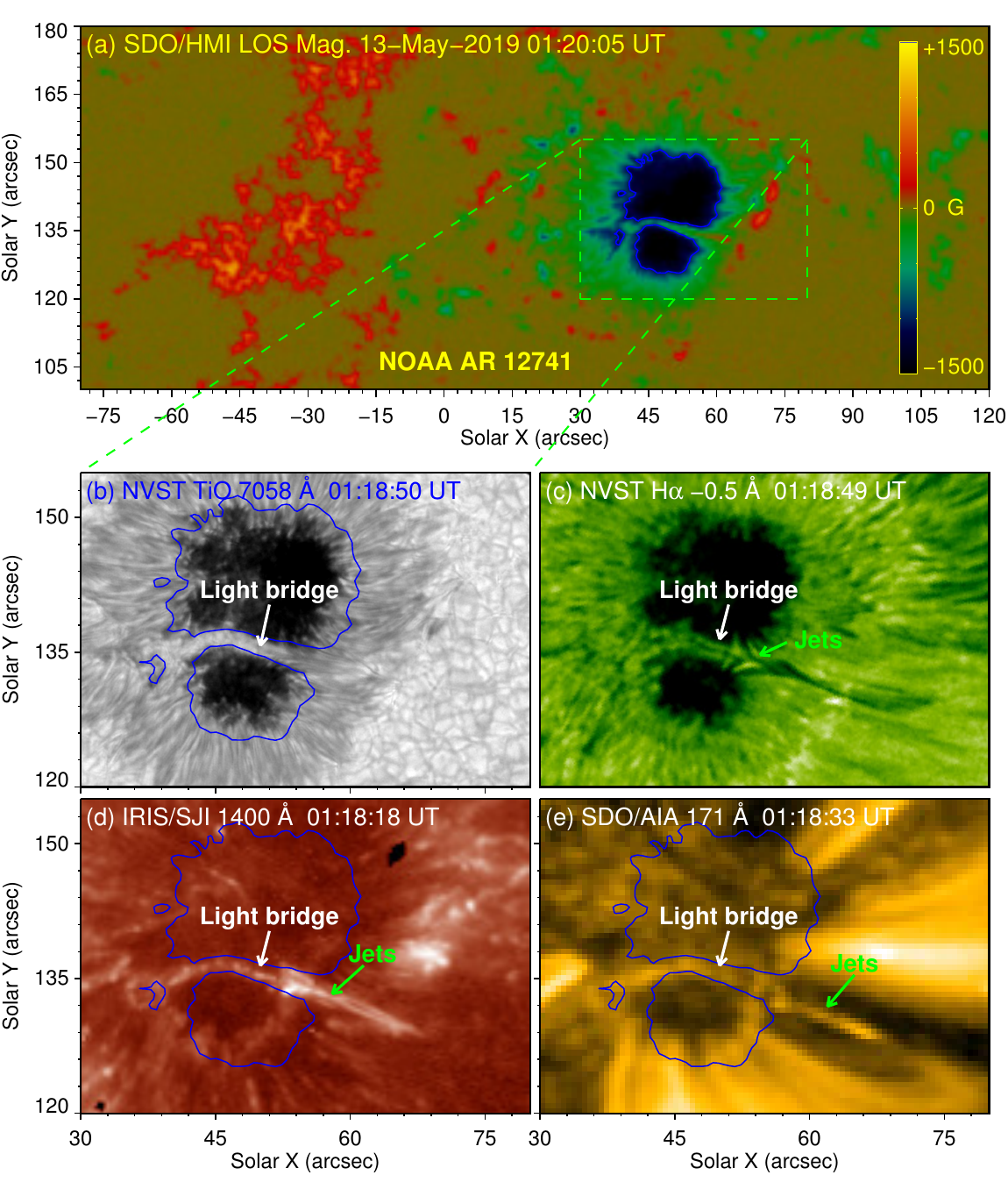}
\caption{Multi-wavelength observations of jets emanating from the sunspot light bridge of AR 12741 on 2019 May 13. Panels (a) and (b) display the sunspot light bridge. Panels (c)--(e) show the jets emanating from the west part of the light bridge within the main sunspot. The blue curves in panel (a) are contours of the line-of-sight magnetogram at -1000 G and are duplicated to panels (b), (d), and (e). Adapted from \citet{HouYJ2020}.}
\label{fig22}
\end{figure}

Sunspots are manifestations of solar magnetic field concentrations, which appear dark in solar photospheric observations because the strong magnetic fields inhibit the normal convective transport of energy from the solar interior to the surface \citep{Solanki2003}. A typical sunspot is characterized by its dark cooler region, i.e., an umbra, surrounded by a less dark filamentary penumbra. 
The sunspot magnetic fields are believed to be almost vertical at the umbra and more inclined in the surrounding penumbra \citep{Thomas2002}. Inside sunspots, some substructures are sometimes detected, for example umbral dots, light bridges, and penumbral filaments. These structures are caused by the interaction between magnetic field and convection. As a result, various  jet-like activities frequently occur; for example surges above light bridges, transient jets above a penumbral filament intrusion into the umbra, and ubiquitous microjets in the penumbral chromosphere.

\begin{figure}[h]
\centering
\includegraphics[width=0.8\textwidth]{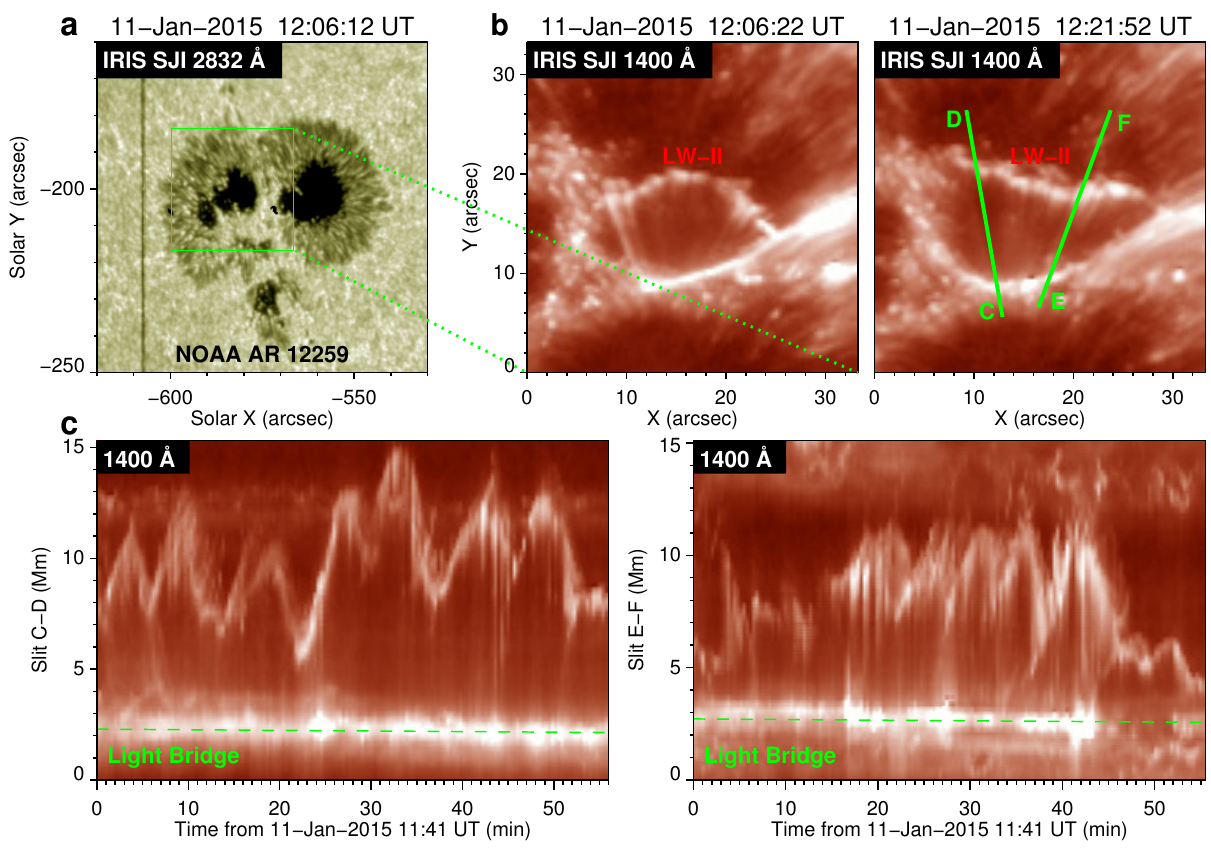}
\caption{UV 1400 {\AA} observations of a type-II light wall above the sunspot light bridge of AR 12259 on 2015 January 11. Adapted from \citet{HouYJ2022}.}
\label{fig23}
\end{figure}

UV imaging and spectroscopic observations with unprecedentedly high resolution from the SDO and IRIS missions, greatly enhanced our knowledge of jet-like activities within sunspots. It has been firmly established that there are two types of jet-like activities within sunspots: one type is oscillatory in nature produced by upward shocked $p$-mode waves, such as bright wall-shaped surges (also named as light walls) above sunspot light bridges \citep{YangSH2015, HouYJ2016b, ZhangJW2017, HouYJ2017}; another type is produced by magnetic reconnection, like transient jets (Figure \ref{fig22}) above light bridges \citep{Toriumi2015, TianH2018, HouYJ2020, LiFY2021}. The most prominent feature of light walls driven by $p$-mode waves is their coherent oscillating bright fronts with almost stationary period of a few minutes in the 1330/1400 {\AA} channels. In addition, there are usually no obvious brightenings around the light wall base. In contrast, the transient jets closely related to reconnection have much larger heights and no consistent periods. Their recurrent base brightenings show significant broadenings and enhancements at both wings in UV line profiles, which strongly indicate the occurrence of magnetic reconnection.

\citet{HouYJ2022} further reported a new type of jet-like activity above sunspot light bridges: type-II light walls, which share a similar wall-like shape with the type-I light walls but show a distinctly different broken or jagged front with a height of $>$10 Mm, a rising velocity of $>$50 km s$^{-1}$, no stable period, and are accompanied by recurrent base brightenings in the entire light bridge (Figure \ref{fig23}). The authors suggested that type-II light walls are more likely to occur in a dynamically evolving sunspot light bridge, where magnetic reconnection driven by vigorous magneto-convection can frequently take place along the whole light bridge. Based on UV observations, these studies have revealed a highly dynamical scenario of jet-like activities within sunspots, which are driven by different basic physical processes, including magneto-convection, magnetic reconnection, and wave leakage.

\section{UV/EUV Observations of Solar Flares}\label{sec3}

UV/EUV observations provide essential diagnostics for unraveling the fundamental physics of solar flares and for assessing their space weather impacts. The UV/EUV emission covers temperatures from 10$^4$ K to 10$^7$ K and is invaluable for probing energy release, plasma heating, and energy transport during flares. Moreover, the huge surge of UV/EUV radiation during flares can significantly disturb Earth's ionosphere, leading to rapid increases in electron density that cause sudden ionospheric disturbances and subsequent disruptions of short-wave radio communications. Consequently, studies of UV/EUV flare emissions are of considerable importance to understand and predict space weather effects. Here, we briefly review how UV/EUV observations have improved our understanding of solar flares, especially their role in diagnosing fundamental physical processes. More comprehensive reviews on solar flares can be found in \citet{Priest2002}, \citet{Fletcher2011}, \citet{Holman2011}, \citet{Shibata2011}, \citet{Milligan2015}, \citet{Janvier2017}, \citet{Benz2017}, and \citet{Fletcher2024}.

\subsection{Observational manifestations}

\subsubsection{Magnetic reconnection}

Magnetic reconnection is widely regarded as the primary process responsible for energy release in solar flares. Therefore, identifying its observational signatures is essential for understanding how energy is released and how flares evolve. The observational manifestations of magnetic reconnection can be broadly classified into the following categories: (1) reconfiguration of coronal loops; (2) current-sheet and cusp-shaped structures; (3) plasma inflows and outflows near the current sheet; (4) supra-arcade downflows; and (5) the evolution of flare ribbons and post-flare loops. UV/EUV observations, which are sensitive to plasma across a wide temperature range, are particularly well suited for detecting these reconnection-related signatures. 

\begin{figure}[ht!]
\centering
\includegraphics[width=0.75\textwidth]{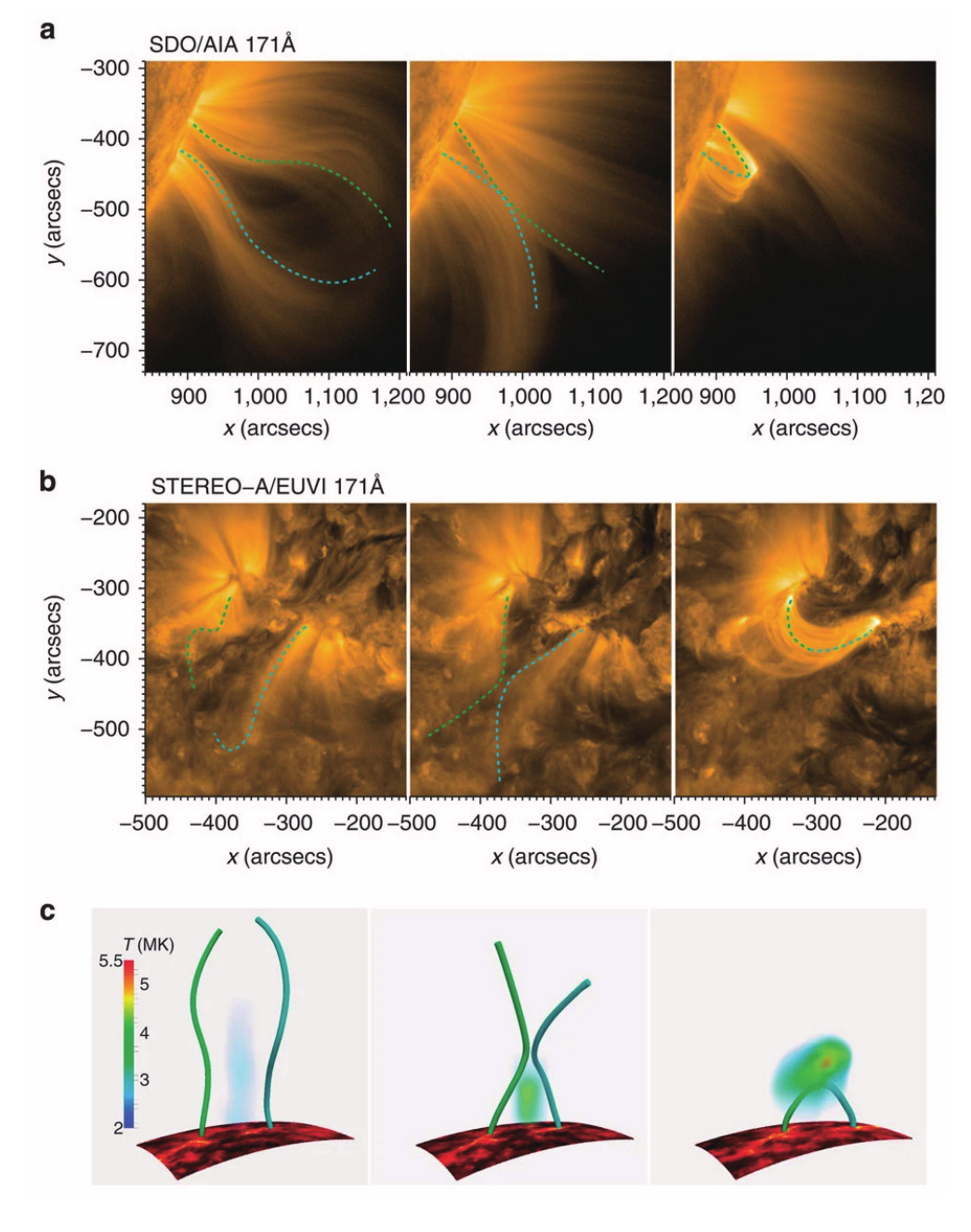}
\caption{Stereoscopic observations of magnetic reconnection in a solar flare from (a) SDO and (b) STEREO--A. Panel (c) displays the change in coronal loop connectivity, with semitransparent contours indicating the temperature evolution during magnetic reconnection. Modified from \citet{Sun2015}.}
\label{fig31}
\end{figure}

The most direct evidence for magnetic reconnection is the change in the connectivity of coronal loops. Using high-cadence SDO/AIA observations, \citet{Su2013} identified inflowing cool loops together with newly formed hot loops, which is consistent with the standard flare model. Using stereoscopic observations from STEREO--A/EUVI and SDO/AIA, as illustrated in Figure~\ref{fig31}, \citet{Sun2015} observed two groups of oppositely directed field lines stretching, approaching each other, and reconnecting to form flare loops, thereby heating the plasma near the reconnection site from $\sim$1 MK to above $\sim$5 MK.

\begin{figure}[ht!]
\centering
\includegraphics[width=0.75\textwidth]{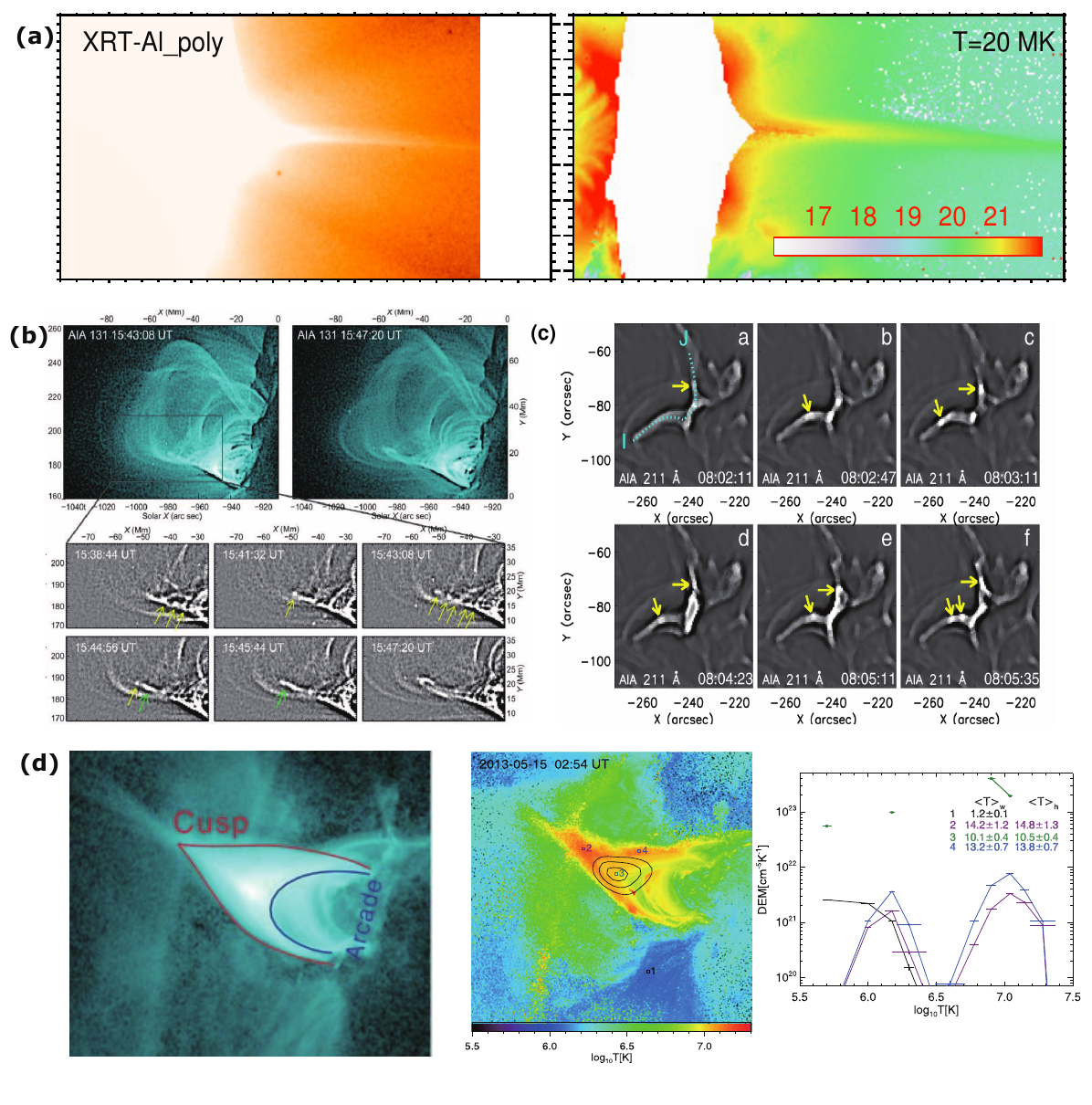}
\caption{Current sheets (Panels a--c) and cusp-shaped structures (Panel d) observed during solar flares. Panel (a): X-ray emission and high-temperature plasma distribution associated with an elongated current-sheet-like structure. The left image shows the structure observed by Hinode/XRT, while the right image presents the DEM temperature distribution at approximately 20 MK. Panels (b, c): Temporal evolution of plasmoids in a reconnecting current sheet and associated separatrices observed in the EUV 131 and 211 \AA\ channels, respectively. Panel (d) Morphology and thermal properties of a cusp-shaped flare structure. The left panel shows the cusp-shaped structure and the underlying flare arcades. The middle panel presents the DEM-derived temperature distribution and the location of a hard X-ray source at 12--25 keV. The right panel shows the DEM profiles at selected locations marked in the middle panel. Adapted from \citet{Cheng2018}, \citet{Yan2022}, \citet{Gou2019} and \citet{Gou2015}.}
\label{fig32}
\end{figure}

During changes in coronal loop connectivity, certain structures indicative of magnetic reconnection are often visible in EUV observations, such as the high-temperature current sheet and the underlying cusp-shaped structure (Figure~\ref{fig32}). For example, \citet{Liu2010} observed a vertical, bright sheet-like structure in the wake of a CME, which was thought to be a current sheet. Subsequent high-resolution observations further revealed more fine structures of the current sheet. \citet{Cheng2018} identified reconnection occurring within a large-scale current sheet and \citet{Yan2022} detected plasmoids within a current sheet in a confined flare. \citet{Gou2019} observed the merging of plasmoids, corresponding to mini flux ropes, within a vertical current sheet that developed before the CME impulsive phase. In particular, the current sheets reported by \citet{Cheng2018} and \citet{Yan2022} were formed during eruption, whereas the current sheet observed by \citet{Gou2019} was formed before eruption. In addition, cusp-shaped structures were also observed below the current sheet, and their temperature increases with height \citep{Gou2015}, favoring the reconnection model.

In addition to the connectivity changes in magnetic field lines participating in reconnection, inflows and outflows near the current sheet also constitute key evidence for magnetic reconnection and can be captured by both imaging and spectroscopic observations (Figure~\ref{fig32}). For example, using SOHO/EIT imaging observations, \citet{Yokoyama2001} reported inflows that persisted for nearly two hours during a flare. Such behavior was reproduced by numerical simulations, although the observed inflow speed is generally much smaller  than the real inflow speed \citep{chen04}. \citet{Cheng2018} observed inflows and outflows within the current sheet formed in the wake of a CME. Moreover, they found that sunward outflow jets from a fragmenting current sheet exhibit power-law distributions in the time variations of both intensity and velocity, indicating the turbulent nature of magnetic reconnection. 

Beyond imaging observations, spectroscopic observations also provided diagnostics for flare reconnection. Based on SOHO/SUMER observations, a significant enhancement of the blue-wing of the Fe~\textsc{xxi} line was detected at the flare-loop top, indicating the presence of high-temperature ($\sim$10~MK) and upward-directed outflows caused by magnetic reconnection \citep{Innes2003b}. \citet{Wang2007} reported both blueshifts and redshifts with velocities of several hundred km~s$^{-1}$ in the Fe~\textsc{xix} line, corresponding to bi-directional flows associated with magnetic reconnection. Using the UV spectra from IRIS, \citet{Tian2014} found high-speed redshifts and pronounced non-thermal broadening in the Fe~\textsc{xxi} line at the loop-top hard X-ray source, corresponding to downward reconnection outflows. 
\citet{Li2017} found that the Si~\textsc{iv} line exhibited broadened wings that extended up to $\sim$200 km~s$^{-1}$, while the line core remained stationary, suggesting bidirectional reconnection flows. \citet{Hara2011} observed reconnection outflows with velocities of 200–400 km s$^{-1}$ in the Ca \textsc{xvii} 192.86~\AA\ and Fe~\textsc{xxiv} 192.03~\AA\ lines, together with inflows at speeds of $\sim$20 km~s$^{-1}$. Based on these measurements, the reconnection rate could  be estimated. In addition, the downward-moving dark, elongated, and finger-like structures were sometimes observed above flare loops. These features, known as supra-arcade downflows, are widely regarded as observational evidence of magnetic reconnection \citep{Mckenzie1999, Innes2003, Chen2017, Li2021, Tan2022, Tan2023, Samanta2021, Xie2025}.

\begin{figure}[ht!]
\centering
\includegraphics[width=0.6\textwidth]{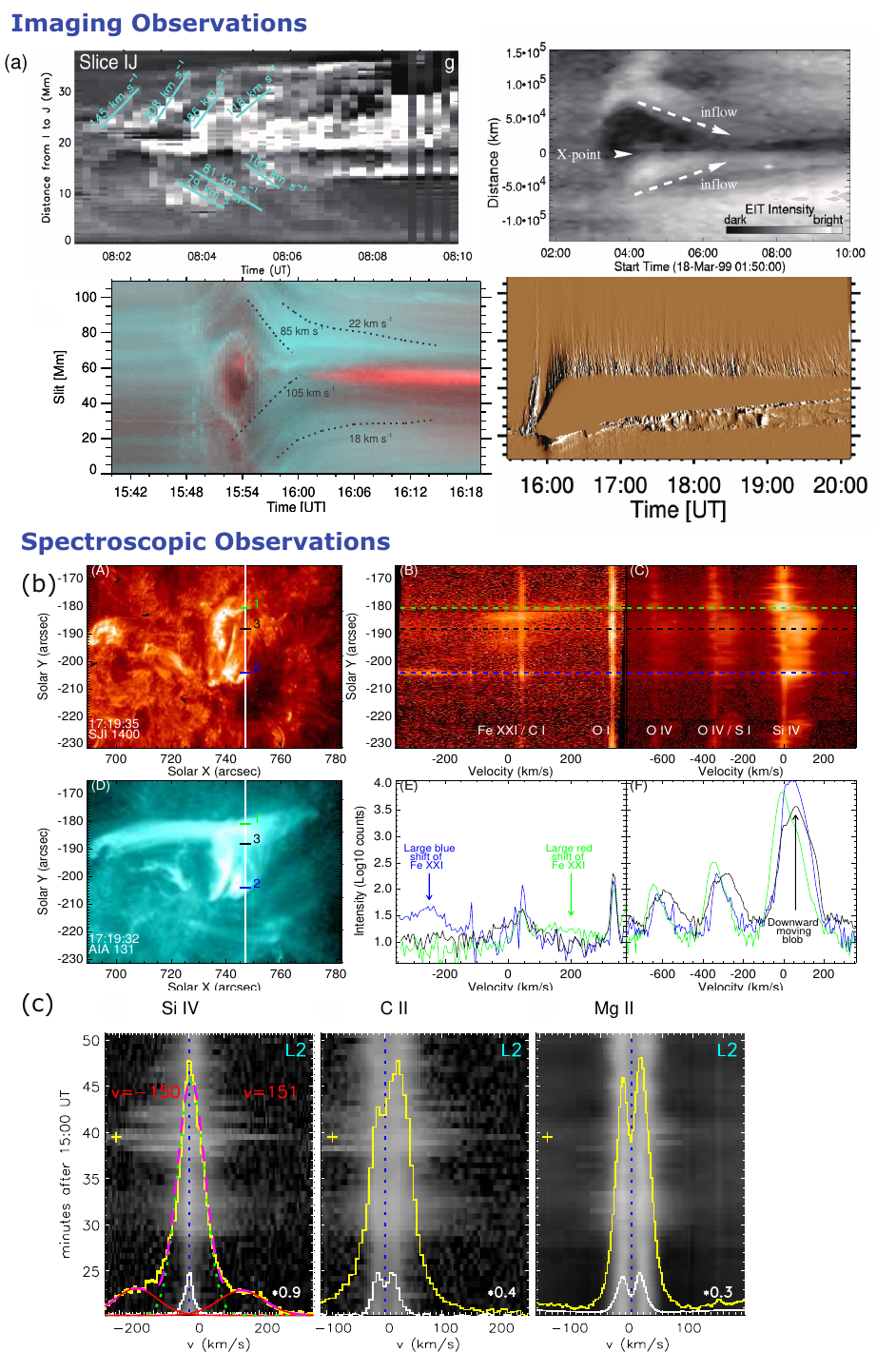}
\caption{Observational evidence of magnetic reconnection inflows and outflows during solar flares. Panel (a): Imaging observations of plasma inflows and outflows associated with magnetic reconnection. The upper-left subpanel shows the motions of multiple plasmoids along a reconnecting current sheet, while the upper-right and lower-left subpanels present the apparent convergence of plasma structures toward reconnection regions. The lower-right subpanel shows intermittent sunward-directed reconnection outflows above flare arcades. Panel (b): Spectroscopic observations of reconnection-related plasma motions from IRIS and SDO/AIA. The left subpanels show the observed flare structures, while the middle and right subpanels present spectral images and line profiles. Panel (c): Evolution of the Si \textsc{iv}, C \textsc{ii}, and Mg \textsc{ii} spectral profiles observed by IRIS. The line broadening and enhanced red- and blue-shifted components provide evidence of high-speed, bidirectional reconnection outflows. Adapted from \citet{Yokoyama2001}, \citet{Cheng2018}, \citet{Yan2022}, \citet{Tian2014}, and \citet{Li2017}.}
\label{fig33}
\end{figure}

UV/EUV observations also revealed the key characteristics of 3D magnetic reconnection, such as slipping reconnection \citep{Aulanier2007, Dudik2014, Li2014} and related flux ropes \citep{Aulanier2019}. Based on high-cadence IRIS observations, \citet{Lorincik2025} demonstrated that the apparent slipping speed of flare kernels can exceed the local Alfv\'en speed, a regime termed slip-running reconnection. For 3D reconnection geometries that involve eruptive flux ropes, several observational signatures have been identified. The 3D magnetic reconnection between an eruptive flux rope and surrounding sheared arcades ($ar$–$rf$ reconnection) is often manifested as the drifting of twin or core dimmings \citep{Aulanier2019, Lorincik2019, Dudik2019, Li2026}, the transformation of filament strands into flare loops \citep{Lorincik2019}, changes in filament footpoints \citep{Dudik2019, Guo2023}, lateral drifting of the erupting filament \citep{Guo2023}, the formation of saddle-like flare loops \citep{Lorincik2021}, cross-active region jets behind the eruptive filament \citep{Guo2025}, and the occurrence of multiple EUV peaks in two-ribbon flares \citep{Li2026}. In contrast, $rr$–$rf$ reconnection is typically evidenced by a decrease in the toroidal flux of the CME flux rope inferred from core-dimming regions \citep{Xing2020, Jiang2021b, Li2026} and by the appearance of prominent supra-arcade downflows resulting from leg–leg reconnection within an erupting filament \citep{Dudik2022}. For a more detailed discussion of the theory and observational manifestations of 3D magnetic reconnection, see \citet{Dudik2025}.

Although magnetic reconnection is the dominant process in solar flares, it is not expected to proceed persistently, as indicated by the limited spatial coverage. 
A key product of simulations are quasi-separatrix layers (QSLs), which are regions where there is a drastic change in field-line topology.
\citet{Chen2012} revealed that in addition to the inner QSLs, which are related to the flare ribbons, there are also  outer QSLs. They proposed that when the flare ribbons approach the outer QSLs, the flare ribbons stop. This model was later validated by data-driven MHD simulations \citep{Guo2023a}. The observed flare ribbons are cospatial with the inner QSLs computed from the MHD simulation, and both terminate at the outer QSLs. This model provides a possible way to predict when and where magnetic reconnection will stop.

\subsubsection{Responses of the lower solar atmosphere}

During solar flares, a large amount of magnetic energy is rapidly released in the corona through magnetic reconnection and almost half is generally thought to be transported downward to the lower solar atmosphere predominantly by waves, non‑thermal electron beams and thermal conduction, thereby driving chromospheric evaporation and condensation \citep{Fisher1985, Milligan2015}. The chromosphere and transition region are now widely recognized not only as major contributors to flare radiation, but also as the mass reservoir that supplies the coronal plasma. Here, we focus on how UV/EUV observations, in particular spectroscopic measurements, have shaped our current understanding of energy transport from the corona to the lower solar atmosphere during flares.

Doppler-shift diagnostics based on spectral lines formed at different atmospheric heights and temperatures play a crucial role in understanding chromospheric evaporation and condensation. Chromospheric evaporation is commonly classified into gentle and explosive regimes. In gentle evaporation, cool lines formed in the chromosphere and transition region usually show small blueshifts corresponding to modest upflows, whereas signatures of downward motion are weak or absent. By contrast, explosive evaporation is characterized by simultaneous red shifts in cool lines, caused by chromospheric condensation due to momentum balance, associated with blueshifts in hot coronal lines \citep{Canfield1999}. Generally, gentle evaporation is more often associated with weaker energy deposition during the pre-flare or gradual phases of the flare and is commonly attributed to thermal conduction and/or relatively weak non-thermal electron heating. In contrast, explosive evaporation typically occurs during the impulsive phase and is closely related to stronger electron-beam heating. Observations have provided strong support for this picture. 

Earlier observations showed puzzling features such as 
asymmetric profiles in flare lines which did not agree 
with theory. The problem was that those observations were not spatially resolved. The  Coronal Diagnostic
Spectrometer (CDS) on board the Solar
and Heliospheric Observatory (SOHO) provided such observations for the first time. An earlier sit-and-stare observation reported by \cite{brosius:2003} showed blue-shifts
during the evaporation at all temperatures, which was not expected.
The first spatially-resolved CDS observation that agrees with theory was presented by \cite{delzanna_etal:2006}: the lower-temperature lines 
were red-shifted and the upflows of the coronal lines were increasing with temperature. Furthermore, the Fe \textsc{xix} flare line
had a large non-thermal width during the impulsive phase, decreasing with time together with the upflows. Finally,
the flare lines were symmetric. A 1D hydrodynamic simulation indicated agreement with observations.

Several authors later presented similar results using CDS data  \citep[cf.][]{Milligan2006}, but also Hinode/EIS and IRIS  \citep[cf.][]{polito_etal:2015,polito_etal:2016}, which have a much better spatial and spectral resolution than CDS.
\citet{Tian2015} reported upflows of about 200 km s$^{-1}$ in the IRIS Fe \textsc{xxi} 1354.08 \AA\ line and found that they correlate with the energy deposition rate inferred from hard X-ray fluxes, implying a close correlation with non-thermal electron beams. Recently, \citet{Ashfield2025} showed that the period of quasi-periodic pulsations in the 25--50 keV hard X-ray light curve agrees well with the period of chromospheric condensation inferred from the Si \textsc{iv} and Mg \textsc{ii} lines, indicating that periodically accelerated electrons can directly modulate the chromospheric response.

One important issue concerns the transition temperature, or flow-reversal temperature between redshifts and blueshifts. In the classical picture, this temperature was expected to lie around 1 MK or lower \citep{Fisher1985}, a prediction later supported by the observations of \citet{Kamio2005} and \cite{delzanna_etal:2006}. However, subsequent measurements indicated a more complex scenario. Using Hinode/EIS observations, \citet{Milligan2009} found that the transition temperature can reach $\sim$1.5 MK, and \citet{Li2011} further reported redshifts in lines formed between 2.5 and 5.0 MK. These results imply that chromospheric condensation may extend to substantially higher temperatures than previously anticipated, suggesting that both the location and extent of energy deposition and the subsequent response of the atmosphere may be more intricate than envisioned in the standard one-dimensional picture. In addition to non‑thermal electron beams and thermal conduction, Alfv\'en waves were also proposed to be able to transport energy down to the lower atmosphere \citep{Fletcher2008, Russell2024}.

\subsubsection{MHD waves during solar flares} 

MHD waves play an important role in solar flares because they are intimately linked to the release and transport of energy within the flaring atmosphere. These waves can be excited impulsively by magnetic reconnection, by pressure perturbations associated with flare heating, or by the rapid magnetic restructuring. In turn, MHD waves may modulate magnetic reconnection, thereby contributing to the quasi-periodic pulsations of both thermal and non-thermal emissions observed in solar flares \citep{Nakariakov2009, McLaughlin2018} and stellar flares \citep{gao08}. Moreover, flare-related waves serve as valuable probes for coronal seismology, enabling key physical parameters of flare loops, such as the magnetic-field strength and plasma density, to be constrained. Here, we briefly review the diagnostic signatures of several types of MHD waves derived from UV/EUV observations, including kink oscillations, sausage oscillations, and Alfv\'en waves.

Kink oscillations in flare loops are of particular importance because they can offer a direct means of deriving the strength of the reconnected magnetic field. \citet{White2012} observed the higher order harmonic kink mode in hot flare loops based on imaging observations. \citet{Lid2018} found decayless oscillations in flaring loops using a combination of imaging and spectroscopic observations, and inferred  magnetic-field strengths of 12--170 G, similar to the results obtained by magnetic extrapolation.

Fast sausage waves are the axisymmetric compressive mode of magnetized loops, wherein periodic variations in the loop cross section induce quasi-periodic perturbations in plasma density, magnetic field strength, and consequently emissivity. In contrast to the typically longer periods of kink modes, sausage modes are widely regarded as a leading MHD candidate for explaining short‑period quasi-periodic pulsations, particularly in dense and comparatively thick flare loops. A representative spectroscopic detection of a sausage oscillation in a flare was reported by \citet{Tian2016}, who observed that the intensity and Doppler shift of Fe~\textsc{xxi} 1354.08 \AA\ exhibit the same short period of $\sim$25 s, with a quarter-period phase shift between the two observables, a pattern strongly indicative of a sausage‑mode oscillation. This finding not only supports the interpretation of short-period quasi-periodic pulsations as sausage oscillations, but also allows the measured phase speed to place a lower limit on the Alfv\'en speed outside the flare loop.

As mentioned in the previous subsection, Alfv\'en waves in solar flares are important because they may transport energy from coronal reconnection regions to the lower atmosphere, and may also contribute to particle acceleration and turbulent dissipation \citep{Fletcher2008, Russell2024}. However, they are difficult to detect directly, because pure linear torsional Alfv\'en waves are largely incompressible and thus produce little intensity variation or resolved transverse displacement. Consequently, current evidence is mainly spectroscopic, especially in the form of non-thermal line broadening observed in current sheets, loop-top sources, and flare ribbons \citep{Fletcher2008, Jeffrey2018, Shen2023}. \citet{Russell2024} emphasized that the widespread non-thermal line broadening observed in flares provides one of the main observational foundations for inferring the presence of Alfv\'en fluctuations. \citet{Jeffrey2018} found that non-thermal broadening in Si \textsc{iv}, interpreted as lower-atmosphere Alfv\'enic fluctuations or turbulence, occurs before the flare brightening and likely contributes directly to heating the flare transition region. This result suggests that Alfv\'en turbulence may play an active role in heating the lower atmosphere during solar flares. However, these signatures are still indirect evidence, and direct evidence needs to be sought in future studies.

\subsection{EUV late phase of solar flares}

The EUV late phase refers to a secondary enhancement of warm coronal EUV emission (e.g., Fe \textsc{xvi} 335 \AA, corresponding to $\sim$3 MK) that occurs tens of minutes to several hours after the main phase of a solar flare. This phenomenon was first reported by \citet{Woods2011} using observations from the EUV Variability Experiment (EVE) onboard SDO. Statistical analyses indicated that the late phase exhibits an average onset delay of 19 minutes, a peak‑to‑peak delay of 88 minutes, and a mean duration of 93 minutes \citep{Harry2025}. In certain events, the intensity of the late-phase peak can even exceed that of the main‑phase peak, forming an ``extremely large EUV late phase".

Two principal mechanisms have been proposed to account for the EUV late phase: (1) the secondary-heating scenario \citep{Woods2011, Hock2012, Dai2013, Sun2013} and (2) the long-loop cooling scenario \citep{Liu2013, Liy2014, Masson2017, Dai2018b}. In the secondary heating scenario,  an additional energy release occurs well after the primary flare. Within a complex quadrupolar magnetic configuration, reconnection may occur not only in the current sheet beneath the erupting flux rope, but also between null-point field lines or between the flux rope and overlying null-point field lines. Such multi-stage reconnection processes can produce multiple peaks in the EUV emission profile beyond the main-phase peak. In contrast, the long-loop cooling scenario is based on the principle that longer coronal loops cool more slowly. If an eruption involves multiple sets of flare loops with differing lengths, their distinct cooling timescales can result in successive temperature crossings through the typical warm coronal range, thereby yielding a delayed EUV enhancement. \citet{Li2024} demonstrated that cross-sectional expansion of flare loops can substantially prolong the cooling timescale, providing a natural explanation for the extended peak-to-peak delays observed in some EUV late-phase flares.

To distinguish between these mechanisms, \citet{Dai2018} performed numerical simulations and showed that light curves characterized by a slow rise followed by a rapid decay are consistent with the long-loop cooling scenario, whereas those exhibiting a rapid rise and slow decay are more likely produced by secondary heating. \citet{Chen2020} examined the magnetic structures and ribbon morphologies of extremely large EUV late-phase flares and found that about half of the events display a two-ribbon morphology, while the remainder exhibit circular or more complex ribbon structures. To explain the multiple peaks observed in two‑ribbon flares, \citet{Li2026} combined observations with data‑constrained MHD simulations and suggested that the longer cooling timescales of flare loops formed via magnetic reconnection involving the flux rope ($ar$–$rf$ and $rr$–$rf$ reconnection), relative to the shorter loops produced by reconnection in the overlying field lines ($aa$–$rf$ reconnection), are responsible for the origin of multiple EUV peaks in such events.

\section{EUV Dimmings}\label{sec4}

\subsection{Formation of coronal dimmings}
Coronal dimmings are transient phenomena commonly observed on the Sun, manifested as localized and abrupt decreases in coronal emission—primarily at EUV and soft X-ray (SXR) wavelengths \citep{Veronig2025}. These phenomena typically occur during the lift-off and early expansion phases of CME events \citep{Hudson1996, Thompson1998, chen11}. Extensive high-cadence and full-disk imaging observations provided by EUV telescopes have allowed for detailed studies of the properties of coronal dimmings. Simultaneous observations of EUV telescopes and coronagraphs have robustly established the connection between coronal dimmings and CMEs. 

The reduced coronal emission during dimmings is attributed to three key factors: mass loss driven by CME expansion and eruption, changes in plasma temperature, or obscuration by cool material suspended in the corona. At EUV wavelengths, mass loss-induced dimmings are prevalent. Unless explicitly stated otherwise, ``coronal dimmings" in this chapter refer specifically to dimmings caused by transient mass loss from the lower corona.

\begin{figure}[ht!]
\centering
\includegraphics[width=0.8\textwidth, trim=0.5 0 2cm 0cm, clip]{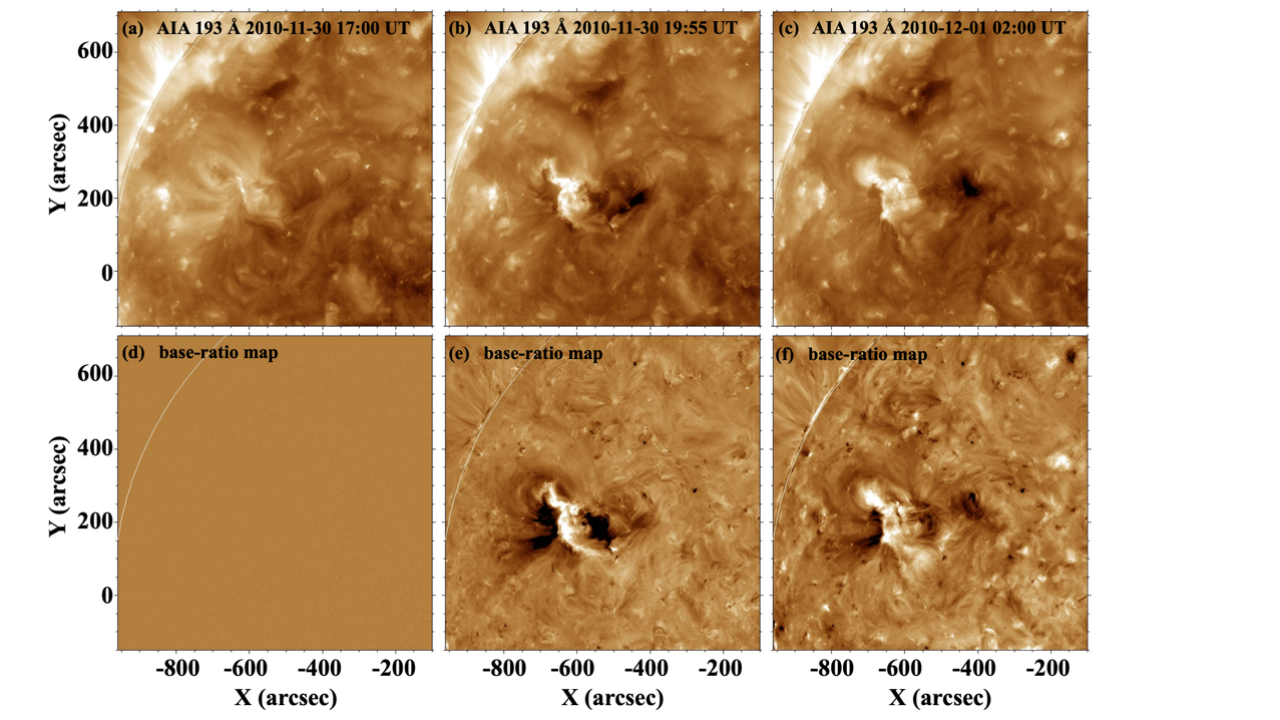}
\caption{Evolution of one dimming event in SDO/AIA 193 \AA\ images. Panels (a) and (d) show a pre-eruption frame. Panels (b) and (e) correspond to the intensity minimum frame. Panels (c) and (f) are images of a frame at the recovery phase.}
\label{fig41}
\end{figure}

Traditionally, coronal dimmings are categorized into the following two types based on their morphological features:

\begin{enumerate}
\item Core (or twin) dimmings: these are localized regions exhibiting a substantial reduction in emission (up to $\sim$90\%), often observed as paired structures on both sides of the magnetic polarity inversion line, corresponding to the hook-like ends of J-shaped flare ribbons. They were interpreted as markers of the footpoints of expanding flux ropes, indicating the opening of magnetic field lines and subsequent loss of coronal mass \citep{Sterling1997, Webb2000}. The dimming regions may exhibit either a shrinking behavior or a quasi-stationary evolution pattern, depending on whether the legs of the flux rope undergo mutual reconnection. Additionally, footpoint drift of a flux rope—resulting from reconnection with the surrounding coronal arcade—can also cause rapid shrinkage \citep{Chen2019} or movement of dimming areas \citep{Gou2023};

 \item Secondary dimmings: these are more extended but less dark regions that correspond to the CME cavity \citep{chenpf05}. They are thought to arise from the stretching of the overlying strapping field during the eruption and/or from reconnection of the erupting magnetic structure with the ambient coronal flux \citep{Thompson2000, Mandrini2007}. Such dimmings can form away from the polarity inversion line or at remote locations.
 \end{enumerate}

Coronal dimmings offer extensive and multifaceted diagnostic potential, largely due to their presence throughout the entire CME evolution—from the pre-eruption phase, through the main eruption, and to the long-duration post-eruption recovery of the corona (Figure~\ref{fig41}). This persistence allows us to explore critical aspects of CME physics, including triggering mechanisms, mass loss, large-scale changes in magnetic connectivity, and post-eruption  replenishment processes. Dimmings could also be used to determine the early propagation direction of CMEs, and are therefore valuable for improving the accuracy of space weather forecasting \citep{Jain2024}. Furthermore, dimmings establish a crucial link between the near-Sun properties of eruptions and their counterparts, interplanetary CMEs (ICMEs). Here we provide a brief overview of coronal dimmings; for a comprehensive review, see \citet{Veronig2025}.

\subsection{Observational characteristics of coronal dimmings}
Coronal dimmings are inherently coupled to the evolution of their associated CMEs and solar flares. The evolution of a typical dimming may be divided into three distinct phases: pre-eruption phase, main phase, and recovery phase.

\subsubsection{Pre-eruption Phase}
In the case of some solar eruptive events, a weak, gradual dimming is observed prior to the main eruption. These pre-eruption dimming signatures may appear tens of minutes to hours before the main eruption. They were interpreted as evidence for the slow expansion of the pre-eruptive structure, such as a flux rope, before its catastrophic loss of equilibrium. These dimmings are often observed as conjugate pairs in regions that later become the core dimmings of the main eruption, mapping the feet of the slowly expanding magnetic structure \citep{Qiu2017, Zhang2017}. The presence of strong, non-neutralized vertical electric currents in these preeruption dimming regions supports the scenario of an expanding, current-carrying flux rope \citep{Wang2023}. While their weak signals make observational detection challenging, pre-eruption dimmings provide valuable insights into the magnetic configuration of CME progenitors and the early stages of CME initiation. It is worth mentioning that in some eruptive events such pre-eruption dimmings are absent or not easily identified.

\subsubsection{Main Phase}
The onset of the dimming is impulsive, coinciding with the CME lift-off and the start of the flare’s impulsive phase. During this phase, the intensity in the dimming region decreases rapidly (by up to $\sim$90\% for the core dimming regions at AIA 193 \AA\ and 211 \AA\ wavelengths), and the dimming area expands. The intensity minimum is typically reached after the peak of the associated X-ray flare.

Early multi-wavelength observations revealed that dimmings appear simultaneously in EUV filters sensitive to a broad temperature range ($\sim$1--3 MK) and in SXRs ($>$3 MK). 
This  indicates that the  decrease in emission is dominated by density depletion (plasma evacuation) rather than temperature changes \citep{Thompson1998}, supported by the fact that radiative and conductive cooling timescales in the corona are too long to explain the rapid dimming \citep{Hudson1996}.

Systematic studies using SDO/AIA data confirmed that dimmings are multi-thermal but most pronounced in the 193 \AA\ and 211 \AA\ passbands (peaking at $\sim$1--2 MK), suggesting that the ejected mass originates primarily from the local corona \citep{Dissauer2018}. This finding was further validated by differential emission measure (DEM) analyses based on EUV spectroscopic observations, which demonstrated that the dimmings were caused by  reduced emission measure at typical coronal temperatures (1--2 MK) \citep{Tian2012}. DEM curves measured before and during dimming events showed minimal changes in peak temperature, ruling out temperature changes as a primary cause. Deep dimmings have occasionally been observed in hotter channels (e.g., AIA 94 \AA), indicating that hot active region loops can also be ejected \citep{Vanninathan2018}. Dimmings have even been detected in chromospheric/transition region lines (e.g., H$\alpha$ 6562.8 \AA, He \textsc{ii} 304 \AA, and Ne \textsc{vii} 465 \AA) in some events, demonstrating that lower atmospheric plasma contributes to CME mass loss \citep{Jiang2003, Xu2022}.

Statistical analyses of dimming parameters showed that dimming areas range from $10^{9}$ to $10^{11}$ km$^{2}$, with underlying magnetic fluxes between $10^{19}$ and $10^{22}$ Mx \citep{Dissauer2018}. The impulsive growth phase of dimmings typically lasts 10--120 minutes, consistent with the timescale of the CME’s main acceleration phase. The total lifetime of dimmings (from onset to full recovery) averages $\sim$7--8 hours \citep{Reinard2008, Veronig2021}. 

Spectroscopic observations provided clear evidence for plasma motions within dimming regions. Hinode/EIS detected persistent outflows (blue shifts) in dimming regions. These outflows were found in both transition region and coronal lines, with velocities on the order of 20 km s$^{-1}$ from single-Gaussian fits \citep{Harra2007}. However, a more sophisticated analysis using red-blue asymmetry analysis and double-Gaussian fitting revealed a more complex picture: the spectra often consist of a nearly stationary background component and a secondary, high-speed outflow component with velocities of 50–150 km s$^{-1}$ \citep{Tian2012, Veronig2019}. This indicates that only a fraction of the plasma along the line of sight is flowing outward at high speeds. These outflows are interpreted as signatures of the plasma evacuation process and are similar to the coronal outflows observed at active region boundaries
\citep[cf.][]{delzanna_etal:2011_outflows}, suggesting that CME-induced dimming regions act as transient coronal holes and may transport mass to solar wind streams \citep{McIntosh2009, Tian2012}.

\subsubsection{Recovery phase}

After reaching minimum intensity, a more gradual and prolonged recovery phase begins, during which the dimming signature gradually fades as the corona is replenished with plasma. This phase can last from several hours to more than a day, with core dimmings typically requiring much longer (e.g., more than 10 hours) to recover than secondary dimmings \citep[$\sim$2 hours,][]{Vanninathan2018}.

Dimming recovery provides insights into post-eruption coronal reconfiguration and replenishment. Recovery occurs via two processes: shrinkage of the dimming area and/or brightness increases within the dimmed volume. Early studies using SXR observations of Yohkoh/SXT suggested that the recovery was driven mainly by area contraction \citep{Kahler2001}, while EUV observations (from SOHO/EIT and SDO/AIA) showed that both processes contribute \citep{Attrill2008, Dissauer2018}. This discrepancy is likely due to the differing temperature sensitivities of the instruments.
The physics behind the recovery is complex and two possible mechanisms have been proposed:

\begin{enumerate}
   \item  Plasma replenishment: mass supply from the lower atmosphere (e.g., lower transition region) in response to the pressure gradient created by the CME eruption \citep{Jin2009};

  \item Interchange reconnection: reconnection between the stretched magnetic field in the dimming region and surrounding small-scale closed loops or emerging flux, which disperses the open flux and facilitates recovery \citep{Attrill2008}.
\end{enumerate}

\subsubsection{Correlation between dimming and CME/flare parameters}

The onset of the dimmings is often synchronized with the start of the flare's impulsive phase and CME lift-off (typically within $\pm$5 minutes), highlighting the tight coupling between these phenomena \citep{Dissauer2018}. Statistical studies confirmed a high association rate between coronal dimmings and CMEs. For example, \citet{bewsher08} found that $\sim$84\% of CMEs are associated with either Fe or Mg line dimmings in their sample. Similarly, \citet{Veronig2021} found that 84\% of CMEs show obvious dimmings in the Sun-as-a-star EUV light curves, and that 97\% of the identified dimmings are linked to CMEs. It has also been found that CMEs associated with dimmings are generally faster, more energetic, and more likely to have flares than those without dimmings \citep{Reinard2009}.

Density diagnostics using density-sensitive line pairs at EUV wavelengths (e.g., Fe {\footnotesize XII} 186.88/195.12 \AA, Si {\footnotesize X} 356/347\AA) show 30--70\% decreases in electron density within dimming regions \citep{Harrison2000, Harrison2003, Tian2012}. CME mass loss can be estimated from dimmings by combining density decreases, dimming area, and assumptions about the depth of the emitting volume. A correlation coefficient of $\sim$0.7 was found between this mass loss and the associated CME mass estimated from coronagraphic observations. Estimates of mass loss typically account for 20--60\% of the CME mass derived from white-light coronagraphs—confirming that dimmings are a main source of CME  mass \citep{Jin2009, Tian2012}. This suggests that  the additional mass must originate from prominences, chromospheric mass replenishment, and/or plasma pile-up. The DEM method has also been applied to AIA images to infer the mass loss from dimming regions \citep{Lopez2019}.

Other dimming parameters were also found to correlate with the associated CME 
features. For example, moderate to strong correlations ($R$$\approx$0.6–-0.8) appear to exist between dimming parameters (area, magnetic flux) and CME mass \citep{Dissauer2019, Chikunova2020}. The growth rate of the dimming area, the rate of total unsigned magnetic flux, and the mean intensity in dimming regions show clear correlations with the maximum speed of the associated CME \citep{Dissauer2019}.

\citet{Dissauer2018} conducted a detailed study of the relationship between dimmings and flares. They found that the total area, magnetic flux, and minimum intensity of the dimmings show strong correlations  with the 1--8~\AA\ SXR flux (total energy). They also found correlations between the peak SXR flux and the rate of change of several dimming parameters such as the maximum growth rate of the dimming area, the maximum rate of magnetic flux in dimming regions, and the maximum rate of change in the dimmings. 

The magnetic flux inferred from secondary dimming regions is often comparable to the magnetic reconnection flux measured from flare ribbons, supporting standard eruption models where reconnection adds poloidal flux to the erupting flux rope \citep{Qiu2007, Xing2020}. The magnetic flux measured in coronal dimming regions can be compared to the flux content of the associated magnetic cloud (MC) observed in situ at 1 AU, supporting the idea that dimmings map the feet of the erupting flux rope \citep{Webb2000, Qiu2007}. 

\section{EUV Waves}\label{sec5}

\subsection{Discovery of EUV waves and a two-wave paradigm}

In the past three decades, one of the most important discoveries of the EUV imaging observations has been large-scale coronal waves, whose extent covers a large portion of the visible solar disk. The interpretation of this phenomenon produced much debate within the community.

\begin{figure}[h]
	\centering
	\includegraphics[width=1.\textwidth]{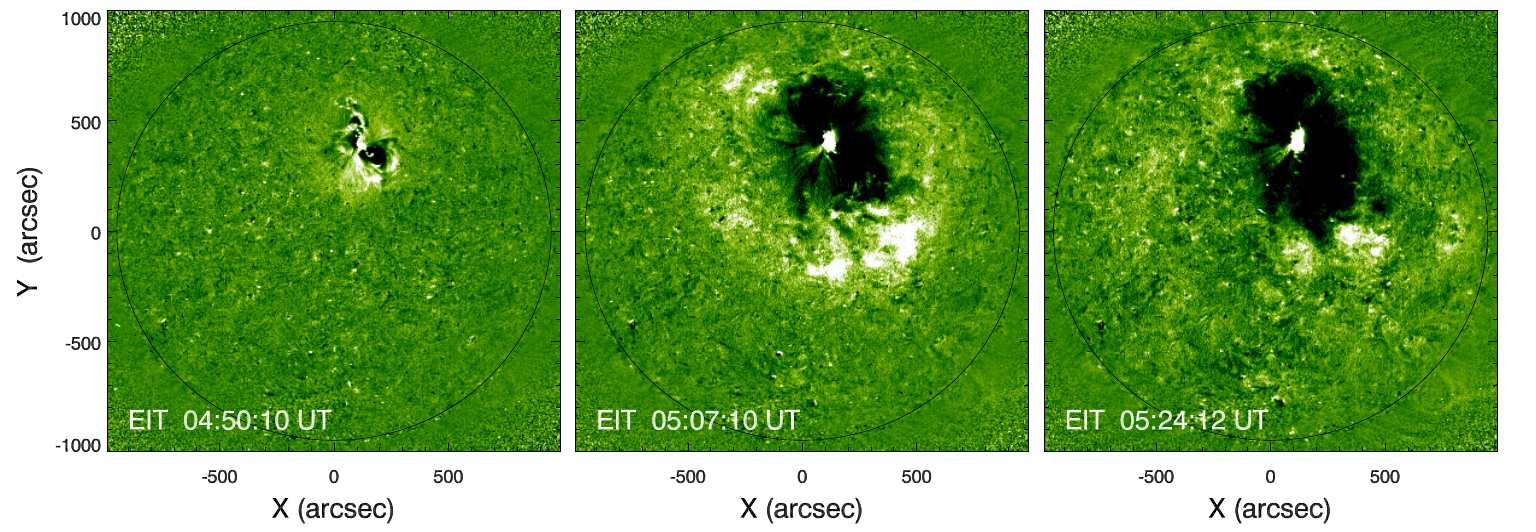}
	\caption{Sequential 195 \AA\ (difference) images of the first EIT wave event observed by the SOHO/EIT telescope on 1997 May 12. A ripple-like bright wave front propagates outward from the flare site. Taken from \citet{chen16}.} \label{fig51}
\end{figure}

On 1997 May 12, the EIT telescope on board the newly launched SOHO satellite revealed that a ripple-like bright front propagated away from the source region of a C-class flare \citep{Thompson1998}, as indicated by Figure \ref{fig51}. The authors named the wave-like phenomenon after the telescope, i.e., EIT waves. The first EIT wave had a propagation speed of 250 \kms. Since solar eruptions are expected to drive fast-mode MHD waves and only an EIT wave was present in the observations, it was immediately claimed that EIT waves are fast-mode MHD waves \citep{Thompson1998, wang00, wust01}.

However, one big problem with the fast-mode wave model is that the mean speed of EIT waves is only 271 \kms, which is about three times smaller than the typical fast-mode wave speed in the corona \citep{klass00, zhan11}. Moreover, \citet{dela99} found that EIT waves would stop near magnetic separatrices, which posed another challenge to the fast-mode wave model since fast-mode waves would not stop at magnetic separatrices. To reconcile these discrepancies, \citet{chen02} proposed a hybrid model, i.e., as a filament erupts, there should be two different types of EUV waves, a faster one and a slower one. The faster one corresponds to the filament piston-driven fast-mode shock wave, whose nose part is responsible for type II radio bursts and the footpoints are responsible for chromospheric Moreton waves \citep{huang25}; on the other hand, the slower one is an apparent propagation: As all magnetic field lines straddling over the erupting filament would be pushed to stretch up one by one, the successive stretching generates an apparent propagation of bright fronts. Both their analytical solution and MHD numerical simulations support that EIT waves have a speed about three times smaller than that of the fast-mode waves in the corona if the magnetic field lines are concentric semicircles. They further demonstrated that, unlike fast-mode waves, EIT waves would stop at magnetic separatrices, consistent with observations \citep{chen05}.

\citet{chen02} pointed out that the EIT telescope failed to detect two waves in one event because of its low cadence of observations, 15 minutes. However, the low-cadence EIT images still  indicated the presence of two types of EUV waves, as $\sim$7\% of the observed wave fronts are sharp, with the slow-component EUV waves being the majority diffuse fronts and the fast-mode waves being the minor sharp fronts \citep{Thompson2000, bies02}. Chen et al. also predicted that future telescopes with high cadence should observe two types of EUV waves simultaneously. Indeed, high-cadence observations from the Transition Region and Coronal Explorer \citep[TRACE;][]{Handy1999} showed two EUV waves appear in one filament eruption event, with speeds are 500 \kms\ and 200 \kms, respectively \citep{harr03}. Later, a statistical study on the kinematics of EUV waves also suggested that there are different types of EUV waves \citep{warm11}.

After the SDO satellite was launched in 2010, AIA has been constantly providing high-quality EUV imaging data with a cadence of 12 s. Since then, two waves have been frequently observed in a single eruption event \citep{chenwu11, asai12, cheng12, kuma13, shen13, lid24, zhangy25}. However, it should be noted that not all CME/flare events have both fast-mode MHD shock wave and EIT wave simultaneously,  as explained in Section \ref{anis}. 

In addition to the magnetic field line stretching model,  several other models have been proposed to explain EIT waves, including the slow-mode wave model \citep{will07, wang09}, the successive reconnection model \citep{attr07}, and the current shell model \citep{dela08}. It remains unclear if these models explain why EIT waves are around three times slower than the fast-mode waves.

\subsection{Relationship between EIT waves and EUV dimmings}

EUV dimmings are a type of phenomena intimately related to EIT waves. 
In fact, even before EIT waves were discovered \citep{Thompson1998}, it was already noticed that expanding EUV dimmings immediately follow propagating EIT waves, i.e., they are related phenomena. In particular, it was proposed that EIT waves correspond to the CME front whereas EUV dimmings correspond to CME cavities \citet{chenpf05, stepan24}. Therefore, any model proposed to explain EIT waves should be able to account for the formation of EUV dimmings \citep{chen02}.

Interestingly, among all the proposed EIT wave models, only the magnetic field line stretching model tried to explain both EIT waves and EUV dimmings simultaneously. 
After a filament erupts, the overlying magnetic field lines would be pushed to stretch up one by one, compressing the outer plasma to form bright wave fronts. The plasma enveloped by the magnetic field lines becomes attenuated as the volume of the magnetic field lines becomes larger, so dimmings are formed naturally \citep{chen02}. Spectroscopic observations indicated that the widths of various lines increase at the outer edge of EUV dimmings \citep{chenf10}, consistent with the prediction of the magnetic field line stretching model.

\subsection{Stationary EUV wave fronts}

\citet{dela99} found that EIT waves stop at magnetic separatrices. This feature is consistent with the typical characteristic of EIT waves, i.e., EIT waves generally stop near the coronal hole boundary \citep{Thompson1998}, since the coronal hole boundaries are magnetic separatrices between closed magnetic field and open field. It was the discovery of stationary EIT wave fronts at magnetic separatrices that inspired \citet{chen02} to propose the magnetic field line stretching model. With MHD numerical simulations, \citet{chen05} demonstrated that the formation of stationary EIT wave fronts can indeed be explained by the magnetic field line stretching model as only the closed magnetic field lines above the erupting filament would be pushed to stretch up, not the magnetic field lines in neighboring magnetic systems, including open magnetic field. Hence, only fast-component EUV waves, not EIT waves (i.e., the slow-component EUV waves), would be seen to penetrate into a coronal hole \citep{piant21, zhang25}.

\begin{figure}[h]
	\centering
	\includegraphics[width=0.7\textwidth]{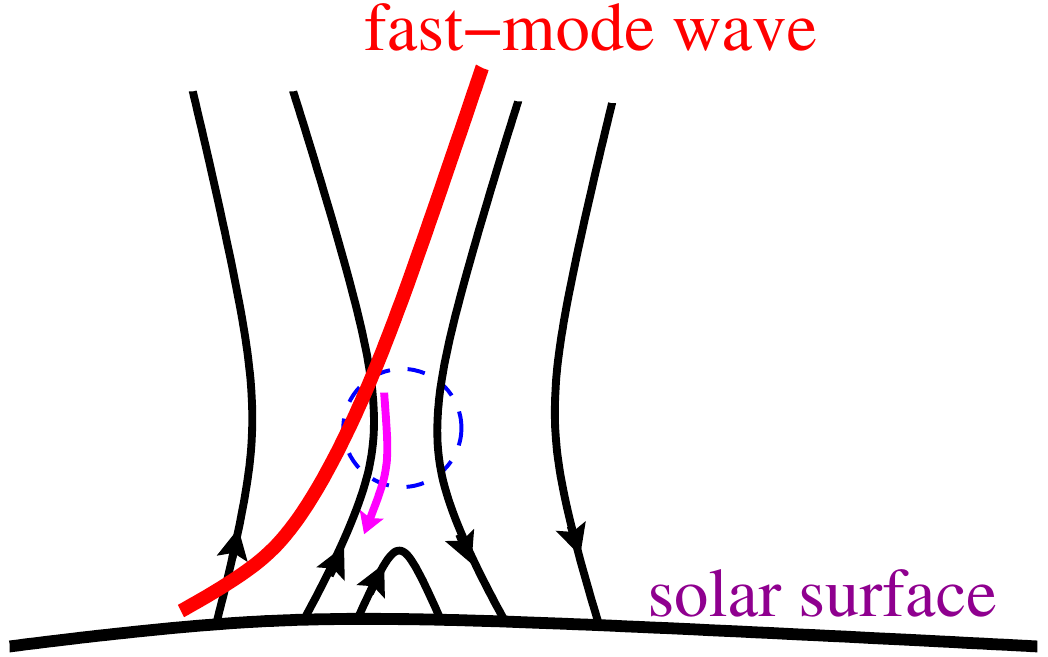}
	\caption{Schematic sketch describing how a fast-mode wave (red line) is converted to a slow-mode wave near a magnetic null point. In this figure, the black lines are magnetic field lines, the dashed circle represents the equipartition layer where Alfv\'en speed equals sound speed, and the magenta arrow indicates that the converted slow-mode wave propagates along magnetic field.} \label{fig52}
\end{figure}

On the other hand, \citet{chan16} discovered another type of stationary EUV wave fronts. It was revealed that when a fast-mode EUV wave passes a magnetic separatrix, a stationary wave front is generated locally. This type of stationary EUV wave fronts has nothing to do with traditional ``EIT waves". \citet{chen16a} performed an MHD numerical simulation and found that when a fast-mode MHD wave interacts with a separatrix, a stationary EUV wave front is generated. After analyzing the numerical results, they proposed that this type of stationary EUV waves is due to mode conversion. 
Near a magnetic null point, there is a layer where the Alfv\'en speed is equal to the sound speed; 
this is called an equipartition layer.
When a fast-mode MHD wave approaches an equipartition layer, the fast-mode wave would be largely converted to a slow-mode MHD wave. As an intrinsic property, slow-mode waves propagate along magnetic field lines parallel to the nearby magnetic separatrices, as illustrated in Figure \ref{fig52}. 
This interpretation was confirmed by later observations \citep{chan18, kuma24}.

\subsection{Anisotropy of EIT waves and its important application}\label{anis}

After the discovery of EIT waves in 1997, an important question remained unanswered for more than two decades, i.e., why are EIT waves anisotropic? 
Another unanswered question is why EIT waves are absent in some EUV wave events, i.e., only a fast-mode wave is present.

\citet{li25} analyzed a solar eruption event with two-component EUV waves. They found  that the EIT wave was strongly anisotropic. By examining the relationship between the EIT wave intensity and the inclination angle of the local magnetic field lines, they proposed that the intensity of the EIT wave fronts depends on the orientation of the magnetic field lines: the more forward-inclined the magnetic field line, the weaker the EIT wave is.  Such a dependence, which is in accordance with the magnetic field line stretching model, was further supported by their MHD numerical simulations. They claimed that forward-inclined magnetic field lines correspond to fainter EIT wave fronts and backward-inclined magnetic field lines correspond to brighter EIT wave fronts. As pointed out in their paper, such a relationship can be applied to resolve the famous $180^\circ$ ambiguity in the measurements of photospheric vector magnetograms. It was further deduced that if the magnetic field lines are strongly forward inclined, the EIT wave front would be too faint to be visible, which explains why only fast-mode waves appear in some events.


\section{First Ionization Potential Effect}\label{sec6}

Most literature incorrectly describes the solar corona as having a first ionization potential (FIP) effect (see Chapter 1 for the definition). This concept was originally based on in-situ observations of the slow solar wind and old remote-sensing observations and approximate analyses, some of which turned out to be incorrect. Within the past 20 years, thanks to the analyses of spectral observations from the X-rays  to the near-infrared, the picture has changed dramatically \citep[cf.][]{delzanna:2013_multithermal, delzanna_mason:2014, delzanna_mason:2018, delzanna_etal:2022_xsm, delzanna_etal:2023_airspec}.

The quiet Sun corona during solar minimum has photospheric abundances, as the transition region does. The 3 MK plasma in  quiescent cores of active regions has a constant FIP bias of about 3.2, independent of age and size of the active region. Note that the FIP bias is defined as the ratio of an element’s abundance in the corona relative to its photospheric abundance, normalized to hydrogen. The 2 MK emission surrounding the active region cores has an FIP bias of about 2. The 1 MK cool loops in active regions show a wide range of FIP biases, but the large body of previous literature over-estimated the abundances by large factors (4 or more), because it was assumed that such loops must have a continuous emission measure distribution whilst in reality SOHO/CDS observations have consistently shown that they are nearly isothermal in their cross-sections \citep[][]{delzanna:2003, delzanna_mason:2003, delzanna_etal:2006_ar}, a result later confirmed by Hinode/EIS \citep[cf.][]{delzanna:2009_fe_7, delzanna_etal:2011_aia, brooks_etal:2012} The other main misconception arose because Sun-as-a-star observations were analyzed when our star was active. In such cases, there is a mix of the FIP effects, with no FIP bias until 1 MK and increasing FIP biases at higher temperatures \citep{delzanna:2019_peve}.

The ponderomotive force model by \citet{laming:2015} is currently one of the most widely accepted theories  to explain both the FIP and inverse FIP effects, as it is able to reproduce some observations. The idea is that coronal Alfv\'en waves either trapped in a closed loop cavity or traveling below the chromosphere (and being there reflected), exert a
ponderomotive force on singly-ionized charge states that are formed within the chromosphere.
As the main process to form the ions is photo-ionization from the neutral hydrogen Ly$\alpha$, it is the elements with FIP lower than 10.2 eV that are more easily ionized and therefore experience the ponderomotive force. Once this happens, it takes a long time for the ions to drift upwards by diffusion  and be further ionized, creating the FIP bias inside the coronal loops. The model predicts an FIP bias of about 3, close to what is observed in active region 3 MK plasma and in solar energetic particle events (SEP), which are thought to originate directly from active regions. One of the problems with the model is that it is static, which is unrealistic. Another problem is that  not all closed loops have this predicted FIP effect: quiet Sun closed structures do not have an FIP effect; cool loops have a variety of FIP biases. A more fundamental problem is that the FIP effect is clearly stronger in the cores of active regions where there is a strong magnetic field, but the model does not depend on the  magnetic field. Finally, once the closed loops are filled with fractionated material, a process such as interchange reconnection \citep{delzanna_etal:2011_outflows} needs to take place to release the fractionated plasma
into the heliosphere. This process would explain the scatter of FIP bias (between 1 and 3) observed in-situ within hotter slow (350 \kms) solar wind originating from active regions \citep{wang_etal:2009}.

\subsection{FIP effect in solar flares}

Skylab EUV observations of low-temperature (transition-region) lines during flares and surges
have shown that the relative abundances of low and high-FIP elements were close
to photospheric  \citep[cf.][]{feldman_widing:1990}. Such observations were then used to
estimate the photospheric abundances of elements that do not produce photospheric
absorption lines such as argon and neon.  The assumption is that due to rapid
chromospheric evaporation, the plasma does not have time to fractionate. 

Several line-to-continuum studies have been carried out in the X-rays mostly on X-class flares, starting from OSO-8 \citep{veck_parkinson:81}, SMM \citep[e.g.,][]{fludra_schmelz:1995}, Yohkoh \citep[e.g.,][]{fludra_schmelz:99}, RESIK \citep[e.g.,][]{sylwester_etal:2014}, finding generally photospheric abundances. Large X-class flares were also observed in the EUV by SDO/EVE. Near photospheric abundances were found by \citet{delzanna_woods:2013}, a result later confirmed by \citet{warren:2014}. The explanation here is that these abundances are caused by the fast chromospheric evaporation filling in the flare loops. 

A significant new finding, later confirmed by several other authors, was obtained by \citet{mondal_etal:2021}. They analyzed Chandrayaan-2 XSM X-Ray Spectral Measurements 
of several  B-class solar flares using CHIANTI atomic data. All the flares showed the same evolution of elemental abundances: a steep (2--3 minutes) decrease from nearly coronal values to nearly photospheric, and a more gradual phase (a few minutes) returning to
coronal values. However, the recovering time varies among different chemical elements, which was explained to be due to the inertia effect \citep{ng24}. Generally, the decrease of the FIP bias happens during the flare impulsive phase. However, \citet{ng25} displayed a flare event where the FIP bias decrease happens before the flare impulsive phase, implying the magnetic reconnection in the standard CME/flare model might start from the chromosphere in some events, rather than in the corona as previously thought.

\citet{feldman_etal:2004_flares} analyzed the gradual phases of two C-class flares with SOHO/SUMER. From the intensity ratio of the Fe \textsc{xix} and Fe \textsc{xvii} UV forbidden lines (at 1118.1 and 1153.2~\AA, respectively), they deduced an approximate temperature (4 MK), and using this value they obtained the absolute iron abundance from the line to UV continuum ratio (assumed to be hydrogen free-free). They found that the iron  abundance was enhanced by a factor of 8–10 relative to the photospheric value, and concluded that the flare plasma must have originated from the corona. Such extremely high FIP bias has never been observed in high-temperature plasma in active regions, though an FIP bias of 6 was recently reported \citep{ng26}, which was explained to be due to enhanced $p$-mode wave leakage as the post-flare magnetic field becomes more inclined. 

Another simplified (line ratio) analysis of a large system of flare loops observed by Skylab 
in EUV lines formed between 2 and 4 MK by \citet{widing_feldman:2008}
resulted in a  scatter of FIP biases around 4. The authors argued that these loops should have been  formed by cooling of the 10--15 MK loops that are emitting the X-rays, and which are known to have photospheric abundances. Given the puzzle in the result,  they suggested that instead the flare loop plasma originated from prominence material. However, prominences typically have very low FIP biases, so that a better explanation is required.

\citet{doschek_etal:2018_x8} analyzed Hinode/EIS EUV spectra of Ca {\footnotesize XIV} and Ar {\footnotesize XIV} lines during the late phase of the famous X8-class limb flare (2017-09-10). They also found what they thought a puzzling result: the FIP bias decreases from the coronal abundances near the top to the footpoint of the flare loops. \citet{to_etal:2024_x8} analyzed the same flare and obtained essentially the same results, but suggested that the observed spatial variation originated from a combination of two processes: calcium-enriched plasma downflowing from the plasma sheet, and chromospheric evaporation filling the loops from the footpoints. The former suggestion is based on the results of \citet{warren_etal:2018}, where a comparison between Hinode/EIS and SDO/AIA images suggested coronal abundances.

\citet{delzanna:2025_eis_flares} analyzed an M-class flare observed by
Hinode/EIS during the beginning of its gradual phase and found relative
elemental abundances in broad agreement with the standard ones in active region
quiescent cores recommended by \citet{delzanna:2013_multithermal}, e.g.,
sulphur as a high-FIP element has an FIP bias of 3.2, but potassium has a different abundance. This element has a very low FIP (4.3 eV) and few measurements
of its coronal abundance are available. The M-class flare observations indicated the need  to increase the K abundance to 5.83 dex, in excellent agreement with the value (5.86 dex) obtained from RESIK X-ray measurements of the potassium He-like lines around 3.5~\AA\  by \citet{sylwester_etal:2014}, but 6 times higher than the photospheric value recommended by \citet{asplund_etal:2009}, 5.03 dex.

The inverse FIP effect has been observed since 1999 from X-ray spectra of active flare stars. It took a long time to analyze years of Hinode/EIS observations
to find just a few small regions near sunspots and near flares that had
an IFIP at around 2 MK, mainly from comparing Ar {\footnotesize XIV} and Ca {\footnotesize XIV} line intensities \citep[cf.][]{doschek_etal:2015,doschek_warren:2016}. Such intensities
would never contribute significantly to the solar irradiance.  
These results are puzzling as one would have expected, according to Laming's model and
the stellar results, very common IFIP effects during solar flares.

\subsection{Comments on the FIP effect}

The moderate-temperature (say, 3--5 MK) plasma in relatively large flares shows consistently coronal abundances during their gradual phase observed in the EUV wavebands, i.e., an FIP bias of about 3--4. On the other hand, all the X-ray observations of much higher temperature flare plasma consistently showed photospheric abundances. These two observational facts should be reconciled. One possibility is that the former component comes from the coronal reconnection outflow and the latter component comes from chromospheric evaporation. B-class flares are very different from M-class or X-class flares, as discussed in \citet{mitra-kraev_delzanna:2019} based on the EUV spectroscopy from Hinode/EIS: a few small flare loops are formed at nearly the same time, which do not typically reach 10 MK, and within minutes drain very rapidly. The first phase can again be explained with chromospheric evaporation, but the second is a bit puzzling.

To make progress, next-generation observations and models are needed. To resolve the apparent discrepancies in the abundances measured in the X-rays and EUV, we would need simultaneous, spatially-resolved spectral observations in key lines/elements of the flare plasma at different temperatures. That would not necessarily require both X-ray and EUV observations: for example, X-ray observations in the 3--25~\AA\ region could provide oxygen, neon vs. iron abundances of 3--5 MK plasma from the H- and He-like O, Ne and the iron lines, and argon, sulphur vs. iron (or silicon, magnesium) in 10--15 MK plasma by measuring the He-like Ar {\footnotesize XVII}  and the S {\footnotesize XV}, S {\footnotesize XVI} lines. The bonus of the X-rays is the availability of strong free-free emission. A variant of the Marshall Grazing Incidence X-ray Spectrometer (MaGIXS) concept, providing large field of views of an active region, could be a viable option \citep[cf.][]{2025ApJ...991..171M}. In terms of modeling, next-generation codes with additional time-dependent physics and atomic processes such as SAMS are needed (see Chapter 1).

\section{UV and EUV Observations from Recent Solar Space Missions}\label{sec7}

\subsection{{\lya} observations by the Advanced Space-based Solar Observatory (ASO-S)}\label{subsec7.1}

The Lyman-alpha ({\lya}) line of neutral hydrogen, centered at 121.6 nm, is the brightest emission line in the solar spectrum. It is a fundamental hydrogen spectral line generated by electron transitions from the $n=2$ to the $n=1$ energy levels in the hydrogen atom. The emission consists of two distinct components associated with different layers of the solar atmosphere: the line wings originate from the chromosphere, where the temperature is on the order of a few thousand Kelvin (K), while the line core is emitted from the transition region at a temperature of $\sim$40,000 K \citep{Vernazza1981}.

Observations of the {\hi} {\lya} provide crucial diagnostics of the solar chromosphere, transition region, and corona. Several space-based instruments have been developed to observe the Sun in {\lya}, employing either direct imaging or raster-scan techniques. The Solar Ultraviolet Measurements of Emitted Radiation \citep[SUMER;][]{Wilhelm1995} and the Ultraviolet Coronagraph Spectrometer \citep[UVCS;][]{Kohl1995} on board SOHO provided spectral and spatial information of the {\lya} radiation from the solar disk to the corona. The TRACE mission carried a {\lya} channel capable of high-resolution imaging of the solar disk. The Multiple XUV Imager \citep[MXUVI;][]{Auchere2001} further obtained full-disk {\lya} images. The Very high Angular resolution ULtraviolet Telescope \citep[VAULT;][]{Korendyke2001, Vourlidas2010} sounding rocket payload was a dedicated {\lya} imaging spectroheliograph that achieved sub-arcsecond resolution of the chromosphere and the low transition region. In addition, the Chromospheric Lyman-Alpha Spectro-Polarimeter experiment \citep[CLASP;][]{Kano2012, Kobayashi2012} measured for the first time the Hanle effect polarization of the {\lya}, providing direct constraints on the magnetic field in the low plasma $\beta$ region. The Ultraviolet Spectro-Coronagraph Pathfinder \citep[UVSC;][]{Strachan2017} was designed to perform narrow-slit measurements of the {\lya} line profile in the off-limb corona at two heights (1.8 and 3.0$R_\odot$), contributing valuable insights into the dynamic properties of non-thermal particles. Most recently, the Extreme Ultraviolet Imager \citep[EUI;][]{Rochus2020AA} and Metis coronagraph \citep{Antonucci2020AA} on board the Solar Orbiter \citep[SolO;][]{Muller2020AA} have produced {\lya} imaging of both the solar disk and the corona, providing an unprecedented connection between small-scale features and large-scale coronal structures.

\begin{figure}[h]
	\centering
	\includegraphics[width=0.63\textwidth]{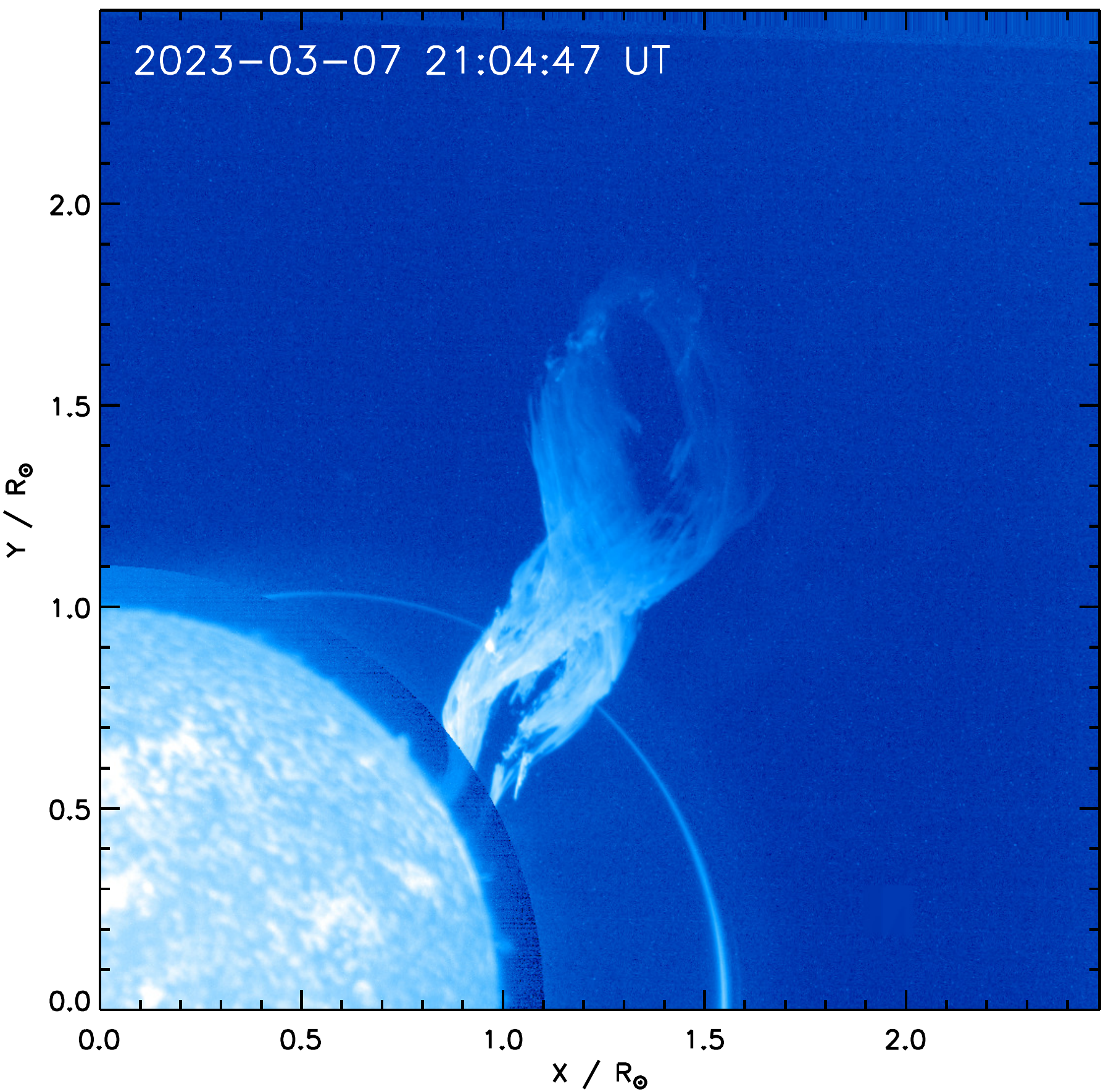}
	\caption{An eruptive prominence observed by SDI and SCIUV on 2023 March 7. The bright ring visible in the image is an instrumental artifact.} \label{fig71}
\end{figure}

The Advanced Space-based Solar Observatory \citep[ASO-S; also known as Kuafu-1;][]{Gan2023SoPh} was launched on 2022 October 9. It is equipped with a Lyman-alpha Solar Telescope \citep[LST;][]{Li2019RAA, Feng2019RAA, Chen2019RAA, Chen2024SoPh}---one of its core payloads dedicated to investigating solar emissions and dynamic activities. The LST integrates three specialized instruments: the Solar Disk Imager (SDI), the Solar Corona Imager (SCI), and the White-light Solar Telescope (WST), designed to observe both the solar disk and the inner corona. Their key parameters are summarized in Table~\ref{tab71.1}. The SCI is a dual-waveband coronagraph that simultaneously and independently observes the inner corona in both {\lya} (SCIUV) and white-light (SCIWL). Among these instruments, the combination of SDI and SCIUV for the first time can seamlessly image the Sun from the solar disk to the inner corona (see Figure~\ref{fig71} for an example).

\begin{table}[h]
	\caption{Key parameters of the LST performance.} \label{tab71.1}
	\begin{tabular*}{\textwidth}{@{\extracolsep\fill}p{1.4cm}<{\centering}p{1.5cm}<{\centering}p{1.8cm}<{\centering}p{3.4cm}<{\centering}p{3.3cm}<{\centering}}
		\toprule%
		Instrument & Waveband (nm) & Field of View (FOV, {\rsun}) & Image Size (${\rm pix^{2}}$) & Cadence \\
		\midrule
		WST  & $360.0 \pm 2.0$ & 0.0 - 1.2 & $4608 \times 4608$ or $1024 \times 1024$ (burst mode) & Routine (2 min); Burst (1s or 2s); User-defined \\
		SDI  & $121.6 \pm 4.5$ & 0.0 - 1.2 & $4608 \times 4608$ or $1024 \times 1024$ (burst mode) & Routine (1 min); Burst (a few s); User-defined \\
		SCIUV  & $122.6 \pm 3.0$ & 1.1 - 2.5 & $2048 \times 2048$ & $\sim$1 min \\
		SCIWL  & $700.0 \pm 32.0$ & 1.1 - 2.5 & $2048 \times 2048$ & $\sim$1 min \\
		\botrule
	\end{tabular*}
\end{table}

The SDI can be used to observe flares and prominences. For a better understanding of the {\lya} emission captured by SDI, see \citet{Yiliang2025}. The contribution of Si {\footnotesize III}  to the {\lya} passband  cannot be neglected, especially during X-class flares. The {\lya} emission exhibits features similar to the He {\footnotesize II} 304 {\AA} emission, particularly along flare ribbons \citep{Shi2024SoPh}. \citet{Xue2024SoPh} investigated the association between a failed prominence eruption and the drainage of mass from another prominence on 2022 October 26, which were the first prominences observed since LST was in operation. \citet{Lu2024SoPh} developed an automatic detection method for {\lya} enhancements observed by SDI. More than 5,000 events were detected within three years after launch.

\begin{figure}[h]
	\centering
	\includegraphics[width=0.8\textwidth]{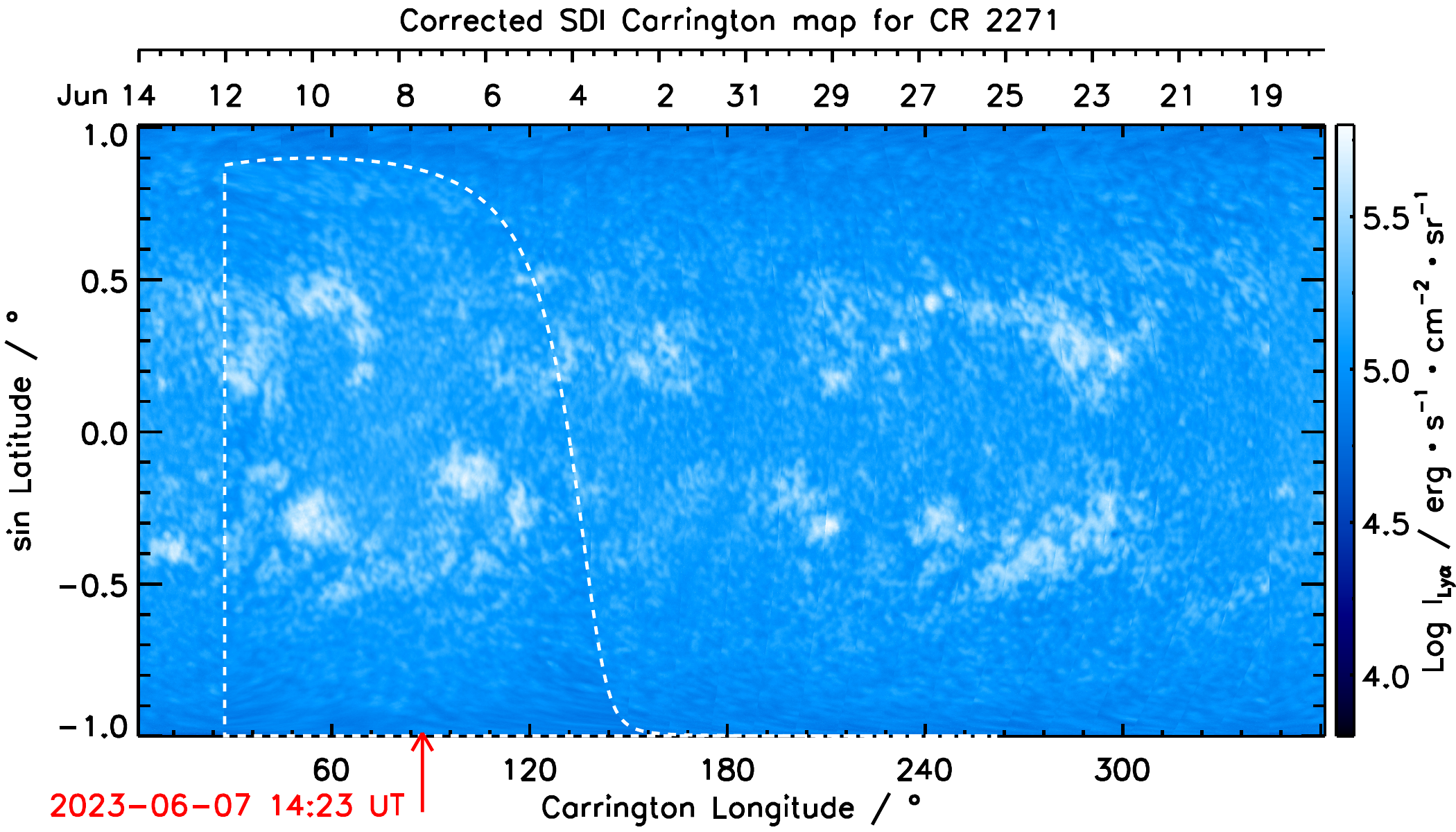}
	\caption{An example of SDI Carrington map considering differential rotation and flare-region replacement.} \label{fig72}
\end{figure}

Coronal {\lya} emission consists of radiative and collisional components \citep[e.g.,][]{Noci1987, Ying2020ApJ, Ying2024SoPh}. The radiative component arises from the resonant scattering of the disk {\lya} photons by neutral hydrogen atoms in the corona, while the collisional component is caused by impact excitation with free electrons. 
For a review of resonant scattering and corrections to errors in the previous literature, see 
\citet{delzanna:2025rpe}. To calculate the radiative component, SDI produces Carrington maps of the disk {\lya} emission. These maps account for solar differential rotation (adjusting for longitude-dependent rotation speeds at a given time) and can be updated with flare-related disk emissions during solar eruptions, ensuring accuracy in coronal emission modeling \citep[][Figure~\ref{fig72}]{Li2024ApJ}. Details of the effects of the solar disk radiation, the geometric scattering function, and the temperature assumptions on the {\lya} intensity in a coronal mass ejection can be found in \citet{Ying2024SoPh}. 

Based on the MHD simulation of an eruptive prominence by \citet{Fan2019}, \citet{Zhao2022AA} synthesized the {\lya} emission under both the optically thin and thick conditions, and found that the optical depth decreases with time, from about 100 to optically thin case. When the prominence is optically thick, the {\lya} emission is one to two orders of magnitude lower than under the optically thin assumption.

\subsection{Near and mid-UV solar observations by Aditya-L1} 

The near- and mid-UV (primarily 200{--}400~nm) radiation is emitted from the lower and middle atmosphere of the Sun, namely the photosphere and the chromosphere. One of the most intriguing problems in solar physics and in astrophysics is to comprehensively understand the existence of the hotter chromosphere and corona above the photosphere. The total energy input required to heat and maintain the chromosphere is enormous compared to the corona. This is primarily due to 2{--}3 orders of magnitude higher density in the chromosphere in comparison to the coronal densities. Thus, to heat and maintain the chromosphere, the Sun needs to provide energy, which is about 4{--}9 orders of magnitude larger than the energy required to heat and maintain the corona, assuming that all the radiation in the chromosphere is due to Bremsstrahlung radiation. However, it is noted that this is a rough assumption and a significant amount of energy goes into line formation, etc. The source of this energy and conversion to heat is not understood at all. Therefore, studying the radiation in the near- and mid-UV that essentially maps the photosphere and chromosphere provides a new window into the understanding of the coupling of these two layers and thereby helps to understand the longstanding question about heating of the magnetized atmosphere.

One of the strongest manifestations of energetic events taking place in the solar atmosphere is 
solar flares. Although we have been observing the Sun across various wavelengths, the spectral energy distribution of the radiated energy from the flares is still poorly understood; 
approximately,  about 77\% of the energy is released in the wavelength range $>$200~nm, and only $\sim$23\% at higher energies  \citep[][]{Nei_1989, NeiK_1993, Kre_2011}. Although the energy content in hard X-ray (HXR) is a tiny fraction of the total energy budget, it  still has a fundamental role in the energization process \citep{Holman2011}.  It has become clear that to develop a comprehensive understanding of the origin and evolution of solar flares, multi-wavelength studies of all kinds of events are needed. Such studies will help us to understand the underlying physical processes,  and to quantify the contribution of solar flares to the solar spectral irradiance (SSI) as a function of the solar activity cycle. 
In fact, one of the major questions in solar physics is how flares affect the solar spectral irradiance and the total solar irradiance variability from a short to much longer solar cycle timescale.

Ideally, it would be essential to observe flares at all wavelengths simultaneously with sufficient spatiotemporal resolution to figure out their spectral energy distribution. Unfortunately, this is generally not the case, and we have to rely on sporadic observations, often made using ground-based instruments in the visible domain. 
An exception is the work by 
 \citet{KreDS_2010} and \citet{Kre_2011}, where they performed statistical studies of a large number of flares across a wide energy range, using observations  recorded from 1996 to 2008 by three visible Solar Photometers, centered on 402~nm, 500~nm, and 862~nm, of SOHO VIRGO, in combination with EUV irradiance in the ranges 0.1--50~nm and 26--34~nm measured by SOHO/Solar EUV Monitor \citep[][]{JudMO_1998} and SXR measurements from GOES satellites. The results show more than $2\sigma$-level variations in the total solar irradiance during the flare peak, indicating flare signals in such measurements.

Moreover, radiation within the near and mid UV range drives global atmospheric dynamics and is believed to play an important role in climate physics. This radiation is absorbed by the O$_2$ continuum and O$_3$ bands and heats the stratosphere and mesosphere of the Earth. Thus, solar variability in this range plays a very important, yet unexplored, role in regulating ozone abundance and in influencing Earth's climate dynamics \citep{Lean_1997, LeaRK_1997}. It is known that the total contribution of UV radiance from the Sun (below 400 nm) to the total solar irradiance is less than 8\%.  However, the variability in UV radiance can reach more than 100\% over the course of the solar cycle, while the total solar irradiance varies by $\sim$0.1\% \citep[][]{KriSF_2006, HarFP_2009, ErmMD_2013}. Therefore, to put tight constraints on the impact of solar UV radiation on the climate, we need to accurately  quantify the irradiance in this band, including its time variation. 

In the  200{--}400~nm wavelength range, there are two emission spectral lines of Mg, namely Mg~{\footnotesize II}~h~\&~k, which map the chromospheric structure and dynamics. These lines are used to derive the Mg~{\footnotesize II} index, which is one of the most widely used proxies of solar variability \citep{HeaS_1986}. 
The Mg~{\footnotesize II} index is derived by taking the ratio of the line cores to the wings. 
The line core forms in the chromosphere and shows strong variability, while the wings, formed in the photosphere, are relatively stable. Therefore, the Mg~{\footnotesize II} index primarily represents the ratio of the strongly variable chromospheric emission to the stable photospheric wings.
Since both the line cores and the wings are measured by the same instrument, the ratio is effectively free of instrumental effects. Therefore, the Mg~{\footnotesize II} index
has become one of the most widely used indices of solar activity, given its 
stability over long time scales, and easier cross-calibration between different instruments. 

However, it is important to note  that so far the Mg~{\footnotesize II} index has been measured using sun-as-a-star spectra. Therefore, it does not provide any information on the contribution of different features to its variation. For a long time we thought that the Mg~{\footnotesize II} index in the quiet Sun and coronal holes is similar; however, using IRIS observations of Mg~{\footnotesize II} lines, \cite{KaySP_2018} found that, given identical magnetic flux, quiet-Sun regions were brighter than coronal holes, whereas there was no such difference in the wing. These results suggest that the Mg~{\footnotesize II} index for coronal holes will be lower than that for the quiet Sun.

The Solar Ultraviolet Imaging Telescope \citep[][]{TriRP_2025, SarNR_2025} (SUIT) onboard the Aditya-L1 \citep[][]{TriCN_2023} mission of the Indian Space Research Organization (ISRO) aims to address the above shortcomings by observing the solar disk in the 200{--}400~nm wavelength range. The Aditya-L1 mission was launched on 2023 Sep. 02 and was subsequently inserted into the halo orbit around the first Lagrange point on the Sun-Earth line on 2024 Jan. 06. After the verification and calibration phase, which was completed by May 2024, SUIT has been continuously observing and providing  data since June 2024. 

SUIT observes the Sun's photosphere and chromosphere using 11 filters which include 8 narrow-band channels and 3 broad-band channels. The narrow-band channels include imaging of Mg~h~\&~k as well as the two wings at 276.7 and 283.2 nm, primarily to derive spatially resolved Mg indices. The three broad-band channels are 200{--}242~nm (Herzberg Continuum), 242{--}300~nm (Hartley Band) and 320{--}360~nm (Huggins Band). While the primary goal of the narrow-band channels is to study the dynamic coupling of the atmosphere in the near- and mid-UV, the broad-band channels were designed to study and characterize the radiation, primarily for climate studies.

In order to study flares, the payload has an automatic detection flare mode implemented,
making use of the two X-ray payloads, the High Energy L1 Orbiting Spectrometer \citep[HEL1OS,][]{NanS_2025} and the SOlar Low Energy Soft X-ray Spectrometer \citep[SoLEXS,][]{SanBS_2025}. These two spectrometers provide full-Sun integrated spectra over the energy range between 8{--}150~keV (with a spectral resolution of $\sim$1 keV at 14 keV (CdTe) and $\sim$7 keV at 60 keV (CZT)) and 1{--}22~keV (spectral resolution of $\sim$170 eV at 5.9 keV), respectively.
While SoLEXS provides a flare flag to SUIT, HEL1OS data are sent to SUIT where the onboard processing unit processes the HEL1OS data and generates a flare flag. As soon as SUIT gets a flag, the payload goes into flare mode. 

\begin{figure}[h]
	\centering
	\includegraphics[width=0.8\textwidth]{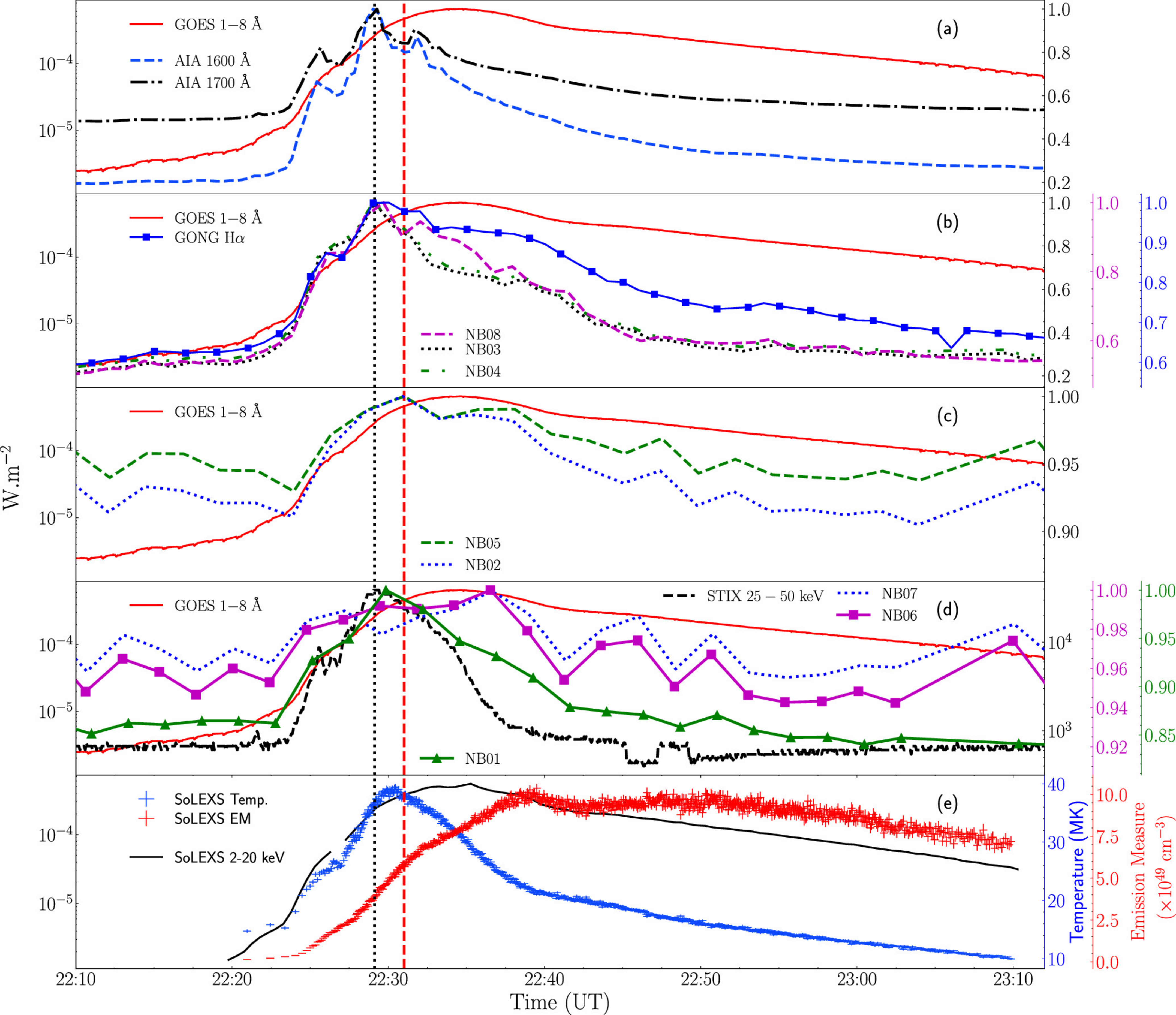}
	\caption{Multi-wavelength light curves obtained over a flare kernel on 2024 Feb 22. Panel (a): GOES 1--8 \AA\ (red), AIA 1600 \AA\ (blue dashed), and AIA 1700 \AA\ (black dotted–dashed); Panel (b): GOES 1--8 \AA\ (red), GONG H$\alpha$ (blue), SUIT Ca~{\footnotesize II}~h (magenta dashed), Mg~{\footnotesize II}~k (black dotted), and Mg~{\footnotesize II}~h (green dotted–dashed); Panel (c): GOES 1--8 \AA\ (red), SUIT 283.2~nm continuum (green dashed), and 276.7~nm continuum (blue dotted); Panel (d): GOES 1--8 \AA\ (red), SUIT 214~nm continuum (green triangles), STIX/HXR (black dashed), SUIT 388~nm continuum (blue dotted), and 300~nm continuum (magenta squares); Panel (e): SoLEXS 2--20 keV (black), the calculated temperature (blue) and emission measure (red). The vertical dotted black line and dashed red line across the panels mark the peak times of [Mg~{\footnotesize II}~h~\&~k, Ca~{\footnotesize II}~h] and [276.7 nm and 283.2 nm continuum], respectively. Taken from \citet{roy25}.} \label{fig73}
\end{figure}

Figure~\ref{fig73} plots the evolution of flare intensity for one of the first flares that was observed by SUIT as well as SoLEXS. The light curves show that before the flare reaches its peak in the soft X-rays (GOES and SOLEXS), it peaks in the SUIT Mg {\footnotesize II} channels, in the AIA 1600 and 1700 \AA\ channels, and H$\alpha$. These peaks match the peak observed in the HXR light curve obtained by the STIX instrument onboard Solar Orbiter. Panel (b) shows that the Mg~{\footnotesize II} and Ca~{\footnotesize II} light curves peak at the same time, similar to H$\alpha$. However, the light curves obtained in the blue and red wings of Mg~{\footnotesize II} using SUIT, peak about 2 min later than Mg~{\footnotesize II} (panel c). The HXR peak matches those on Mg~{\footnotesize II}, Ca~{\footnotesize II}, and H$\alpha$ (see panel d). Moreover, unlike other light curves, during the impulsive phase of the flare, the 300 nm and 388 nm continuum light curves exhibit a very small ($\sim$0.03\%) sharp increase in intensity. The authors also compared the 2--20 keV SoLEXS light curve (black), temperature (blue), and emission measure (red) (panel e). It is important to note that the temperature peak of the flare matches the HXR peak and the emission measure peaks about 9 min later. This is one of the first observations to cover such a broad wavelength, and demonstrates the potential of SUIT to study solar flares, when combined with other instruments.

\section{Lessons for Stellar Activity}\label{sec8}

During the ISSI-Beijing workshop, several colleagues from the solar community presented talks on various topics, including solar jets, flares, EUV dimmings, and EUV waves. Similar eruptions are expected to occur on other stars and even accretion disks. With the advantage of high spatial resolution, it is anticipated that the solar observations could shed light on our understanding of stellar activities \citep{degr26}; for example, \citet{Namekata2022} estimated the mass of a stellar filament based on the scaling law of solar filaments.

Ubiquitous solar jets and sporadic flares are the natural result of magnetic reconnection accompanied by various waves, both contributing to coronal heating on the Sun. 
They should be present on other stellar atmospheres and accretion disk coronae, contributing to the heating. While stellar flares can be detected by their emissions in X-ray, EUV, or white light, the presence of stellar jets and accretion disk jets can be inferred from their associated type III radio bursts.

Coronal dimmings are defined by abrupt, localized decreases in coronal emission—primarily at EUV and X-ray wavelengths. These complex, dynamic phenomena exhibit distinct observational characteristics, including multi-thermal visibility, mass depletion, associated plasma outflows, and structured temporal evolution. Together, these features make dimmings a powerful diagnostic tool for investigating the physical processes of magnetic reconnection and flux rope evolution during solar eruptions. More importantly, observations of full-Sun integrated EUV light curves (``Sun-as-a-star" observations) from instruments like SDO/EVE have detected dimmings associated with solar CMEs \citep{Mason2014}. Through analysis of EVE-observed dimming events, \citet{Mason2016} concluded that CME mass and speed can be inferred from Sun-as-a-star EUV light curves of dimmings. This provides a direct methodological framework for detecting and characterizing CMEs on other solar-like stars—a field where direct imaging is impossible and detections remain rare. Unlike other techniques for stellar CMEs detection, e.g., time-domain XUV and radio spectroscopy, searching for signatures of coronal dimmings caused by stellar CMEs is much easier to achieve, as it only requires photometric monitoring of target stars at X-ray or EUV wavelengths. For example, \citet{Veronig2021} identified 21 dimming events associated with stellar flares using broadband X-ray and EUV light curves of stars. The typical rise time, recovery time, and total duration of these stellar coronal dimmings are similar to those of solar dimmings. However, the depth (percentage flux decrease) of stellar dimmings ranges from 5\% to 56\%—an order of magnitude larger than that of solar coronal dimmings. While this larger depth may be due to selection bias, it also raises questions about whether these stellar dimmings are analogous to solar CME-induced coronal dimmings.

In addition, \citet{Loyd2022} identified one coronal dimming event from far-ultraviolet observations of a K2-type dwarf star with the Hubble Space Telescope. This event revealed a decrease in intensity of the Fe {\footnotesize XXI} 1354 \AA\ spectral line by $\sim$80\% immediately after a flare on the star. Such a strong dimming suggests a possible ejection of some 10$^{16}$ g plasma with a temperature of $\sim$10 MK.

Notably, intensity dimming caused by obscuration from erupting cool filaments has also been detected in Sun-as-a-star observations of He II 304 \AA. This suggests that stellar He \textsc{ii} 304 \AA\ dimming signatures could be used to detect stellar filament eruptions during CMEs \citep{Xu2024}. It should be pointed out that much of the EUV spectral range is  obscured by interstellar absorption; for example, the He {\footnotesize II} 30.4 nm line can only be observable for some nearby stars \citep{linsky17}.

Understanding stellar CMEs is critical to assess stellar evolution, angular momentum loss, and—most importantly—the habitability of orbiting exoplanets,  as frequent CME impacts can erode planetary atmospheres \citep{Airapetian2020}. Although progress has been made in the detection of stellar CMEs \citep{Veronig2021, Argiroffi2019, Namekata2022, Chen2022, Lu2025}, systematic observations remain elusive.

Waves provide an alternative approach to diagnose physical parameters which cannot be measured directly, such as the coronal magnetic field. This approach is called coronal seismology. Indeed, EUV waves have been utilized to derive the magnetic field strength in the solar corona. Since there are two types of EUV waves with different nature, their application to coronal seismology should be treated with great care. If the slow-component EUV wave, i.e., an EIT wave, is treated as a fast-mode wave, as done frequently in the literature during the SOHO era, the coronal magnetic field strength would be significantly under-estimated. Only when the fast-component EUV wave is treated as a fast-mode wave, the magnetic field in the solar corona can be precisely derived \citep{long13, miao26}. On the other hand, it was proposed that the slow-component EUV waves, i.e., EIT waves, can also be applied to coronal seismology under the framework of the magnetic field line stretching model \citep{chen09}. More importantly, EIT waves have the advantage that they can be used to estimate the magnetic configuration, i.e., if the speed of the slow-component EUV wave is about one-third that of the fast-component EUV wave, the coronal magnetic field should be like concentric semicircles; if the speed of the slow-component EUV wave is smaller or larger than one-third that of the fast-component EUV wave, the coronal magnetic field should be more or less stretched in the vertical direction compared to concentric semicircles.

Coronal seismology could be useful for studying stellar coronae. As solar eruptions produce two types of EUV waves, stellar eruptions should generate two types of EUV waves as well. The speed of the fast-component EUV wave can be approximately inferred from the type II radio bursts in the stellar radio dynamic spectra, and the speed of the slow-component EUV waves can be derived from the slope of the stellar EUV dimmings. Based on whether their speed ratio is larger or smaller than 3, we could then infer whether the  magnetic field in the stellar corona is more or less elongated radially.

The FIP effect provides a useful diagnostic tool for studying coronal heating in solar/stellar coronae. In principle, if the physics of all these abundance variations is understood, it could be used to understand which processes might occur in the coronae of active stars and during flares. However, the fact that the inverse FIP effect is rarely observed on the Sun is difficult to reconcile with stellar observations. Future observations and modeling 
are required.

In this review, we have introduced recent UV/EUV telescopes from China and India for solar observations. Interestingly, several other  EUV satellite missions have been proposed 
to systematically search for stellar CMEs and characterize their properties using methods such as coronal dimming and Doppler shifts. Among them, the ESA proposal 
for Stellar \& ISM Research via In-orbit Ultraviolet Spectroscopy (SIRIUS, 
\citealt{barstow+2026}), 
the NASA proposal for an 
Extreme-ultraviolet Stellar Characterization for Atmospheric Physics and Evolution (ESCAPE) mission \citep{France2019} and the Coronal Explorer for the Sun and Nearby Stars (CESS) mission \citep{Tian2025}. 

\bmhead{Acknowledgements}
PFC was supported by NSFC (12127901). GDZ acknowledges support from STFC (UK) via the consolidated grant to the atomic astrophysics group at DAMTP, University of Cambridge (ST/T000481/1), the STFC large Award UKRI1165: the Solar Atmospheric Modeling Suite (SAMS),
and his late wife M.B. Mahler Araujo. LF is supported by the Strategic Priority Research Program of the Chinese Academy of Sciences (XDB0560000), National Key R$\&$D Program of China (2022YFF0503003). JHG was supported by NSFC (1250030413) and the China National Postdoctoral Program for Innovative Talents fellowship (BX20240159). YJH was supported by NSFC (12273060, 12673070), the Strategic Priority Research Program of CAS (XDB0560000), and the Specialized Research Fund for State Key Laboratory of Solar Activity and Space Weather.

\bibliography{chap2}

@ARTICLE{polito_etal:2015,
   author = {{Polito}, V. and {Reeves}, K.~K. and {Del Zanna}, G. and {Golub}, L. and 
	{Mason}, H.~E.},
    title = "{Joint High Temperature Observation of a Small C6.5 Solar Flare With Iris/Eis/Aia}",
  journal = {\apj},
     year = 2015,
    month = apr,
   volume = 803,
      eid = {84},
    pages = {84},
      doi = {10.1088/0004-637X/803/2/84},
   adsurl = {http://adsabs.harvard.edu/abs/2015ApJ...803...84P}
}

@INPROCEEDINGS{barstow+2026,
       author = {{Barstow}, Martin and {Alameri}, Noora and {Argiroffi}, Costanza and {d'Avillez}, Miguel and {Bannister}, Nigel and {Barstow}, Joanna and {Bird}, Rachel and {Burleigh}, Matthew and {Casewell}, Sarah and {Clements}, Elizabeth and {Deliporanidou}, Evangelia and {Del Zanna}, Giulio and {Diebold}, Sebastian and {Drake}, Jeremy and {Fraux}, Vincent and {Gillon}, Michael and {de Castro}, Ana Inez Gomez and {de Grijs}, Richard and {Hampson}, Richard and {Harper}, Graham and {Hermans}, Aline and {Harra}, Louise and {Jacobs}, Jerome and {Jacques}, Lionel and {Kintziger}, Christian and {Lapington}, Jon and {de Martino}, Domitilla and {Naze}, Yael and {Al Naimiy}, Hamid and {Nichols}, Jonathan and {Owen}, James and {Rogers}, James and {Speight}, Roisin and {Stelzer}, Beate and {Sundararaman}, Venkatesh and {Al Wardat}, Mashhoor and {Townend}, Martin and {Werner}, Klaus and {Youngblood}, Allison},
        title = "{SIRIUS: stellar and ISM research via in-orbit ultraviolet spectroscopy}",
    booktitle = {Society of Photo-Optical Instrumentation Engineers (SPIE) Conference Series},
         year = 2026,
       series = {Society of Photo-Optical Instrumentation Engineers (SPIE) Conference Series},
       volume = {14146},
        month = aug,
    publisher = {SPIE},
          eid = {1414608},
        pages = {1414608},
          doi = {10.1117/12.3104810},
       adsurl = {https://ui.adsabs.harvard.edu/abs/2026SPIE14146E..08B}
}

@ARTICLE{delzanna:2025rpe,
       author = {{Del Zanna}, G.},
        title = "{The middle corona {\textendash} resonant photoexcitation in EUV lines}",
      journal = {\mnras},
         year = 2025,
        month = oct,
       volume = {543},
       number = {1},
        pages = {390-406},
          doi = {10.1093/mnras/staf1401},
       adsurl = {https://ui.adsabs.harvard.edu/abs/2025MNRAS.543..390D}
}

@ARTICLE{brosius:2003,
   author = {{Brosius}, J.~W.},
    title = "{Chromospheric Evaporation and Warm Rain during a Solar Flare Observed in High Time Resolution with the Coronal Diagnostic Spectrometer aboard the Solar and Heliospheric Observatory}",
  journal = {\apj},
     year = 2003,
    month = apr,
   volume = 586,
    pages = {1417-1429},
      doi = {10.1086/367958},
   adsurl = {http://adsabs.harvard.edu/abs/2003ApJ...586.1417B}
}

@ARTICLE{delzanna_etal:2006,
   author = {{Del Zanna}, G. and {Berlicki}, A. and {Schmieder}, B. and {Mason}, H.~E.
	},
    title = "{A Multi-Wavelength Study of the Compact M1 Flare on October 22, 2002}",
  journal = {\solphys},
     year = 2006,
    month = mar,
   volume = 234,
    pages = {95-113},
      doi = {10.1007/s11207-006-0016-6},
   adsurl = {http://ukads.nottingham.ac.uk/abs/2006SoPh..234...95Z}
}

@ARTICLE{polito_etal:2016,
   author = {{Polito}, V. and {Reep}, J.~W. and {Reeves}, K.~K. and {Sim{\~o}es}, P.~J.~A. and 
	{Dud{\'{\i}}k}, J. and {Del Zanna}, G. and {Mason}, H.~E. and 
	{Golub}, L.},
    title = "{Simultaneous IRIS and Hinode/EIS Observations and Modelling of the 2014 October 27 X2.0 Class Flare}",
  journal = {\apj},
archivePrefix = "arXiv",
 primaryClass = "astro-ph.SR",
     year = 2016,
    month = jan,
   volume = 816,
      eid = {89},
    pages = {89},
      doi = {10.3847/0004-637X/816/2/89},
   adsurl = {http://adsabs.harvard.edu/abs/2016ApJ...816...89P}
}

@ARTICLE{UpenT_2022,
       author = {{Upendran}, Vishal and {Tripathi}, Durgesh},
        title = "{On the Formation of Solar Wind and Switchbacks, and Quiet Sun Heating}",
      journal = {\apj},
         year = 2022,
        month = feb,
       volume = {926},
       number = {2},
          eid = {138},
        pages = {138},
          doi = {10.3847/1538-4357/ac3d88},
archivePrefix = {arXiv},
       eprint = {2111.11668},
 primaryClass = {astro-ph.SR},
       adsurl = {https://ui.adsabs.harvard.edu/abs/2022ApJ...926..138U}
}

@ARTICLE{TriNS_2021,
       author = {{Tripathi}, Durgesh and {Nived}, V.~N. and {Solanki}, Sami K.},
        title = "{Coronal Heating and Solar Wind Formation in Quiet Sun and Coronal Holes: A Unified Scenario}",
      journal = {\apj},
         year = 2021,
        month = feb,
       volume = {908},
       number = {1},
          eid = {28},
        pages = {28},
          doi = {10.3847/1538-4357/abcc6b},
archivePrefix = {arXiv},
       eprint = {2011.09803},
 primaryClass = {astro-ph.SR},
       adsurl = {https://ui.adsabs.harvard.edu/abs/2021ApJ...908...28T}
}

@ARTICLE{MalUT_2024,
       author = {{Malaker}, Biswanath and {Upendran}, Vishal and {Tripathi}, Durgesh},
        title = "{Thermodynamic Evolution of Plumes}",
      journal = {\apj},
         year = 2024,
        month = oct,
       volume = {974},
       number = {2},
          eid = {163},
        pages = {163},
          doi = {10.3847/1538-4357/ad6c4b},
archivePrefix = {arXiv},
       eprint = {2408.11771},
 primaryClass = {astro-ph.SR},
       adsurl = {https://ui.adsabs.harvard.edu/abs/2024ApJ...974..163M}
}

@ARTICLE{Tri_2021,
       author = {{Tripathi}, Durgesh},
        title = "{Transient Formation of Loops in the Core of an Active Region}",
      journal = {\apj},
         year = 2021,
        month = mar,
       volume = {909},
       number = {2},
          eid = {105},
        pages = {105},
          doi = {10.3847/1538-4357/abdd2e},
archivePrefix = {arXiv},
       eprint = {2101.06622},
 primaryClass = {astro-ph.SR},
       adsurl = {https://ui.adsabs.harvard.edu/abs/2021ApJ...909..105T}
}

@ARTICLE{ChaGMT_2015,
       author = {{Chandra}, R. and {Gupta}, G.~R. and {Mulay}, Sargam and {Tripathi}, Durgesh},
        title = "{Sunspot waves and triggering of homologous active region jets}",
      journal = {\mnras},
         year = 2015,
        month = feb,
       volume = {446},
       number = {4},
        pages = {3741-3748},
          doi = {10.1093/mnras/stu2305},
archivePrefix = {arXiv},
       eprint = {1410.8315},
 primaryClass = {astro-ph.SR},
       adsurl = {https://ui.adsabs.harvard.edu/abs/2015MNRAS.446.3741C}
}

@ARTICLE{KaySMT_2013,
       author = {{Kayshap}, P. and {Srivastava}, Abhishek K. and {Murawski}, K. and {Tripathi}, Durgesh},
        title = "{Origin of Macrospicule and Jet in Polar Corona by a Small-scale Kinked Flux Tube}",
      journal = {\apjl},
         year = 2013,
        month = jun,
       volume = {770},
       number = {1},
          eid = {L3},
        pages = {L3},
          doi = {10.1088/2041-8205/770/1/L3},
archivePrefix = {arXiv},
       eprint = {1305.0338},
 primaryClass = {astro-ph.SR},
       adsurl = {https://ui.adsabs.harvard.edu/abs/2013ApJ...770L...3K}
}

@ARTICLE{MulTDM_2006,
       author = {{Mulay}, Sargam M. and {Tripathi}, Durgesh and {Del Zanna}, Giulio and {Mason}, Helen},
        title = "{Multiwavelength study of 20 jets that emanate from the periphery of active regions}",
      journal = {\aap},
         year = 2016,
        month = may,
       volume = {589},
          eid = {A79},
        pages = {A79},
          doi = {10.1051/0004-6361/201527473},
archivePrefix = {arXiv},
       eprint = {1602.00151},
 primaryClass = {astro-ph.SR},
       adsurl = {https://ui.adsabs.harvard.edu/abs/2016A&A...589A..79M}
}

@ARTICLE{MulTMD_2023,
       author = {{Mulay}, Sargam M. and {Tripathi}, Durgesh and {Mason}, Helen and {Del Zanna}, Giulio and {Archontis}, Vasilis},
        title = "{Formation and thermodynamic evolution of plasmoids in active region jets}",
      journal = {\mnras},
         year = 2023,
        month = jan,
       volume = {518},
       number = {2},
        pages = {2287-2299},
          doi = {10.1093/mnras/stac3035},
archivePrefix = {arXiv},
       eprint = {2211.01740},
 primaryClass = {astro-ph.SR},
       adsurl = {https://ui.adsabs.harvard.edu/abs/2023MNRAS.518.2287M}
}

@ARTICLE{ChiYI_2008,
       author = {{Chifor}, C. and {Young}, P.~R. and {Isobe}, H. and {Mason}, H.~E. and {Tripathi}, D. and {Hara}, H. and {Yokoyama}, T.},
        title = "{An active region jet observed with Hinode}",
      journal = {\aap},
         year = 2008,
        month = apr,
       volume = {481},
       number = {1},
        pages = {L57-L60},
          doi = {10.1051/0004-6361:20079081},
       adsurl = {https://ui.adsabs.harvard.edu/abs/2008A&A...481L..57C}
}

@ARTICLE{Vernazza1981,
       author = {{Vernazza}, J.~E. and {Avrett}, E.~H. and {Loeser}, R.},
        title = "{Structure of the solar chromosphere. III. Models of the EUV brightness components of the quiet sun.}",
      journal = {\apjs},
         year = 1981,
        month = apr,
       volume = {45},
        pages = {635-725},
          doi = {10.1086/190731},
       adsurl = {https://ui.adsabs.harvard.edu/abs/1981ApJS...45..635V}
}

@ARTICLE{Handy1999,
       author = {{Handy}, B.~N. and {Acton}, L.~W. and {Kankelborg}, C.~C. and {Wolfson}, C.~J. and {Akin}, D.~J. and {Bruner}, M.~E. and {Caravalho}, R. and {Catura}, R.~C. and {Chevalier}, R. and {Duncan}, D.~W. and {Edwards}, C.~G. and {Feinstein}, C.~N. and {Freeland}, S.~L. and {Friedlaender}, F.~M. and {Hoffmann}, C.~H. and {Hurlburt}, N.~E. and {Jurcevich}, B.~K. and {Katz}, N.~L. and {Kelly}, G.~A. and {Lemen}, J.~R. and {Levay}, M. and {Lindgren}, R.~W. and {Mathur}, D.~P. and {Meyer}, S.~B. and {Morrison}, S.~J. and {Morrison}, M.~D. and {Nightingale}, R.~W. and {Pope}, T.~P. and {Rehse}, R.~A. and {Schrijver}, C.~J. and {Shine}, R.~A. and {Shing}, L. and {Strong}, K.~T. and {Tarbell}, T.~D. and {Title}, A.~M. and {Torgerson}, D.~D. and {Golub}, L. and {Bookbinder}, J.~A. and {Caldwell}, D. and {Cheimets}, P.~N. and {Davis}, W.~N. and {Deluca}, E.~E. and {McMullen}, R.~A. and {Warren}, H.~P. and {Amato}, D. and {Fisher}, R. and {Maldonado}, H. and {Parkinson}, C.},
        title = "{The transition region and coronal explorer}",
      journal = {\solphys},
         year = 1999,
        month = jul,
       volume = {187},
       number = {2},
        pages = {229-260},
          doi = {10.1023/A:1005166902804},
       adsurl = {https://ui.adsabs.harvard.edu/abs/1999SoPh..187..229H}
}

@ARTICLE{Wilhelm1995,
       author = {{Wilhelm}, K. and {Curdt}, W. and {Marsch}, E. and {Sch{\"u}hle}, U. and {Lemaire}, P. and {Gabriel}, A. and {Vial}, J. -C. and {Grewing}, M. and {Huber}, M.~C.~E. and {Jordan}, S.~D. and {Poland}, A.~I. and {Thomas}, R.~J. and {K{\"u}hne}, M. and {Timothy}, J.~G. and {Hassler}, D.~M. and {Siegmund}, O.~H.~W.},
        title = "{SUMER - Solar Ultraviolet Measurements of Emitted Radiation}",
      journal = {\solphys},
         year = 1995,
        month = dec,
       volume = {162},
       number = {1-2},
        pages = {189-231},
          doi = {10.1007/BF00733430},
       adsurl = {https://ui.adsabs.harvard.edu/abs/1995SoPh..162..189W}
}

@ARTICLE{Kohl1995,
       author = {{Kohl}, J.~L. and {Esser}, R. and {Gardner}, L.~D. and {Habbal}, S. and {Daigneau}, P.~S. and {Dennis}, E.~F. and {Nystrom}, G.~U. and {Panasyuk}, A. and {Raymond}, J.~C. and {Smith}, P.~L. and {Strachan}, L. and {Van Ballegooijen}, A.~A. and {Noci}, G. and {Fineschi}, S. and {Romoli}, M. and {Ciaravella}, A. and {Modigliani}, A. and {Huber}, M.~C.~E. and {Antonucci}, E. and {Benna}, C. and {Giordano}, S. and {Tondello}, G. and {Nicolosi}, P. and {Naletto}, G. and {Pernechele}, C. and {Spadaro}, D. and {Poletto}, G. and {Livi}, S. and {Von Der L{\"u}he}, O. and {Geiss}, J. and {Timothy}, J.~G. and {Gloeckler}, G. and {Allegra}, A. and {Basile}, G. and {Brusa}, R. and {Wood}, B. and {Siegmund}, O.~H.~W. and {Fowler}, W. and {Fisher}, R. and {Jhabvala}, M.},
        title = "{The Ultraviolet Coronagraph Spectrometer for the Solar and Heliospheric Observatory}",
      journal = {\solphys},
         year = 1995,
        month = dec,
       volume = {162},
       number = {1-2},
        pages = {313-356},
          doi = {10.1007/BF00733433},
       adsurl = {https://ui.adsabs.harvard.edu/abs/1995SoPh..162..313K}
}

@ARTICLE{Domingo1995,
       author = {{Domingo}, V. and {Fleck}, B. and {Poland}, A.~I.},
        title = "{The SOHO Mission: an Overview}",
      journal = {\solphys},
         year = 1995,
        month = dec,
       volume = {162},
       number = {1-2},
        pages = {1-37},
          doi = {10.1007/BF00733425},
       adsurl = {https://ui.adsabs.harvard.edu/abs/1995SoPh..162....1D}
}

@ARTICLE{Auchere2001,
       author = {{Auch{\`e}re}, Fr{\'e}d{\'e}ric and {Hassler}, Donald M. and {Slater}, David C. and {Woods}, Thomas N.},
        title = "{SWRI/LASP Sounding Rocket Inter-Calibration With The Eit Instrument On Board SOHO}",
      journal = {\solphys},
         year = 2001,
        month = sep,
       volume = {202},
       number = {2},
        pages = {269-280},
          doi = {10.1023/A:1012237825048},
       adsurl = {https://ui.adsabs.harvard.edu/abs/2001SoPh..202..269A}
}

@ARTICLE{Korendyke2001,
       author = {{Korendyke}, C.~M. and {Vourlidas}, A. and {Cook}, J.~W. and {Dere}, K.~P. and {Howard}, R.~A. and {Morrill}, J.~S. and {Moses}, J.~D. and {Moulton}, N.~E. and {Socker}, D.~G.},
        title = "{High-resolution Imaging of the Upper Solar Chromosphere: First Light Performance of the Very-high-Resolution Advanced ULtraviolet Telescope}",
      journal = {\solphys},
         year = 2001,
        month = may,
       volume = {200},
        pages = {63-73},
          doi = {10.1023/A:1010310217570},
       adsurl = {https://ui.adsabs.harvard.edu/abs/2001SoPh..200...63K}
}

@ARTICLE{Vourlidas2010,
       author = {{Vourlidas}, A. and {Sanchez Andrade-Nu{\~n}o}, B. and {Landi}, E. and {Patsourakos}, S. and {Teriaca}, L. and {Sch{\"u}hle}, U. and {Korendyke}, C.~M. and {Nestoras}, I.},
        title = "{The Structure and Dynamics of the Upper Chromosphere and Lower Transition Region as Revealed by the Subarcsecond VAULT Observations}",
      journal = {\solphys},
         year = 2010,
        month = jan,
       volume = {261},
       number = {1},
          eid = {53},
        pages = {53},
          doi = {10.1007/s11207-009-9475-x},
archivePrefix = {arXiv},
       eprint = {0912.2272},
 primaryClass = {astro-ph.SR},
       adsurl = {https://ui.adsabs.harvard.edu/abs/2010SoPh..261...53V}
}

@ARTICLE{Rochus2020AA,
       author = {{Rochus}, P. and {Auch{\`e}re}, F. and {Berghmans}, D. and {Harra}, L. and {Schmutz}, W. and {Sch{\"u}hle}, U. and {Addison}, P. and {Appourchaux}, T. and {Aznar Cuadrado}, R. and {Baker}, D. and {Barbay}, J. and {Bates}, D. and {BenMoussa}, A. and {Bergmann}, M. and {Beurthe}, C. and {Borgo}, B. and {Bonte}, K. and {Bouzit}, M. and {Bradley}, L. and {B{\"u}chel}, V. and {Buchlin}, E. and {B{\"u}chner}, J. and {Cab{\'e}}, F. and {Cadiergues}, L. and {Chaigneau}, M. and {Chares}, B. and {Choque Cortez}, C. and {Coker}, P. and {Condamin}, M. and {Coumar}, S. and {Curdt}, W. and {Cutler}, J. and {Davies}, D. and {Davison}, G. and {Defise}, J. -M. and {Del Zanna}, G. and {Delmotte}, F. and {Delouille}, V. and {Dolla}, L. and {Dumesnil}, C. and {D{\"u}rig}, F. and {Enge}, R. and {Fran{\c{c}}ois}, S. and {Fourmond}, J. -J. and {Gillis}, J. -M. and {Giordanengo}, B. and {Gissot}, S. and {Green}, L.~M. and {Guerreiro}, N. and {Guilbaud}, A. and {Gyo}, M. and {Haberreiter}, M. and {Hafiz}, A. and {Hailey}, M. and {Halain}, J. -P. and {Hansotte}, J. and {Hecquet}, C. and {Heerlein}, K. and {Hellin}, M. -L. and {Hemsley}, S. and {Hermans}, A. and {Hervier}, V. and {Hochedez}, J. -F. and {Houbrechts}, Y. and {Ihsan}, K. and {Jacques}, L. and {J{\'e}r{\^o}me}, A. and {Jones}, J. and {Kahle}, M. and {Kennedy}, T. and {Klaproth}, M. and {Kolleck}, M. and {Koller}, S. and {Kotsialos}, E. and {Kraaikamp}, E. and {Langer}, P. and {Lawrenson}, A. and {Le Clech'}, J. -C. and {Lenaerts}, C. and {Liebecq}, S. and {Linder}, D. and {Long}, D.~M. and {Mampaey}, B. and {Markiewicz-Innes}, D. and {Marquet}, B. and {Marsch}, E. and {Matthews}, S. and {Mazy}, E. and {Mazzoli}, A. and {Meining}, S. and {Meltchakov}, E. and {Mercier}, R. and {Meyer}, S. and {Monecke}, M. and {Monfort}, F. and {Morinaud}, G. and {Moron}, F. and {Mountney}, L. and {M{\"u}ller}, R. and {Nicula}, B. and {Parenti}, S. and {Peter}, H. and {Pfiffner}, D. and {Philippon}, A. and {Phillips}, I. and {Plesseria}, J. -Y. and {Pylyser}, E. and {Rabecki}, F. and {Ravet-Krill}, M. -F. and {Rebellato}, J. and {Renotte}, E. and {Rodriguez}, L. and {Roose}, S. and {Rosin}, J. and {Rossi}, L. and {Roth}, P. and {Rouesnel}, F. and {Roulliay}, M. and {Rousseau}, A. and {Ruane}, K. and {Scanlan}, J. and {Schlatter}, P. and {Seaton}, D.~B. and {Silliman}, K. and {Smit}, S. and {Smith}, P.~J. and {Solanki}, S.~K. and {Spescha}, M. and {Spencer}, A. and {Stegen}, K. and {Stockman}, Y. and {Szwec}, N. and {Tamiatto}, C. and {Tandy}, J. and {Teriaca}, L. and {Theobald}, C. and {Tychon}, I. and {van Driel-Gesztelyi}, L. and {Verbeeck}, C. and {Vial}, J. -C. and {Werner}, S. and {West}, M.~J. and {Westwood}, D. and {Wiegelmann}, T. and {Willis}, G. and {Winter}, B. and {Zerr}, A. and {Zhang}, X. and {Zhukov}, A.~N.},
        title = "{The Solar Orbiter EUI instrument: The Extreme Ultraviolet Imager}",
      journal = {\aap},
         year = 2020,
        month = oct,
       volume = {642},
          eid = {A8},
        pages = {A8},
          doi = {10.1051/0004-6361/201936663},
       adsurl = {https://ui.adsabs.harvard.edu/abs/2020A&A...642A...8R}
}

@ARTICLE{Antonucci2020AA,
       author = {{Antonucci}, Ester and {Romoli}, Marco and {Andretta}, Vincenzo and {Fineschi}, Silvano and {Heinzel}, Petr and {Moses}, J. Daniel and {Naletto}, Giampiero and {Nicolini}, Gianalfredo and {Spadaro}, Daniele and {Teriaca}, Luca and {Berlicki}, Arkadiusz and {Capobianco}, Gerardo and {Crescenzio}, Giuseppe and {Da Deppo}, Vania and {Focardi}, Mauro and {Frassetto}, Fabio and {Heerlein}, Klaus and {Landini}, Federico and {Magli}, Enrico and {Marco Malvezzi}, Andrea and {Massone}, Giuseppe and {Melich}, Radek and {Nicolosi}, Piergiorgio and {Noci}, Giancarlo and {Pancrazzi}, Maurizio and {Pelizzo}, Maria G. and {Poletto}, Luca and {Sasso}, Clementina and {Sch{\"u}hle}, Udo and {Solanki}, Sami K. and {Strachan}, Leonard and {Susino}, Roberto and {Tondello}, Giuseppe and {Uslenghi}, Michela and {Woch}, Joachim and {Abbo}, Lucia and {Bemporad}, Alessandro and {Casti}, Marta and {Dolei}, Sergio and {Grimani}, Catia and {Messerotti}, Mauro and {Ricci}, Marco and {Straus}, Thomas and {Telloni}, Daniele and {Zuppella}, Paola and {Auch{\`e}re}, Frederic and {Bruno}, Roberto and {Ciaravella}, Angela and {Corso}, Alain J. and {Alvarez Copano}, Miguel and {Aznar Cuadrado}, Regina and {D'Amicis}, Raffaella and {Enge}, Reiner and {Gravina}, Alessio and {Jej{\v{c}}i{\v{c}}}, Sonja and {Lamy}, Philippe and {Lanzafame}, Alessandro and {Meierdierks}, Thimo and {Papagiannaki}, Ioanna and {Peter}, Hardi and {Fernandez Rico}, German and {Giday Sertsu}, Mewael and {Staub}, Jan and {Tsinganos}, Kanaris and {Velli}, Marco and {Ventura}, Rita and {Verroi}, Enrico and {Vial}, Jean~Claude and {Vives}, Sebastien and {Volpicelli}, Antonio and {Werner}, Stephan and {Zerr}, Andreas and {Negri}, Barbara and {Castronuovo}, Marco and {Gabrielli}, Alessandro and {Bertacin}, Roberto and {Carpentiero}, Rita and {Natalucci}, Silvia and {Marliani}, Filippo and {Cesa}, Marco and {Laget}, Philippe and {Morea}, Danilo and {Pieraccini}, Stefano and {Radaelli}, Paolo and {Sandri}, Paolo and {Sarra}, Paolo and {Cesare}, Stefano and {Del Forno}, Felice and {Massa}, Ernesto and {Montabone}, Mauro and {Mottini}, Sergio and {Quattropani}, Daniele and {Schillaci}, Tiziano and {Boccardo}, Roberto and {Brando}, Rosario and {Pandi}, Arianna and {Baietto}, Cristian and {Bertone}, Riccardo and {Alvarez-Herrero}, Alberto and {Garc{\'\i}a Parejo}, Pilar and {Cebollero}, Mar{\'\i}a and {Amoruso}, Mauro and {Centonze}, Vito},
        title = "{Metis: the Solar Orbiter visible light and ultraviolet coronal imager}",
      journal = {\aap},
         year = 2020,
        month = oct,
       volume = {642},
          eid = {A10},
        pages = {A10},
          doi = {10.1051/0004-6361/201935338},
archivePrefix = {arXiv},
       eprint = {1911.08462},
 primaryClass = {astro-ph.SR},
       adsurl = {https://ui.adsabs.harvard.edu/abs/2020A&A...642A..10A}
}

@ARTICLE{Muller2020AA,
       author = {{M{\"u}ller}, D. and {St. Cyr}, O.~C. and {Zouganelis}, I. and {Gilbert}, H.~R. and {Marsden}, R. and {Nieves-Chinchilla}, T. and {Antonucci}, E. and {Auch{\`e}re}, F. and {Berghmans}, D. and {Horbury}, T.~S. and {Howard}, R.~A. and {Krucker}, S. and {Maksimovic}, M. and {Owen}, C.~J. and {Rochus}, P. and {Rodriguez-Pacheco}, J. and {Romoli}, M. and {Solanki}, S.~K. and {Bruno}, R. and {Carlsson}, M. and {Fludra}, A. and {Harra}, L. and {Hassler}, D.~M. and {Livi}, S. and {Louarn}, P. and {Peter}, H. and {Sch{\"u}hle}, U. and {Teriaca}, L. and {del Toro Iniesta}, J.~C. and {Wimmer-Schweingruber}, R.~F. and {Marsch}, E. and {Velli}, M. and {De Groof}, A. and {Walsh}, A. and {Williams}, D.},
        title = "{The Solar Orbiter mission. Science overview}",
      journal = {\aap},
         year = 2020,
        month = oct,
       volume = {642},
          eid = {A1},
        pages = {A1},
          doi = {10.1051/0004-6361/202038467},
archivePrefix = {arXiv},
       eprint = {2009.00861},
 primaryClass = {astro-ph.SR},
       adsurl = {https://ui.adsabs.harvard.edu/abs/2020A&A...642A...1M}
}

@INPROCEEDINGS{Kano2012,
       author = {{Kano}, Ryouhei and {Bando}, Takamasa and {Narukage}, Noriyuki and {Ishikawa}, Ryoko and {Tsuneta}, Saku and {Katsukawa}, Yukio and {Kubo}, Masahito and {Ishikawa}, Shin-nosuke and {Hara}, Hirohisa and {Shimizu}, Toshifumi and {Suematsu}, Yoshinori and {Ichimoto}, Kiyoshi and {Sakao}, Taro and {Goto}, Motoshi and {Kato}, Yoshiaki and {Imada}, Shinsuke and {Kobayashi}, Ken and {Holloway}, Todd and {Winebarger}, Amy and {Cirtain}, Jonathan and {De Pontieu}, Bart and {Casini}, Roberto and {Trujillo Bueno}, Javier and {{\v{S}}tep{\'a}n}, Ji{\v{r}}{\'\i} and {Manso Sainz}, Rafael and {Belluzzi}, Luca and {Asensio Ramos}, Andres and {Auch{\`e}re}, Fr{\'e}d{\'e}ric and {Carlsson}, Mats},
        title = "{Chromospheric Lyman-alpha spectro-polarimeter (CLASP)}",
    booktitle = {Space Telescopes and Instrumentation 2012: Ultraviolet to Gamma Ray},
         year = 2012,
       editor = {{Takahashi}, Tadayuki and {Murray}, Stephen S. and {den Herder}, Jan-Willem A.},
       series = {Society of Photo-Optical Instrumentation Engineers (SPIE) Conference Series},
       volume = {8443},
        month = sep,
          eid = {84434F},
        pages = {84434F},
          doi = {10.1117/12.925991},
       adsurl = {https://ui.adsabs.harvard.edu/abs/2012SPIE.8443E..4FK}
}

@INPROCEEDINGS{Kobayashi2012,
       author = {{Kobayashi}, K. and {Kano}, R. and {Trujillo-Bueno}, J. and {Asensio Ramos}, A. and {Bando}, T. and {Belluzzi}, L. and {Carlsson}, M. and {De Pontieu}, R.~C.~B. and {Hara}, H. and {Ichimoto}, K. and {Ishikawa}, R. and {Katsukawa}, Y. and {Kubo}, M. and {Manso Sainz}, R. and {Narukage}, N. and {Sakao}, T. and {Stepan}, J. and {Suematsu}, Y. and {Tsuneta}, S. and {Watanabe}, H. and {Winebarger}, A.},
        title = "{The Chromospheric Lyman-Alpha SpectroPolarimeter: CLASP}",
    booktitle = {Fifth Hinode Science Meeting},
         year = 2012,
       editor = {{Golub}, L. and {De Moortel}, I. and {Shimizu}, T.},
       series = {Astronomical Society of the Pacific Conference Series},
       volume = {456},
        month = may,
        pages = {233},
       adsurl = {https://ui.adsabs.harvard.edu/abs/2012ASPC..456..233K}
}

@INPROCEEDINGS{Strachan2017,
       author = {{Strachan}, Leonard and {Laming}, J. Martin and {Ko}, Yuan-Kuen and {Tun Beltran}, Samuel and {Korendyke}, Clarence M. and {Brown}, Charles M. and {Socker}, Dennis G. and {Galysh}, Ivan J. and {Finne}, Theodore T. and {Eisenhower}, Kevin C. and {Brechbiel}, David J. and {Noya}, Mario and {Provornikova}, Elena and {Gardner}, Larry D.},
        title = "{The Ultraviolet Spectro-Coronagraph (UVSC) Pathfinder Experiment for the Remote Detection of Suprathermal Seed Particles: Instrument Status}",
    booktitle = {AAS/Solar Physics Division Abstracts \#48},
         year = 2017,
       series = {AAS/Solar Physics Division Meeting},
       volume = {48},
        month = aug,
          eid = {110.07},
        pages = {110.07},
       adsurl = {https://ui.adsabs.harvard.edu/abs/2017SPD....4811007S}
}

@ARTICLE{Gan2023SoPh,
       author = {{Gan}, Weiqun and {Zhu}, Cheng and {Deng}, Yuanyong and {Zhang}, Zhe and {Chen}, Bo and {Huang}, Yu and {Deng}, Lei and {Wu}, Haiyan and {Zhang}, Haiying and {Li}, Hui and {Su}, Yang and {Su}, Jiangtao and {Feng}, Li and {Wu}, Jian and {Cui}, Jijun and {Wang}, Chi and {Chang}, Jin and {Yin}, Zengshan and {Xiong}, Weiming and {Chen}, Bin and {Yang}, Jianfeng and {Li}, Fu and {Lin}, Jiaben and {Hou}, Junfeng and {Bai}, Xianyong and {Chen}, Dengyi and {Zhang}, Yan and {Hu}, Yiming and {Liang}, Yaoming and {Wang}, Jianping and {Song}, Kefei and {Guo}, Quanfeng and {He}, Lingping and {Zhang}, Guang and {Wang}, Peng and {Bao}, Haicao and {Cao}, Caixia and {Bai}, Yanping and {Chen}, Binglong and {He}, Tao and {Li}, Xinyu and {Zhang}, Ye and {Liao}, Xing and {Jiang}, Hu and {Li}, Youping and {Su}, Yingna and {Lei}, Shijun and {Chen}, Wei and {Li}, Ying and {Zhao}, Jie and {Li}, Jingwei and {Ge}, Yunyi and {Zou}, Ziming and {Hu}, Tai and {Su}, Miao and {Ji}, Haidong and {Gu}, Mei and {Zheng}, Yonghuang and {Xu}, Dezhen and {Wang}, Xing},
        title = "{The Advanced Space-Based Solar Observatory (ASO-S)}",
      journal = {\solphys},
         year = 2023,
        month = may,
       volume = {298},
       number = {5},
          eid = {68},
        pages = {68},
          doi = {10.1007/s11207-023-02166-x},
       adsurl = {https://ui.adsabs.harvard.edu/abs/2023SoPh..298...68G}
}

@ARTICLE{Li2019RAA,
       author = {{Li}, Hui and {Chen}, Bo and {Feng}, Li and {Li}, Ying and {Huang}, Yu and {Li}, Jing~Wei and {Lu}, Lei and {Xue}, Jian~Chao and {Ying}, Bei~Li and {Zhao}, Jie and {Yang}, Yu~Tong and {Gan}, Wei~Qun and {Fang}, Cheng and {Song}, Ke~Fei and {Wang}, Hong and {Guo}, Quan~Feng and {He}, Ling~Ping and {Zhu}, Bo and {Zhu}, Cheng and {Deng}, Lei and {Bao}, Hai~Chao and {Cao}, Cai~Xia and {Yang}, Zhong~Guang},
        title = "{The Lyman-alpha Solar Telescope (LST) for the ASO-S mission {\textemdash} I. Scientific objectives and overview}",
      journal = {\raa},
         year = 2019,
        month = nov,
       volume = {19},
       number = {11},
          eid = {158},
        pages = {158},
          doi = {10.1088/1674-4527/19/11/158},
       adsurl = {https://ui.adsabs.harvard.edu/abs/2019RAA....19..158L}
}

@ARTICLE{Chen2019RAA,
       author = {{Chen}, Bo and {Li}, Hui and {Song}, Ke~Fei and {Guo}, Quan~Feng and {Zhang}, Pei~Jie and {He}, Ling~Ping and {Dai}, Shuang and {Wang}, Xiao~Dong and {Wang}, Hai~Feng and {Liu}, Chun~Long and {Zhang}, Hong~Ji and {Zhang}, Guang and {Wang}, Yunqi and {Liu}, Shi~Jie and {Zhang}, Hong~Xin and {Liu}, Lei and {Mao}, Shi~Lei and {Liu}, Yang and {Peng}, Jia~Hao and {Wang}, Peng and {Sun}, Liang and {Liu}, Yang and {Han}, Zhen~Wei and {Wang}, Yan~Long and {Wu}, Kun and {Ding}, Guang~Xing and {Zhou}, Peng and {Zheng}, Xin and {Xia}, Ming~Yi and {Wu}, Qing~Wen and {Xie}, Jin~Jiang and {Chen}, Ya and {Song}, Shu~Mei and {Wang}, Hong and {Zhu}, Bo and {Chu}, Chang~Bo and {Yang}, Wen~Gang and {Feng}, Li and {Huang}, Yu and {Gan}, Wei~Qun and {Li}, Ying and {Li}, Jing~Wei and {Lu}, Lei and {Xue}, Jian~Chao and {Ying}, Bei~Li and {Sun}, Ming~Zhe and {Zhu}, Cheng and {Bao}, Wei~Min and {Deng}, Lei and {Yin}, Zeng~Shan},
        title = "{The Lyman-alpha Solar Telescope (LST) for the ASO-S mission - II. design of LST}",
      journal = {\raa},
         year = 2019,
        month = nov,
       volume = {19},
       number = {11},
          eid = {159},
        pages = {159},
          doi = {10.1088/1674-4527/19/11/159},
       adsurl = {https://ui.adsabs.harvard.edu/abs/2019RAA....19..159C}
}

@ARTICLE{Feng2019RAA,
       author = {{Feng}, Li and {Li}, Hui and {Chen}, Bo and {Li}, Ying and {Susino}, Roberto and {Huang}, Yu and {Lu}, Lei and {Ying}, Bei~Li and {Li}, Jing~Wei and {Xue}, Jian~Chao and {Yang}, Yu~Tong and {Hong}, Jie and {Li}, Jian~Ping and {Zhao}, Jie and {Gan}, Wei~Qun and {Zhang}, Yan},
        title = "{The Lyman-alpha Solar Telescope (LST) for the ASO-S mission - III. data and potential diagnostics}",
      journal = {\raa},
         year = 2019,
        month = nov,
       volume = {19},
       number = {11},
          eid = {162},
        pages = {162},
          doi = {10.1088/1674-4527/19/11/162},
       adsurl = {https://ui.adsabs.harvard.edu/abs/2019RAA....19..162F}
}

@ARTICLE{Chen2024SoPh,
       author = {{Chen}, Bo and {Feng}, Li and {Zhang}, Guang and {Li}, Hui and {He}, Lingping and {Song}, Kefei and {Guo}, Quanfeng and {Li}, Ying and {Huang}, Yu and {Li}, Jingwei and {Zhao}, Jie and {Xue}, Jianchao and {Li}, Gen and {Shi}, Guanglu and {Song}, Dechao and {Lu}, Lei and {Ying}, Beili and {Wang}, Haifeng and {Dai}, Shuang and {Wang}, Xiaodong and {Mao}, Shilei and {Wang}, Peng and {Wu}, Kun and {Ren}, Shuai and {Sun}, Liang and {Yang}, Xianwei and {Xia}, Mingyi and {Zhang}, Xiaoxue and {Zhou}, Peng and {Tao}, Chen and {Liu}, Yang and {Yu}, Sibo and {Li}, Xinkai and {Li}, Shuting and {Zhang}, Ping and {Li}, Qiao and {Tian}, Zhengyuan and {Zhou}, Yue and {Tian}, Jun and {Shan}, Jiahui and {Liu}, Xiaofeng and {Jing}, Zhichen and {Gan}, Weiqun},
        title = "{Inflight Performance and Calibrations of the Lyman-Alpha Solar Telescope on Board the Advanced Space-Based Solar Observatory}",
      journal = {\solphys},
         year = 2024,
        month = aug,
       volume = {299},
       number = {8},
          eid = {118},
        pages = {118},
          doi = {10.1007/s11207-024-02354-3},
archivePrefix = {arXiv},
       eprint = {2408.01937},
 primaryClass = {astro-ph.SR},
       adsurl = {https://ui.adsabs.harvard.edu/abs/2024SoPh..299..118C}
}

@ARTICLE{Yiliang2025,
       author = {{Li}, Yiliang and {Zhang}, Ping and {Tian}, Zhengyuan and {Feng}, Li and {Shi}, Guanglu and {Xue}, Jianchao and {Li}, Ying and {Tian}, Jun and {Ji}, Kaifan and {Ying}, Beili and {Lu}, Lei and {Li}, Shuting and {Shan}, Jiahui and {Li}, Hui and {Gan}, Weiqun},
        title = "{Understanding the Ly{\ensuremath{\alpha}} Emission Observed by the Solar Disk Imager Aboard the Advanced Space-Based Solar Observatory}",
      journal = {\solphys},
         year = 2025,
        month = jun,
       volume = {300},
       number = {6},
          eid = {78},
        pages = {78},
          doi = {10.1007/s11207-025-02491-3},
archivePrefix = {arXiv},
       eprint = {2505.16218},
 primaryClass = {astro-ph.SR},
       adsurl = {https://ui.adsabs.harvard.edu/abs/2025SoPh..300...78L}
}

@ARTICLE{Shi2024SoPh,
       author = {{Shi}, Guanglu and {Feng}, Li and {Chen}, Jun and {Ying}, Beili and {Li}, Shuting and {Li}, Qiao and {Li}, Hui and {Li}, Ying and {Ji}, Kaifan and {Huang}, Yu and {Li}, Youping and {Li}, Jingwei and {Zhao}, Jie and {Lu}, Lei and {Xue}, Jianchao and {Zhang}, Ping and {Song}, Dechao and {Tian}, Zhengyuan and {Su}, Yingna and {Zhang}, Qingmin and {Ge}, Yunyi and {Shan}, Jiahui and {Zhou}, Yue and {Tian}, Jun and {Li}, Gen and {Liu}, Xiaofeng and {Jing}, Zhichen and {Lei}, Shijun and {Gan}, Weiqun},
        title = "{Asymmetric Hard X-ray Radiation of Two Ribbons in a Thermal-Dominated C-Class Flare}",
      journal = {\solphys},
         year = 2024,
        month = jul,
       volume = {299},
       number = {7},
          eid = {104},
        pages = {104},
          doi = {10.1007/s11207-024-02349-0},
archivePrefix = {arXiv},
       eprint = {2407.13099},
 primaryClass = {astro-ph.SR},
       adsurl = {https://ui.adsabs.harvard.edu/abs/2024SoPh..299..104S}
}

@ARTICLE{Xue2024SoPh,
       author = {{Xue}, Jian~Chao and {Feng}, Li and {Li}, Hui and {Zhang}, Ping and {Chen}, Jun and {Shi}, Guang~Lu and {Ji}, Kai~Fan and {Qiu}, Ye and {Li}, Chuan and {Lu}, Lei and {Ying}, Bei~Li and {Li}, Ying and {Huang}, Yu and {Li}, You~Ping and {Li}, Jing~Wei and {Zhao}, Jie and {Song}, De~Chao and {Li}, Shu~Ting and {Tian}, Zheng~Yuan and {Su}, Ying~Na and {Zhang}, Qing~Min and {Ge}, Yun~Yi and {Shan}, Jia~Hui and {Li}, Qiao and {Li}, Gen and {Zhou}, Yue and {Tian}, Jun and {Liu}, Xiao~Feng and {Jing}, Zhi~Chen and {Chen}, Bo and {Song}, Ke~Fei and {He}, Ling~Ping and {Lei}, Shi~Jun and {Gan}, Wei~Qun},
        title = "{Association Between a Failed Prominence Eruption and the Drainage of Mass from Another Prominence}",
      journal = {\solphys},
         year = 2024,
        month = jun,
       volume = {299},
       number = {6},
          eid = {89},
        pages = {89},
          doi = {10.1007/s11207-024-02336-5},
archivePrefix = {arXiv},
       eprint = {2406.11602},
 primaryClass = {astro-ph.SR},
       adsurl = {https://ui.adsabs.harvard.edu/abs/2024SoPh..299...89X}
}

@ARTICLE{Lu2024SoPh,
       author = {{Lu}, Lei and {Tian}, Zhengyuan and {Feng}, Li and {Shan}, Jiahui and {Li}, Hui and {Su}, Yang and {Li}, Ying and {Huang}, Yu and {Li}, Youping and {Li}, Jingwei and {Zhao}, Jie and {Ying}, Beili and {Xue}, Jianchao and {Zhang}, Ping and {Song}, Dechao and {Li}, Shuting and {Shi}, Guanglu and {Su}, Yingna and {Zhang}, Qingmin and {Ge}, Yunyi and {Chen}, Bo and {Li}, Qiao and {Li}, Gen and {Zhou}, Yue and {Tian}, Jun and {Liu}, Xiaofeng and {Jing}, Zhichen and {Gan}, Weiqun and {Song}, Kefei and {He}, Lingping and {Lei}, Shijun},
        title = "{Automatic Solar Flare Detection Using the Solar Disk Imager Onboard the ASO-S Mission}",
      journal = {\solphys},
         year = 2024,
        month = may,
       volume = {299},
       number = {5},
          eid = {72},
        pages = {72},
          doi = {10.1007/s11207-024-02310-1},
       adsurl = {https://ui.adsabs.harvard.edu/abs/2024SoPh..299...72L}
}

@ARTICLE{Noci1987,
       author = {{Noci}, Giancarlo and {Kohl}, John L. and {Withbroe}, George L.},
        title = "{Solar Wind Diagnostics from Doppler-enhanced Scattering}",
      journal = {\apj},
         year = 1987,
        month = apr,
       volume = {315},
        pages = {706},
          doi = {10.1086/165172},
       adsurl = {https://ui.adsabs.harvard.edu/abs/1987ApJ...315..706N}
}

@ARTICLE{Ying2020ApJ,
       author = {{Ying}, Beili and {Bemporad}, Alessandro and {Feng}, Li and {Lu}, Lei and {Gan}, Weiqun and {Li}, Hui},
        title = "{Extensive Study of a Coronal Mass Ejection with UV and White-light Coronagraphs: The Need for Multiwavelength Observations}",
      journal = {\apj},
         year = 2020,
        month = aug,
       volume = {899},
       number = {1},
          eid = {12},
        pages = {12},
          doi = {10.3847/1538-4357/aba431},
archivePrefix = {arXiv},
       eprint = {2007.04575},
 primaryClass = {astro-ph.SR},
       adsurl = {https://ui.adsabs.harvard.edu/abs/2020ApJ...899...12Y}
}

@ARTICLE{Ying2024SoPh,
       author = {{Ying}, Beili and {Shi}, Guanglu and {Feng}, Li and {Lu}, Lei and {Xue}, Jianchao and {Li}, Shuting and {Gan}, Weiqun and {Li}, Hui},
        title = "{Parameter Effects on the Total Intensity of H I Ly{\ensuremath{\alpha}} Line for a Modeled Coronal Mass Ejection and Its Driven Shock}",
      journal = {\solphys},
         year = 2024,
        month = jun,
       volume = {299},
       number = {6},
          eid = {92},
        pages = {92},
          doi = {10.1007/s11207-024-02330-x},
archivePrefix = {arXiv},
       eprint = {2406.11297},
 primaryClass = {astro-ph.SR},
       adsurl = {https://ui.adsabs.harvard.edu/abs/2024SoPh..299...92Y}
}

@ARTICLE{Li2024ApJ,
       author = {{Li}, Shuting and {Feng}, Li and {Ying}, Beili and {Lu}, Lei and {Ji}, Kaifan and {Li}, Ying and {Shi}, Guanglu and {Li}, Hui and {Huang}, Yu and {Gan}, Weiqun},
        title = "{Carrington Maps in H I Ly{\ensuremath{\alpha}} and Their Relationships with Extreme-ultraviolet and Magnetic-field Maps}",
      journal = {\apjl},
         year = 2024,
        month = jul,
       volume = {969},
       number = {1},
          eid = {L16},
        pages = {L16},
          doi = {10.3847/2041-8213/ad5679},
       adsurl = {https://ui.adsabs.harvard.edu/abs/2024ApJ...969L..16L}
}

@ARTICLE{Fan2019,
       author = {{Fan}, Yuhong and {Liu}, Tie},
        title = "{MHD simulation of prominence-cavity system}",
      journal = {Frontiers in Astronomy and Space Sciences},
         year = 2019,
        month = apr,
       volume = {6},
          eid = {27},
        pages = {27},
          doi = {10.3389/fspas.2019.00027},
archivePrefix = {arXiv},
       eprint = {1905.08226},
 primaryClass = {astro-ph.SR},
       adsurl = {https://ui.adsabs.harvard.edu/abs/2019FrASS...6...27F}
}

@ARTICLE{Zhao2022AA,
       author = {{Zhao}, J. and {Zhang}, P. and {Gibson}, S.~E. and {Fan}, Y. and {Feng}, L. and {Yu}, F. and {Li}, H. and {Gan}, W.~Q.},
        title = "{Synthetic Lyman-{\ensuremath{\alpha}} emissions for the coronagraph aboard the ASO-S mission. I. An eruptive prominence-cavity system}",
      journal = {\aap},
         year = 2022,
        month = sep,
       volume = {665},
          eid = {A39},
        pages = {A39},
          doi = {10.1051/0004-6361/202243029},
       adsurl = {https://ui.adsabs.harvard.edu/abs/2022A&A...665A..39Z}
}

@ARTICLE{Dissauer2018,
       author = {{Dissauer}, K. and {Veronig}, A.~M. and {Temmer}, M. and {Podladchikova}, T. and {Vanninathan}, K.},
        title = "{Statistics of Coronal Dimmings Associated with Coronal Mass Ejections. I. Characteristic Dimming Properties and Flare Association}",
      journal = {\apj},
         year = 2018,
        month = aug,
       volume = {863},
       number = {2},
          eid = {169},
        pages = {169},
          doi = {10.3847/1538-4357/aad3c6},
archivePrefix = {arXiv},
       eprint = {1807.05056},
 primaryClass = {astro-ph.SR},
       adsurl = {https://ui.adsabs.harvard.edu/abs/2018ApJ...863..169D}
}

@ARTICLE{Dissauer2019,
       author = {{Dissauer}, K. and {Veronig}, A.~M. and {Temmer}, M. and {Podladchikova}, T.},
        title = "{Statistics of Coronal Dimmings Associated with Coronal Mass Ejections. II. Relationship between Coronal Dimmings and Their Associated CMEs}",
      journal = {\apj},
         year = 2019,
        month = apr,
       volume = {874},
       number = {2},
          eid = {123},
        pages = {123},
          doi = {10.3847/1538-4357/ab0962},
archivePrefix = {arXiv},
       eprint = {1810.01589},
 primaryClass = {astro-ph.SR},
       adsurl = {https://ui.adsabs.harvard.edu/abs/2019ApJ...874..123D}
}

@ARTICLE{Hudson1996,
       author = {{Hudson}, H.~S. and {Acton}, L.~W. and {Freeland}, S.~L.},
        title = "{A Long-Duration Solar Flare with Mass Ejection and Global Consequences}",
      journal = {\apj},
         year = 1996,
        month = oct,
       volume = {470},
        pages = {629},
          doi = {10.1086/177894},
       adsurl = {https://ui.adsabs.harvard.edu/abs/1996ApJ...470..629H}
}

@ARTICLE{Thompson1998,
       author = {{Thompson}, B.~J. and {Plunkett}, S.~P. and {Gurman}, J.~B. and {Newmark}, J.~S. and {St. Cyr}, O.~C. and {Michels}, D.~J.},
        title = "{SOHO/EIT observations of an Earth-directed coronal mass ejection on May 12, 1997}",
      journal = {\grl},
         year = 1998,
        month = jul,
       volume = {25},
       number = {14},
        pages = {2465-2468},
          doi = {10.1029/98GL50429},
       adsurl = {https://ui.adsabs.harvard.edu/abs/1998GeoRL..25.2465T}
}

@ARTICLE{Sterling1997,
       author = {{Sterling}, Alphonse C. and {Hudson}, Hugh S.},
        title = "{Yohkoh SXT Observations of X-Ray ``Dimming'' Associated with a Halo Coronal Mass Ejection}",
      journal = {\apjl},
         year = 1997,
        month = dec,
       volume = {491},
       number = {1},
        pages = {L55-L58},
          doi = {10.1086/311043},
       adsurl = {https://ui.adsabs.harvard.edu/abs/1997ApJ...491L..55S}
}

@ARTICLE{Chen2019,
       author = {{Chen}, Hechao and {Yang}, Jiayan and {Ji}, Kaifan and {Duan}, Yadan},
        title = "{Observational Analysis on the Early Evolution of a CME Flux Rope: Preflare Reconnection and Flux Rope{\textquoteright}s Footpoint Drift}",
      journal = {\apj},
         year = 2019,
        month = dec,
       volume = {887},
       number = {2},
          eid = {118},
        pages = {118},
          doi = {10.3847/1538-4357/ab527e},
archivePrefix = {arXiv},
       eprint = {1911.00257},
 primaryClass = {astro-ph.SR},
       adsurl = {https://ui.adsabs.harvard.edu/abs/2019ApJ...887..118C}
}

@ARTICLE{Gou2019,
       author = {{Gou}, Tingyu and {Liu}, Rui and {Kliem}, Bernhard and {Wang}, Yuming and {Veronig}, Astrid M.},
        title = "{The Birth of A Coronal Mass Ejection}",
      journal = {Science Advances},
         year = 2019,
        month = mar,
       volume = {5},
       number = {3},
        pages = {7004},
          doi = {10.1126/sciadv.aau7004},
archivePrefix = {arXiv},
       eprint = {1811.04707},
 primaryClass = {astro-ph.SR},
       adsurl = {https://ui.adsabs.harvard.edu/abs/2019SciA....5.7004G}
}

@ARTICLE{Thompson2000,
       author = {{Thompson}, B.~J. and {Cliver}, E.~W. and {Nitta}, N. and {Delann{\'e}e}, C. and {Delaboudini{\`e}re}, J.-P.},
        title = "{Coronal dimmings and energetic CMEs in April-May 1998}",
      journal = {\grl},
         year = 2000,
        month = may,
       volume = {27},
       number = {10},
        pages = {1431-1434},
          doi = {10.1029/1999GL003668},
       adsurl = {https://ui.adsabs.harvard.edu/abs/2000GeoRL..27.1431T}
}

@ARTICLE{Mandrini2007,
       author = {{Mandrini}, C.~H. and {Nakwacki}, M.~S. and {Attrill}, G. and {van Driel-Gesztelyi}, L. and {D{\'e}moulin}, P. and {Dasso}, S. and {Elliott}, H.},
        title = "{Are CME-Related Dimmings Always a Simple Signature of Interplanetary Magnetic Cloud Footpoints?}",
      journal = {\solphys},
         year = 2007,
        month = aug,
       volume = {244},
       number = {1-2},
        pages = {25-43},
          doi = {10.1007/s11207-007-9020-8},
       adsurl = {https://ui.adsabs.harvard.edu/abs/2007SoPh..244...25M}
}

@ARTICLE{Qiu2017,
       author = {{Qiu}, Jiong and {Cheng}, Jianxia},
        title = "{Gradual Solar Coronal Dimming and Evolution of Coronal Mass Ejection in the Early Phase}",
      journal = {\apjl},
         year = 2017,
        month = mar,
       volume = {838},
       number = {1},
          eid = {L6},
        pages = {L6},
          doi = {10.3847/2041-8213/aa6798},
archivePrefix = {arXiv},
       eprint = {1707.02480},
 primaryClass = {astro-ph.SR},
       adsurl = {https://ui.adsabs.harvard.edu/abs/2017ApJ...838L...6Q}
}

@ARTICLE{Wang2023,
       author = {{Wang}, Wensi and {Qiu}, Jiong and {Liu}, Rui and {Zhu}, Chunming and {Yang}, Kai E. and {Hu}, Qiang and {Wang}, Yuming},
        title = "{Investigating Pre-eruptive Magnetic Properties at the Footprints of Erupting Magnetic Flux Ropes}",
      journal = {\apj},
         year = 2023,
        month = feb,
       volume = {943},
       number = {2},
          eid = {80},
        pages = {80},
          doi = {10.3847/1538-4357/aca6e1},
archivePrefix = {arXiv},
       eprint = {2211.15909},
 primaryClass = {astro-ph.SR},
       adsurl = {https://ui.adsabs.harvard.edu/abs/2023ApJ...943...80W}
}

@ARTICLE{Vanninathan2018,
       author = {{Vanninathan}, Kamalam and {Veronig}, Astrid M. and {Dissauer}, Karin and {Temmer}, Manuela},
        title = "{Plasma Diagnostics of Coronal Dimming Events}",
      journal = {\apj},
         year = 2018,
        month = apr,
       volume = {857},
       number = {1},
          eid = {62},
        pages = {62},
          doi = {10.3847/1538-4357/aab09a},
archivePrefix = {arXiv},
       eprint = {1802.06152},
 primaryClass = {astro-ph.SR},
       adsurl = {https://ui.adsabs.harvard.edu/abs/2018ApJ...857...62V}
}

@ARTICLE{Tian2012,
       author = {{Tian}, Hui and {McIntosh}, Scott W. and {Xia}, Lidong and {He}, Jiansen and {Wang}, Xin},
        title = "{What can We Learn about Solar Coronal Mass Ejections, Coronal Dimmings, and Extreme-ultraviolet Jets through Spectroscopic Observations?}",
      journal = {\apj},
         year = 2012,
        month = apr,
       volume = {748},
       number = {2},
          eid = {106},
        pages = {106},
          doi = {10.1088/0004-637X/748/2/106},
archivePrefix = {arXiv},
       eprint = {1201.2204},
 primaryClass = {astro-ph.SR},
       adsurl = {https://ui.adsabs.harvard.edu/abs/2012ApJ...748..106T}
}

@ARTICLE{Reinard2008,
       author = {{Reinard}, A.~A. and {Biesecker}, D.~A.},
        title = "{Coronal Mass Ejection-Associated Coronal Dimmings}",
      journal = {\apj},
         year = 2008,
        month = feb,
       volume = {674},
       number = {1},
        pages = {576-585},
          doi = {10.1086/525269},
       adsurl = {https://ui.adsabs.harvard.edu/abs/2008ApJ...674..576R}
}

@ARTICLE{Veronig2021,
       author = {{Veronig}, Astrid M. and {Odert}, Petra and {Leitzinger}, Martin and {Dissauer}, Karin and {Fleck}, Nikolaus C. and {Hudson}, Hugh S.},
        title = "{Indications of stellar coronal mass ejections through coronal dimmings}",
      journal = {Nature Astronomy},
         year = 2021,
        month = jan,
       volume = {5},
        pages = {697-706},
          doi = {10.1038/s41550-021-01345-9},
archivePrefix = {arXiv},
       eprint = {2110.12029},
 primaryClass = {astro-ph.SR},
       adsurl = {https://ui.adsabs.harvard.edu/abs/2021NatAs...5..697V}
}

@ARTICLE{Harra2007,
       author = {{Harra}, Louise K. and {Hara}, Hirohisa and {Imada}, Shinsuke and {Young}, Peter R. and {Williams}, David R. and {Sterling}, Alphonse C. and {Korendyke}, Clarence and {Attrill}, Gemma D.~R.},
        title = "{Coronal Dimming Observed with Hinode: Outflows Related to a Coronal Mass Ejection}",
      journal = {\pasj},
         year = 2007,
        month = nov,
       volume = {59},
        pages = {S801},
          doi = {10.1093/pasj/59.sp3.S801},
       adsurl = {https://ui.adsabs.harvard.edu/abs/2007PASJ...59S.801H}
}

@ARTICLE{Veronig2019,
       author = {{Veronig}, Astrid M. and {G{\"o}m{\"o}ry}, Peter and {Dissauer}, Karin and {Temmer}, Manuela and {Vanninathan}, Kamalam},
        title = "{Spectroscopy and Differential Emission Measure Diagnostics of a Coronal Dimming Associated with a Fast Halo CME}",
      journal = {\apj},
         year = 2019,
        month = jul,
       volume = {879},
       number = {2},
          eid = {85},
        pages = {85},
          doi = {10.3847/1538-4357/ab2712},
archivePrefix = {arXiv},
       eprint = {1906.01517},
 primaryClass = {astro-ph.SR},
       adsurl = {https://ui.adsabs.harvard.edu/abs/2019ApJ...879...85V}
}

@ARTICLE{McIntosh2009,
       author = {{McIntosh}, Scott W.},
        title = "{The Inconvenient Truth About Coronal Dimmings}",
      journal = {\apj},
         year = 2009,
        month = mar,
       volume = {693},
       number = {2},
        pages = {1306-1309},
          doi = {10.1088/0004-637X/693/2/1306},
archivePrefix = {arXiv},
       eprint = {0809.4024},
 primaryClass = {astro-ph},
       adsurl = {https://ui.adsabs.harvard.edu/abs/2009ApJ...693.1306M}
}

@ARTICLE{Kahler2001,
       author = {{Kahler}, S.~W. and {Hudson}, H.~S.},
        title = "{Origin and development of transient coronal holes}",
      journal = {\jgr},
         year = 2001,
        month = dec,
       volume = {106},
       number = {A12},
        pages = {29239-29248},
          doi = {10.1029/2001JA000127},
       adsurl = {https://ui.adsabs.harvard.edu/abs/2001JGR...10629239K}
}

@ARTICLE{Attrill2008,
       author = {{Attrill}, G.~D.~R. and {van Driel-Gesztelyi}, L. and {D{\'e}moulin}, P. and {Zhukov}, A.~N. and {Steed}, K. and {Harra}, L.~K. and {Mandrini}, C.~H. and {Linker}, J.},
        title = "{The Recovery of CME-Related Dimmings and the ICME's Enduring Magnetic Connection to the Sun}",
      journal = {\solphys},
         year = 2008,
        month = nov,
       volume = {252},
       number = {2},
        pages = {349-372},
          doi = {10.1007/s11207-008-9255-z},
       adsurl = {https://ui.adsabs.harvard.edu/abs/2008SoPh..252..349A}
}

@ARTICLE{Jin2009,
       author = {{Jin}, M. and {Ding}, M.~D. and {Chen}, P.~F. and {Fang}, C. and {Imada}, S.},
        title = "{Coronal Mass Ejection Induced Outflows Observed with Hinode/EIS}",
      journal = {\apj},
         year = 2009,
        month = sep,
       volume = {702},
       number = {1},
        pages = {27-38},
          doi = {10.1088/0004-637X/702/1/27},
       adsurl = {https://ui.adsabs.harvard.edu/abs/2009ApJ...702...27J}
}

@ARTICLE{Jiang2003,
       author = {{Jiang}, Yunchun and {Ji}, Haisheng and {Wang}, Haimin and {Chen}, Huadong},
        title = "{H{\ensuremath{\alpha}} Dimmings Associated with the X1.6 Flare and Halo Coronal Mass Ejection on 2001 October 19}",
      journal = {\apjl},
         year = 2003,
        month = nov,
       volume = {597},
       number = {2},
        pages = {L161-L164},
          doi = {10.1086/379756},
       adsurl = {https://ui.adsabs.harvard.edu/abs/2003ApJ...597L.161J}
}

@ARTICLE{Xu2022,
       author = {{Xu}, Yu and {Tian}, Hui and {Hou}, Zhenyong and {Yang}, Zihao and {Gao}, Yuhang and {Bai}, Xianyong},
        title = "{Sun-as-a-star Spectroscopic Observations of the Line-of-sight Velocity of a Solar Eruption on 2021 October 28}",
      journal = {\apj},
         year = 2022,
        month = jun,
       volume = {931},
       number = {2},
          eid = {76},
        pages = {76},
          doi = {10.3847/1538-4357/ac69d5},
archivePrefix = {arXiv},
       eprint = {2204.11722},
 primaryClass = {astro-ph.SR},
       adsurl = {https://ui.adsabs.harvard.edu/abs/2022ApJ...931...76X}
}

@ARTICLE{Xu2024,
       author = {{Xu}, Yu and {Tian}, Hui and {Veronig}, Astrid M. and {Dissauer}, Karin},
        title = "{Sun-as-a-star Observations of Obscuration Dimmings Caused by Filament Eruptions}",
      journal = {\apj},
         year = 2024,
        month = jul,
       volume = {970},
       number = {1},
          eid = {60},
        pages = {60},
          doi = {10.3847/1538-4357/ad500b},
archivePrefix = {arXiv},
       eprint = {2405.13671},
 primaryClass = {astro-ph.SR},
       adsurl = {https://ui.adsabs.harvard.edu/abs/2024ApJ...970...60X}
}

@ARTICLE{Mason2016,
       author = {{Mason}, James Paul and {Woods}, Thomas N. and {Webb}, David F. and {Thompson}, Barbara J. and {Colaninno}, Robin C. and {Vourlidas}, Angelos},
        title = "{Relationship of EUV Irradiance Coronal Dimming Slope and Depth to Coronal Mass Ejection Speed and Mass}",
      journal = {\apj},
         year = 2016,
        month = oct,
       volume = {830},
       number = {1},
          eid = {20},
        pages = {20},
          doi = {10.3847/0004-637X/830/1/20},
archivePrefix = {arXiv},
       eprint = {1607.05284},
 primaryClass = {astro-ph.SR},
       adsurl = {https://ui.adsabs.harvard.edu/abs/2016ApJ...830...20M}
}

@ARTICLE{Mason2014,
       author = {{Mason}, James Paul and {Woods}, T.~N. and {Caspi}, A. and {Thompson}, B.~J. and {Hock}, R.~A.},
        title = "{Mechanisms and Observations of Coronal Dimming for the 2010 August 7 Event}",
      journal = {\apj},
         year = 2014,
        month = jul,
       volume = {789},
       number = {1},
          eid = {61},
        pages = {61},
          doi = {10.1088/0004-637X/789/1/61},
archivePrefix = {arXiv},
       eprint = {1404.1364},
 primaryClass = {astro-ph.SR},
       adsurl = {https://ui.adsabs.harvard.edu/abs/2014ApJ...789...61M}
}

@ARTICLE{Loyd2022,
       author = {{Loyd}, R.~O. Parke and {Mason}, James Paul and {Jin}, Meng and {Shkolnik}, Evgenya L. and {France}, Kevin and {Youngblood}, Allison and {Villadsen}, Jackie and {Schneider}, Christian and {Schneider}, Adam C. and {Llama}, Joe and {Ramiaramanantsoa}, Tahina and {Richey-Yowell}, Tyler},
        title = "{Constraining the Physical Properties of Stellar Coronal Mass Ejections with Coronal Dimming: Application to Far-ultraviolet Data of Eridani}",
      journal = {\apj},
         year = 2022,
        month = sep,
       volume = {936},
       number = {2},
          eid = {170},
        pages = {170},
          doi = {10.3847/1538-4357/ac80c1},
archivePrefix = {arXiv},
       eprint = {2207.05115},
 primaryClass = {astro-ph.SR},
       adsurl = {https://ui.adsabs.harvard.edu/abs/2022ApJ...936..170L}
}

@ARTICLE{Airapetian2020,
       author = {{Airapetian}, V.~S. and {Barnes}, R. and {Cohen}, O. and {Collinson}, G.~A. and {Danchi}, W.~C. and {Dong}, C.~F. and {Del Genio}, A.~D. and {France}, K. and {Garcia-Sage}, K. and {Glocer}, A. and {Gopalswamy}, N. and {Grenfell}, J.~L. and {Gronoff}, G. and {G{\"u}del}, M. and {Herbst}, K. and {Henning}, W.~G. and {Jackman}, C.~H. and {Jin}, M. and {Johnstone}, C.~P. and {Kaltenegger}, L. and {Kay}, C.~D. and {Kobayashi}, K. and {Kuang}, W. and {Li}, G. and {Lynch}, B.~J. and {L{\"u}ftinger}, T. and {Luhmann}, J.~G. and {Maehara}, H. and {Mlynczak}, M.~G. and {Notsu}, Y. and {Osten}, R.~A. and {Ramirez}, R.~M. and {Rugheimer}, S. and {Scheucher}, M. and {Schlieder}, J.~E. and {Shibata}, K. and {Sousa-Silva}, C. and {Stamenkovi{\'c}}, V. and {Strangeway}, R.~J. and {Usmanov}, A.~V. and {Vergados}, P. and {Verkhoglyadova}, O.~P. and {Vidotto}, A.~A. and {Voytek}, M. and {Way}, M.~J. and {Zank}, G.~P. and {Yamashiki}, Y.},
        title = "{Impact of space weather on climate and habitability of terrestrial-type exoplanets}",
      journal = {International Journal of Astrobiology},
         year = 2020,
        month = apr,
       volume = {19},
       number = {2},
        pages = {136-194},
          doi = {10.1017/S1473550419000132},
archivePrefix = {arXiv},
       eprint = {1905.05093},
 primaryClass = {astro-ph.EP},
       adsurl = {https://ui.adsabs.harvard.edu/abs/2020IJAsB..19..136A}
}

@ARTICLE{Argiroffi2019,
       author = {{Argiroffi}, C. and {Reale}, F. and {Drake}, J.~J. and {Ciaravella}, A. and {Testa}, P. and {Bonito}, R. and {Miceli}, M. and {Orlando}, S. and {Peres}, G.},
        title = "{A stellar flare-coronal mass ejection event revealed by X-ray plasma motions}",
      journal = {Nature Astronomy},
         year = 2019,
        month = may,
       volume = {3},
        pages = {742-748},
          doi = {10.1038/s41550-019-0781-4},
archivePrefix = {arXiv},
       eprint = {1905.11325},
 primaryClass = {astro-ph.SR},
       adsurl = {https://ui.adsabs.harvard.edu/abs/2019NatAs...3..742A}
}

@ARTICLE{Chen2022,
       author = {{Chen}, Hechao and {Tian}, Hui and {Li}, Hao and {Wang}, Jianguo and {Lu}, Hongpeng and {Xu}, Yu and {Hou}, Zhenyong and {Wu}, Yuchuan},
        title = "{Detection of Flare-induced Plasma Flows in the Corona of EV Lac with X-Ray Spectroscopy}",
      journal = {\apj},
         year = 2022,
        month = jul,
       volume = {933},
       number = {1},
          eid = {92},
        pages = {92},
          doi = {10.3847/1538-4357/ac739b},
archivePrefix = {arXiv},
       eprint = {2205.14293},
 primaryClass = {astro-ph.SR},
       adsurl = {https://ui.adsabs.harvard.edu/abs/2022ApJ...933...92C}
}

@ARTICLE{Namekata2022,
       author = {{Namekata}, Kosuke and {Maehara}, Hiroyuki and {Honda}, Satoshi and {Notsu}, Yuta and {Okamoto}, Soshi and {Takahashi}, Jun and {Takayama}, Masaki and {Ohshima}, Tomohito and {Saito}, Tomoki and {Katoh}, Noriyuki and {Tozuka}, Miyako and {Murata}, Katsuhiro L. and {Ogawa}, Futa and {Niwano}, Masafumi and {Adachi}, Ryo and {Oeda}, Motoki and {Shiraishi}, Kazuki and {Isogai}, Keisuke and {Seki}, Daikichi and {Ishii}, Takako T. and {Ichimoto}, Kiyoshi and {Nogami}, Daisaku and {Shibata}, Kazunari},
        title = "{Probable detection of an eruptive filament from a superflare on a solar-type star}",
      journal = {Nature Astronomy},
         year = 2021,
        month = dec,
       volume = {6},
        pages = {241-248},
          doi = {10.1038/s41550-021-01532-8},
archivePrefix = {arXiv},
       eprint = {2112.04808},
 primaryClass = {astro-ph.SR},
       adsurl = {https://ui.adsabs.harvard.edu/abs/2022NatAs...6..241N}
}

@ARTICLE{Lu2025,
       author = {{Lu}, Hong-Peng and {Tian}, Hui and {Zhang}, Li-Yun and {Chen}, He-Chao and {Li}, Ying and {Yang}, Zi-Hao and {Wang}, Jia-Sheng and {Zhang}, Jia-Le and {Sun}, Zheng},
        title = "{An Extreme Stellar Prominence Eruption Observed by LAMOST Time-domain Spectroscopy}",
      journal = {\apjl},
         year = 2025,
        month = jan,
       volume = {978},
       number = {2},
          eid = {L32},
        pages = {L32},
          doi = {10.3847/2041-8213/ad93cc},
archivePrefix = {arXiv},
       eprint = {2411.11076},
 primaryClass = {astro-ph.SR},
       adsurl = {https://ui.adsabs.harvard.edu/abs/2025ApJ...978L..32L}
}

@ARTICLE{Harrison2000,
       author = {{Harrison}, R.~A. and {Lyons}, M.},
        title = "{A spectroscopic study of coronal dimming associated with a coronal mass ejection}",
      journal = {\aap},
         year = 2000,
        month = jun,
       volume = {358},
        pages = {1097-1108},
       adsurl = {https://ui.adsabs.harvard.edu/abs/2000A&A...358.1097H}
}

@ARTICLE{Harrison2003,
       author = {{Harrison}, R.~A. and {Bryans}, P. and {Simnett}, G.~M. and {Lyons}, M.},
        title = "{Coronal dimming and the coronal mass ejection onset}",
      journal = {\aap},
         year = 2003,
        month = mar,
       volume = {400},
        pages = {1071-1083},
          doi = {10.1051/0004-6361:20030088},
       adsurl = {https://ui.adsabs.harvard.edu/abs/2003A&A...400.1071H}
}

@ARTICLE{Lopez2019,
       author = {{L{\'o}pez}, F.~M. and {Cremades}, H. and {Balmaceda}, L.~A. and {Nuevo}, F.~A. and {V{\'a}squez}, A.~M.},
        title = "{Estimating the mass of CMEs from the analysis of EUV dimmings}",
      journal = {\aap},
         year = 2019,
        month = jul,
       volume = {627},
          eid = {A8},
        pages = {A8},
          doi = {10.1051/0004-6361/201834163},
       adsurl = {https://ui.adsabs.harvard.edu/abs/2019A&A...627A...8L}
}

@ARTICLE{Chikunova2020,
       author = {{Chikunova}, Galina and {Dissauer}, Karin and {Podladchikova}, Tatiana and {Veronig}, Astrid M.},
        title = "{Coronal Dimmings Associated with Coronal Mass Ejections on the Solar Limb}",
      journal = {\apj},
         year = 2020,
        month = jun,
       volume = {896},
       number = {1},
          eid = {17},
        pages = {17},
          doi = {10.3847/1538-4357/ab9105},
archivePrefix = {arXiv},
       eprint = {2005.03348},
 primaryClass = {astro-ph.SR},
       adsurl = {https://ui.adsabs.harvard.edu/abs/2020ApJ...896...17C}
}

@ARTICLE{Qiu2007,
       author = {{Qiu}, Jiong and {Hu}, Qiang and {Howard}, Timothy A. and {Yurchyshyn}, Vasyl B.},
        title = "{On the Magnetic Flux Budget in Low-Corona Magnetic Reconnection and Interplanetary Coronal Mass Ejections}",
      journal = {\apj},
         year = 2007,
        month = apr,
       volume = {659},
       number = {1},
        pages = {758-772},
          doi = {10.1086/512060},
       adsurl = {https://ui.adsabs.harvard.edu/abs/2007ApJ...659..758Q}
}

@ARTICLE{Zhang2017,
       author = {{Zhang}, Q.~M. and {Su}, Y.~N. and {Ji}, H.~S.},
        title = "{Pre-flare coronal dimmings}",
      journal = {\aap},
         year = 2017,
        month = feb,
       volume = {598},
          eid = {A3},
        pages = {A3},
          doi = {10.1051/0004-6361/201629477},
archivePrefix = {arXiv},
       eprint = {1611.08371},
 primaryClass = {astro-ph.SR},
       adsurl = {https://ui.adsabs.harvard.edu/abs/2017A&A...598A...3Z}
}

@ARTICLE{Jain2024,
       author = {{Jain}, Shantanu and {Podladchikova}, Tatiana and {Chikunova}, Galina and {Dissauer}, Karin and {Veronig}, Astrid M.},
        title = "{Coronal dimmings as indicators of the direction of early coronal mass ejection propagation}",
      journal = {\aap},
         year = 2024,
        month = mar,
       volume = {683},
          eid = {A15},
        pages = {A15},
          doi = {10.1051/0004-6361/202347927},
archivePrefix = {arXiv},
       eprint = {2311.13942},
 primaryClass = {astro-ph.SR},
       adsurl = {https://ui.adsabs.harvard.edu/abs/2024A&A...683A..15J}
}

@ARTICLE{Veronig2025,
       author = {{Veronig}, Astrid M. and {Dissauer}, Karin and {Kliem}, Bernhard and {Downs}, Cooper and {Hudson}, Hugh S. and {Jin}, Meng and {Osten}, Rachel and {Podladchikova}, Tatiana and {Prasad}, Avijeet and {Qiu}, Jiong and {Thompson}, Barbara and {Tian}, Hui and {Vourlidas}, Angelos},
        title = "{Coronal dimmings and what they tell us about solar and stellar coronal mass ejections}",
      journal = {Living Reviews in Solar Physics},
         year = 2025,
        month = jul,
       volume = {22},
       number = {1},
          eid = {2},
        pages = {2},
          doi = {10.1007/s41116-025-00041-4},
archivePrefix = {arXiv},
       eprint = {2505.19228},
 primaryClass = {astro-ph.SR},
       adsurl = {https://ui.adsabs.harvard.edu/abs/2025LRSP...22....2V}
}

@article{Tian2025,
  title   = {Coronal Explorer for the Sun and Nearby Stars},
  author  = {Tian, Hui and Bai, Xianyong and Feng, Li and Xiong, Ming and Chen, Yajie and Hou, Zhenyong and Wang, Yamin},
  journal = {Chinese Journal of Space Science},
  year    = {2025},
  volume  = {45},
  number  = {4},
  pages   = {881--898},
  issn    = {0254-6124},
  doi     = {10.11728/cjss2025.04.2025-0060},
  url     = {https://www.cjss.ac.cn/article/doi/10.11728/cjss2025.04.2025-0060}
}

@INPROCEEDINGS{France2019,
       author = {{France}, Kevin and {Fleming}, Brian T. and {Drake}, Jeremy J. and {Mason}, James P. and {Youngblood}, Allison and {Bourrier}, Vincent and {Fossati}, Luca and {Froning}, Cynthia S. and {Koskinen}, Tommi and {Kruczek}, Nicholas and {Lipscy}, Sarah and {McEntaffer}, Randall and {Romaine}, Suzanne and {Siegmund}, Oswald H.~W. and {Wilkinson}, Erik},
        title = "{The extreme-ultraviolet stellar characterization for atmospheric physics and evolution (ESCAPE) mission concept}",
    booktitle = {UV, X-Ray, and Gamma-Ray Space Instrumentation for Astronomy XXI},
         year = 2019,
       editor = {{Siegmund}, Oswald H.},
       series = {Society of Photo-Optical Instrumentation Engineers (SPIE) Conference Series},
       volume = {11118},
        month = sep,
          eid = {1111808},
        pages = {1111808},
          doi = {10.1117/12.2526859},
       adsurl = {https://ui.adsabs.harvard.edu/abs/2019SPIE11118E..08F}
}

@ARTICLE{Webb2000,
       author = {{Webb}, D.~F. and {Lepping}, R.~P. and {Burlaga}, L.~F. and {DeForest}, C.~E. and {Larson}, D.~E. and {Martin}, S.~F. and {Plunkett}, S.~P. and {Rust}, D.~M.},
        title = "{The origin and development of the May 1997 magnetic cloud}",
      journal = {\jgr},
         year = 2000,
        month = dec,
       volume = {105},
       number = {A12},
        pages = {27251-27260},
          doi = {10.1029/2000JA000021},
       adsurl = {https://ui.adsabs.harvard.edu/abs/2000JGR...10527251W}
}

@ARTICLE{chen16,
       author = {{Chen}, P.~F.},
        title = "{Global Coronal Waves}",
      journal = {Geophysical Monograph Series},
         year = 2016,
        month = feb,
       volume = {216},
        pages = {381-394},
          doi = {10.1002/9781119055006.ch22},
archivePrefix = {arXiv},
       eprint = {1604.07991},
 primaryClass = {astro-ph.SR},
       adsurl = {https://ui.adsabs.harvard.edu/abs/2016GMS...216..381C}
}

@ARTICLE{wust01,
       author = {{Wu}, S.~T. and {Zheng}, Huinan and {Wang}, S. and {Thompson}, B.~J. and {Plunkett}, S.~P. and {Zhao}, X.~P. and {Dryer}, M.},
        title = "{Three-dimensional numerical simulation of MHD waves observed by the Extreme Ultraviolet Imaging Telescope}",
      journal = {\jgr},
         year = 2001,
        month = nov,
       volume = {106},
       number = {A11},
        pages = {25089-25102},
          doi = {10.1029/2000JA000447},
       adsurl = {https://ui.adsabs.harvard.edu/abs/2001JGR...10625089W}
}

@ARTICLE{wang00,
       author = {{Wang}, Y.-M.},
        title = "{EIT Waves and Fast-Mode Propagation in the Solar Corona}",
      journal = {\apjl},
         year = 2000,
        month = nov,
       volume = {543},
       number = {1},
        pages = {L89-L93},
          doi = {10.1086/318178},
       adsurl = {https://ui.adsabs.harvard.edu/abs/2000ApJ...543L..89W}
}

@ARTICLE{klass00,
       author = {{Klassen}, A. and {Aurass}, H. and {Mann}, G. and {Thompson}, B.~J.},
        title = "{Catalogue of the 1997 SOHO-EIT coronal transient waves and associated type II radio burst spectra}",
      journal = {\aaps},
         year = 2000,
        month = feb,
       volume = {141},
        pages = {357-369},
          doi = {10.1051/aas:2000125},
       adsurl = {https://ui.adsabs.harvard.edu/abs/2000A&AS..141..357K}
}

@ARTICLE{zhan11,
       author = {{Zhang}, Yuzong and {Kitai}, Reizaburo and {Narukage}, Noriyuki and {Matsumoto}, Takuma and {Ueno}, Satoru and {Shibata}, Kazunari and {Wang}, Jingxiu},
        title = "{Propagation of Moreton Waves}",
      journal = {\pasj},
         year = 2011,
        month = jun,
       volume = {63},
       number = {3},
        pages = {685-696},
          doi = {10.1093/pasj/63.3.685},
       adsurl = {https://ui.adsabs.harvard.edu/abs/2011PASJ...63..685Z}
}

@ARTICLE{dela99,
       author = {{Delann{\'e}e}, C. and {Aulanier}, G.},
        title = "{Cme Associated with Transequatorial Loops and a Bald Patch Flare}",
      journal = {\solphys},
         year = 1999,
        month = dec,
       volume = {190},
        pages = {107-129},
          doi = {10.1023/A:1005249416605},
       adsurl = {https://ui.adsabs.harvard.edu/abs/1999SoPh..190..107D}
}

@ARTICLE{chen02,
       author = {{Chen}, P.~F. and {Wu}, S.~T. and {Shibata}, K. and {Fang}, C.},
        title = "{Evidence of EIT and Moreton Waves in Numerical Simulations}",
      journal = {\apjl},
         year = 2002,
        month = jun,
       volume = {572},
       number = {1},
        pages = {L99-L102},
          doi = {10.1086/341486},
       adsurl = {https://ui.adsabs.harvard.edu/abs/2002ApJ...572L..99C}
}

@ARTICLE{chen05,
       author = {{Chen}, P.~F. and {Fang}, C. and {Shibata}, K.},
        title = "{A Full View of EIT Waves}",
      journal = {\apj},
         year = 2005,
        month = apr,
       volume = {622},
       number = {2},
        pages = {1202-1210},
          doi = {10.1086/428084},
       adsurl = {https://ui.adsabs.harvard.edu/abs/2005ApJ...622.1202C}
}

@ARTICLE{bies02,
       author = {{Biesecker}, D.~A. and {Myers}, D.~C. and {Thompson}, B.~J. and {Hammer}, D.~M. and {Vourlidas}, A.},
        title = "{Solar Phenomena Associated with ``EIT Waves''}",
      journal = {\apj},
         year = 2002,
        month = apr,
       volume = {569},
       number = {2},
        pages = {1009-1015},
          doi = {10.1086/339402},
       adsurl = {https://ui.adsabs.harvard.edu/abs/2002ApJ...569.1009B}
}

@ARTICLE{harr03,
       author = {{Harra}, Louise K. and {Sterling}, Alphonse C.},
        title = "{Imaging and Spectroscopic Investigations of a Solar Coronal Wave: Properties of the Wave Front and Associated Erupting Material}",
      journal = {\apj},
         year = 2003,
        month = apr,
       volume = {587},
       number = {1},
        pages = {429-438},
          doi = {10.1086/368079},
       adsurl = {https://ui.adsabs.harvard.edu/abs/2003ApJ...587..429H}
}

@ARTICLE{warm11,
       author = {{Warmuth}, A. and {Mann}, G.},
        title = "{Kinematical evidence for physically different classes of large-scale coronal EUV waves}",
      journal = {\aap},
         year = 2011,
        month = aug,
       volume = {532},
          eid = {A151},
        pages = {A151},
          doi = {10.1051/0004-6361/201116685},
       adsurl = {https://ui.adsabs.harvard.edu/abs/2011A&A...532A.151W}
}

@ARTICLE{will07,
       author = {{Wills-Davey}, M.~J. and {DeForest}, C.~E. and {Stenflo}, J.~O.},
        title = "{Are ``EIT Waves'' Fast-Mode MHD Waves?}",
      journal = {\apj},
         year = 2007,
        month = jul,
       volume = {664},
       number = {1},
        pages = {556-562},
          doi = {10.1086/519013},
archivePrefix = {arXiv},
       eprint = {0704.2828},
 primaryClass = {astro-ph},
       adsurl = {https://ui.adsabs.harvard.edu/abs/2007ApJ...664..556W}
}

@ARTICLE{wang09,
       author = {{Wang}, Hongjuan and {Shen}, Chengcai and {Lin}, Jun},
        title = "{Numerical Experiments of Wave-like Phenomena Caused by the Disruption of an Unstable Magnetic Configuration}",
      journal = {\apj},
         year = 2009,
        month = aug,
       volume = {700},
       number = {2},
        pages = {1716-1731},
          doi = {10.1088/0004-637X/700/2/1716},
archivePrefix = {arXiv},
       eprint = {0906.2677},
 primaryClass = {astro-ph.SR},
       adsurl = {https://ui.adsabs.harvard.edu/abs/2009ApJ...700.1716W}
}

@ARTICLE{attr07,
       author = {{Attrill}, G.~D.~R. and {Harra}, L.~K. and {van Driel-Gesztelyi}, L. and {D{\'e}moulin}, P. and {W{\"u}lser}, J.-P.},
        title = "{Coronal ``wave'': A signature of the mechanism making CMEs large-scale in the low corona?}",
      journal = {Astronomische Nachrichten},
         year = 2007,
        month = oct,
       volume = {328},
       number = {8},
        pages = {760},
          doi = {10.1002/asna.200710794},
       adsurl = {https://ui.adsabs.harvard.edu/abs/2007AN....328..760A}
}

@ARTICLE{dela08,
       author = {{Delann{\'e}e}, C. and {T{\"o}r{\"o}k}, T. and {Aulanier}, G. and {Hochedez}, J.-F.},
        title = "{A New Model for Propagating Parts of EIT Waves: A Current Shell in a CME}",
      journal = {\solphys},
         year = 2008,
        month = jan,
       volume = {247},
       number = {1},
        pages = {123-150},
          doi = {10.1007/s11207-007-9085-4},
       adsurl = {https://ui.adsabs.harvard.edu/abs/2008SoPh..247..123D}
}

@ARTICLE{asai12,
       author = {{Asai}, Ayumi and {Ishii}, Takako T. and {Isobe}, Hiroaki and {Kitai}, Reizaburo and {Ichimoto}, Kiyoshi and {UeNo}, Satoru and {Nagata}, Shin'ichi and {Morita}, Satoshi and {Nishida}, Keisuke and {Shiota}, Daikou and {Oi}, Akihito and {Akioka}, Maki and {Shibata}, Kazunari},
        title = "{First Simultaneous Observation of an H{\ensuremath{\alpha}} Moreton Wave, EUV Wave, and Filament/Prominence Oscillations}",
      journal = {\apjl},
         year = 2012,
        month = feb,
       volume = {745},
       number = {2},
          eid = {L18},
        pages = {L18},
          doi = {10.1088/2041-8205/745/2/L18},
archivePrefix = {arXiv},
       eprint = {1112.5915},
 primaryClass = {astro-ph.SR},
       adsurl = {https://ui.adsabs.harvard.edu/abs/2012ApJ...745L..18A}
}

@ARTICLE{chenwu11,
       author = {{Chen}, P.~F. and {Wu}, Y.},
        title = "{First Evidence of Coexisting EIT Wave and Coronal Moreton Wave from SDO/AIA Observations}",
      journal = {\apjl},
         year = 2011,
        month = may,
       volume = {732},
       number = {2},
          eid = {L20},
        pages = {L20},
          doi = {10.1088/2041-8205/732/2/L20},
archivePrefix = {arXiv},
       eprint = {1103.0871},
 primaryClass = {astro-ph.SR},
       adsurl = {https://ui.adsabs.harvard.edu/abs/2011ApJ...732L..20C}
}

@ARTICLE{cheng12,
       author = {{Cheng}, X. and {Zhang}, J. and {Olmedo}, O. and {Vourlidas}, A. and {Ding}, M.~D. and {Liu}, Y.},
        title = "{Investigation of the Formation and Separation of an Extreme-ultraviolet Wave from the Expansion of a Coronal Mass Ejection}",
      journal = {\apjl},
         year = 2012,
        month = jan,
       volume = {745},
       number = {1},
          eid = {L5},
        pages = {L5},
          doi = {10.1088/2041-8205/745/1/L5},
archivePrefix = {arXiv},
       eprint = {1112.4540},
 primaryClass = {astro-ph.SR},
       adsurl = {https://ui.adsabs.harvard.edu/abs/2012ApJ...745L...5C}
}

@ARTICLE{kuma13,
       author = {{Kumar}, Pankaj and {Cho}, K.-S. and {Chen}, P.~F. and {Bong}, S.-C. and {Park}, Sung-Hong},
        title = "{Multiwavelength Study of a Solar Eruption from AR NOAA 11112: II. Large-Scale Coronal Wave and Loop Oscillation}",
      journal = {\solphys},
         year = 2013,
        month = feb,
       volume = {282},
       number = {2},
        pages = {523-541},
          doi = {10.1007/s11207-012-0158-7},
archivePrefix = {arXiv},
       eprint = {1210.3417},
 primaryClass = {astro-ph.SR},
       adsurl = {https://ui.adsabs.harvard.edu/abs/2013SoPh..282..523K}
}

@ARTICLE{shen13,
       author = {{Shen}, Yuandeng and {Liu}, Yu and {Su}, Jiangtao and {Li}, Hui and {Zhao}, Ruijuan and {Tian}, Zhanjun and {Ichimoto}, Kiyoshi and {Shibata}, Kazunari},
        title = "{Diffraction, Refraction, and Reflection of an Extreme-ultraviolet Wave Observed during Its Interactions with Remote Active Regions}",
      journal = {\apjl},
         year = 2013,
        month = aug,
       volume = {773},
       number = {2},
          eid = {L33},
        pages = {L33},
          doi = {10.1088/2041-8205/773/2/L33},
archivePrefix = {arXiv},
       eprint = {1307.6098},
 primaryClass = {astro-ph.SR},
       adsurl = {https://ui.adsabs.harvard.edu/abs/2013ApJ...773L..33S}
}

@ARTICLE{chenf10,
       author = {{Chen}, F. and {Ding}, M.~D. and {Chen}, P.~F.},
        title = "{Spectroscopic Analysis of an EIT Wave/dimming Observed by Hinode/EIS}",
      journal = {\apj},
         year = 2010,
        month = sep,
       volume = {720},
       number = {2},
        pages = {1254-1261},
          doi = {10.1088/0004-637X/720/2/1254},
archivePrefix = {arXiv},
       eprint = {1009.3190},
 primaryClass = {astro-ph.SR},
       adsurl = {https://ui.adsabs.harvard.edu/abs/2010ApJ...720.1254C}
}

@ARTICLE{chan16,
       author = {{Chandra}, R. and {Chen}, P.~F. and {Fulara}, A. and {Srivastava}, A.~K. and {Uddin}, W.},
        title = "{Peculiar Stationary EUV Wave Fronts in the Eruption on 2011 May 11}",
      journal = {\apj},
         year = 2016,
        month = may,
       volume = {822},
       number = {2},
          eid = {106},
        pages = {106},
          doi = {10.3847/0004-637X/822/2/106},
archivePrefix = {arXiv},
       eprint = {1602.08693},
 primaryClass = {astro-ph.SR},
       adsurl = {https://ui.adsabs.harvard.edu/abs/2016ApJ...822..106C}
}

@ARTICLE{chen16a,
       author = {{Chen}, P.~F. and {Fang}, C. and {Chandra}, R. and {Srivastava}, A.~K.},
        title = "{Can a Fast-Mode EUV Wave Generate a Stationary Front?}",
      journal = {\solphys},
         year = 2016,
        month = nov,
       volume = {291},
       number = {11},
        pages = {3195-3206},
          doi = {10.1007/s11207-016-0920-3},
archivePrefix = {arXiv},
       eprint = {1604.07982},
 primaryClass = {astro-ph.SR},
       adsurl = {https://ui.adsabs.harvard.edu/abs/2016SoPh..291.3195C}
}

@ARTICLE{chan18,
       author = {{Chandra}, Ramesh and {Chen}, P.~F. and {Joshi}, Reetika and {Joshi}, Bhuwan and {Schmieder}, Brigitte},
        title = "{Observations of Two Successive EUV Waves and Their Mode Conversion}",
      journal = {\apj},
         year = 2018,
        month = aug,
       volume = {863},
       number = {1},
          eid = {101},
        pages = {101},
          doi = {10.3847/1538-4357/aad097},
archivePrefix = {arXiv},
       eprint = {1806.11350},
 primaryClass = {astro-ph.SR},
       adsurl = {https://ui.adsabs.harvard.edu/abs/2018ApJ...863..101C}
}

@ARTICLE{kuma24,
       author = {{Kumar}, Pankaj and {Nakariakov}, Valery M. and {Karpen}, Judith T. and {Cho}, Kyung-Suk},
        title = "{Direct imaging of magnetohydrodynamic wave mode conversion near a 3D null point on the sun}",
      journal = {Nature Communications},
         year = 2024,
        month = mar,
       volume = {15},
          eid = {2667},
        pages = {2667},
          doi = {10.1038/s41467-024-46736-4},
archivePrefix = {arXiv},
       eprint = {2403.02250},
 primaryClass = {astro-ph.SR},
       adsurl = {https://ui.adsabs.harvard.edu/abs/2024NatCo..15.2667K}
}

@ARTICLE{li25,
       author = {{Li}, Y.~W. and {Guo}, J.~H. and {Ni}, Y.~W. and {Zhang}, Z.~Y. and {Chen}, P.~F.},
        title = "{Dependence of the intensity of the non-wave component of extreme UV waves on the coronal magnetic field configuration}",
      journal = {\aap},
         year = 2025,
        month = jun,
       volume = {698},
          eid = {A316},
        pages = {A316},
          doi = {10.1051/0004-6361/202554363},
archivePrefix = {arXiv},
       eprint = {2505.11115},
 primaryClass = {astro-ph.SR},
       adsurl = {https://ui.adsabs.harvard.edu/abs/2025A&A...698A.316L}
}

@ARTICLE{long13,
       author = {{Long}, D.~M. and {Williams}, D.~R. and {R{\'e}gnier}, S. and {Harra}, L.~K.},
        title = "{Measuring the Magnetic-Field Strength of the Quiet Solar Corona Using ``EIT Waves''}",
      journal = {\solphys},
         year = 2013,
        month = dec,
       volume = {288},
       number = {2},
        pages = {567-583},
          doi = {10.1007/s11207-013-0331-7},
archivePrefix = {arXiv},
       eprint = {1305.5169},
 primaryClass = {astro-ph.SR},
       adsurl = {https://ui.adsabs.harvard.edu/abs/2013SoPh..288..567L}
}

@ARTICLE{chen09,
       author = {{Chen}, P.~F.},
        title = "{EIT waves and coronal magnetic field diagnostics}",
      journal = {Science in China: Physics, Mechanics and Astronomy},
         year = 2009,
        month = nov,
       volume = {52},
       number = {11},
        pages = {1785-1789},
          doi = {10.1007/s11433-009-0240-9},
archivePrefix = {arXiv},
       eprint = {1001.0065},
 primaryClass = {astro-ph.SR},
       adsurl = {https://ui.adsabs.harvard.edu/abs/2009ScChG..52.1785C}
}

@ARTICLE{Asai2001,
       author = {{Asai}, Ayumi and {Ishii}, Takako T. and {Kurokawa}, Hiroki},
        title = "{Plasma Ejections from a Light Bridge in a Sunspot Umbra}",
      journal = {\apjl},
         year = 2001,
        month = jul,
       volume = {555},
       number = {1},
        pages = {L65-L68},
          doi = {10.1086/321738},
archivePrefix = {arXiv},
       eprint = {astro-ph/0111021},
 primaryClass = {astro-ph},
       adsurl = {https://ui.adsabs.harvard.edu/abs/2001ApJ...555L..65A}
}

@ARTICLE{Beckers1968,
       author = {{Beckers}, J.~M.},
        title = "{Solar Spicules (Invited Review Paper)}",
      journal = {\solphys},
         year = 1968,
        month = mar,
       volume = {3},
       number = {3},
        pages = {367-433},
          doi = {10.1007/BF00171614},
       adsurl = {https://ui.adsabs.harvard.edu/abs/1968SoPh....3..367B}
}

@ARTICLE{ChenPF2000,
       author = {{Chen}, P.~F. and {Shibata}, K.},
        title = "{An Emerging Flux Trigger Mechanism for Coronal Mass Ejections}",
      journal = {\apj},
         year = 2000,
        month = dec,
       volume = {545},
       number = {1},
        pages = {524-531},
          doi = {10.1086/317803},
       adsurl = {https://ui.adsabs.harvard.edu/abs/2000ApJ...545..524C}
}

@ARTICLE{ChenY2016,
       author = {{Chen}, Yao and {Du}, Guohui and {Zhao}, Di and {Wu}, Zhao and {Liu}, Wei and {Wang}, Bing and {Ruan}, Guiping and {Feng}, Shiwei and {Song}, Hongqiang},
        title = "{Imaging a Magnetic-breakout Solar Eruption}",
      journal = {\apjl},
         year = 2016,
        month = apr,
       volume = {820},
       number = {2},
          eid = {L37},
        pages = {L37},
          doi = {10.3847/2041-8205/820/2/L37},
       adsurl = {https://ui.adsabs.harvard.edu/abs/2016ApJ...820L..37C}
}

@ARTICLE{ChengX2023,
       author = {{Cheng}, X. and {Priest}, E.~R. and {Li}, H.~T. and {Chen}, J. and {Aulanier}, G. and {Chitta}, L.~P. and {Wang}, Y.~L. and {Peter}, H. and {Zhu}, X.~S. and {Xing}, C. and {Ding}, M.~D. and {Solanki}, S.~K. and {Berghmans}, D. and {Teriaca}, L. and {Aznar Cuadrado}, R. and {Zhukov}, A.~N. and {Guo}, Y. and {Long}, D. and {Harra}, L. and {Smith}, P.~J. and {Rodriguez}, L. and {Verbeeck}, C. and {Barczynski}, K. and {Parenti}, S.},
        title = "{Ultra-high-resolution observations of persistent null-point reconnection in the solar corona}",
      journal = {Nature Communications},
         year = 2023,
        month = apr,
       volume = {14},
          eid = {2107},
        pages = {2107},
          doi = {10.1038/s41467-023-37888-w},
archivePrefix = {arXiv},
       eprint = {2304.08725},
 primaryClass = {astro-ph.SR},
       adsurl = {https://ui.adsabs.harvard.edu/abs/2023NatCo..14.2107C}
}

@ARTICLE{DePontieu2014,
       author = {{De Pontieu}, B. and {Title}, A.~M. and {Lemen}, J.~R. and {Kushner}, G.~D. and {Akin}, D.~J. and {Allard}, B. and {Berger}, T. and {Boerner}, P. and {Cheung}, M. and {Chou}, C. and {Drake}, J.~F. and {Duncan}, D.~W. and {Freeland}, S. and {Heyman}, G.~F. and {Hoffman}, C. and {Hurlburt}, N.~E. and {Lindgren}, R.~W. and {Mathur}, D. and {Rehse}, R. and {Sabolish}, D. and {Seguin}, R. and {Schrijver}, C.~J. and {Tarbell}, T.~D. and {W{\"u}lser}, J. -P. and {Wolfson}, C.~J. and {Yanari}, C. and {Mudge}, J. and {Nguyen-Phuc}, N. and {Timmons}, R. and {van Bezooijen}, R. and {Weingrod}, I. and {Brookner}, R. and {Butcher}, G. and {Dougherty}, B. and {Eder}, J. and {Knagenhjelm}, V. and {Larsen}, S. and {Mansir}, D. and {Phan}, L. and {Boyle}, P. and {Cheimets}, P.~N. and {DeLuca}, E.~E. and {Golub}, L. and {Gates}, R. and {Hertz}, E. and {McKillop}, S. and {Park}, S. and {Perry}, T. and {Podgorski}, W.~A. and {Reeves}, K. and {Saar}, S. and {Testa}, P. and {Tian}, H. and {Weber}, M. and {Dunn}, C. and {Eccles}, S. and {Jaeggli}, S.~A. and {Kankelborg}, C.~C. and {Mashburn}, K. and {Pust}, N. and {Springer}, L. and {Carvalho}, R. and {Kleint}, L. and {Marmie}, J. and {Mazmanian}, E. and {Pereira}, T.~M.~D. and {Sawyer}, S. and {Strong}, J. and {Worden}, S.~P. and {Carlsson}, M. and {Hansteen}, V.~H. and {Leenaarts}, J. and {Wiesmann}, M. and {Aloise}, J. and {Chu}, K. -C. and {Bush}, R.~I. and {Scherrer}, P.~H. and {Brekke}, P. and {Martinez-Sykora}, J. and {Lites}, B.~W. and {McIntosh}, S.~W. and {Uitenbroek}, H. and {Okamoto}, T.~J. and {Gummin}, M.~A. and {Auker}, G. and {Jerram}, P. and {Pool}, P. and {Waltham}, N.},
        title = "{The Interface Region Imaging Spectrograph (IRIS)}",
      journal = {\solphys},
         year = 2014,
        month = jul,
       volume = {289},
       number = {7},
        pages = {2733-2779},
          doi = {10.1007/s11207-014-0485-y},
archivePrefix = {arXiv},
       eprint = {1401.2491},
 primaryClass = {astro-ph.SR},
       adsurl = {https://ui.adsabs.harvard.edu/abs/2014SoPh..289.2733D}
}

@ARTICLE{DuanYD2024,
       author = {{Duan}, Yadan and {Tian}, Hui and {Chen}, Hechao and {Shen}, Yuandeng and {Sun}, Zheng and {Hou}, Zhenyong and {Li}, Chuan},
        title = "{Formation of Fan-spine Magnetic Topology through Flux Emergence and Subsequent Jet Production}",
      journal = {\apjl},
         year = 2024,
        month = feb,
       volume = {962},
       number = {2},
          eid = {L38},
        pages = {L38},
          doi = {10.3847/2041-8213/ad24f3},
archivePrefix = {arXiv},
       eprint = {2402.02087},
 primaryClass = {astro-ph.SR},
       adsurl = {https://ui.adsabs.harvard.edu/abs/2024ApJ...962L..38D}
}

@ARTICLE{Heyvaerts1977,
       author = {{Heyvaerts}, J. and {Priest}, E.~R. and {Rust}, D.~M.},
        title = "{An emerging flux model for the solar phenomenon.}",
      journal = {\apj},
         year = 1977,
        month = aug,
       volume = {216},
        pages = {123-137},
          doi = {10.1086/155453},
       adsurl = {https://ui.adsabs.harvard.edu/abs/1977ApJ...216..123H}
}

@ARTICLE{HongJC2019,
       author = {{Hong}, Junchao and {Yang}, Jiayan and {Chen}, Huadong and {Bi}, Yi and {Yang}, Bo and {Chen}, Hechao},
        title = "{Observation of a Reversal of Breakout Reconnection Preceding a Jet: Evidence of Oscillatory Magnetic Reconnection?}",
      journal = {\apj},
         year = 2019,
        month = apr,
       volume = {874},
       number = {2},
          eid = {146},
        pages = {146},
          doi = {10.3847/1538-4357/ab0c9d},
archivePrefix = {arXiv},
       eprint = {1903.01201},
 primaryClass = {astro-ph.SR},
       adsurl = {https://ui.adsabs.harvard.edu/abs/2019ApJ...874..146H}
}

@ARTICLE{HouYJ2016a,
       author = {{Hou}, Y.~J. and {Li}, T. and {Zhang}, J.},
        title = "{Flux rope proxies and fan-spine structures in active region NOAA 11897}",
      journal = {\aap},
         year = 2016,
        month = aug,
       volume = {592},
          eid = {A138},
        pages = {A138},
          doi = {10.1051/0004-6361/201628851},
archivePrefix = {arXiv},
       eprint = {1606.09322},
 primaryClass = {astro-ph.SR},
       adsurl = {https://ui.adsabs.harvard.edu/abs/2016A&A...592A.138H}
}

@ARTICLE{HouYJ2016b,
       author = {{Hou}, Yijun and {Zhang}, Jun and {Li}, Ting and {Yang}, Shuhong and {Li}, Leping and {Li}, Xiaohong},
        title = "{A Solar Flare Disturbing a Light Wall above a Sunspot Light Bridge}",
      journal = {\apjl},
         year = 2016,
        month = oct,
       volume = {829},
       number = {2},
          eid = {L29},
        pages = {L29},
          doi = {10.3847/2041-8205/829/2/L29},
archivePrefix = {arXiv},
       eprint = {1609.05412},
 primaryClass = {astro-ph.SR},
       adsurl = {https://ui.adsabs.harvard.edu/abs/2016ApJ...829L..29H}
}

@ARTICLE{HouYJ2019,
       author = {{Hou}, Yijun and {Li}, Ting and {Yang}, Shuhong and {Zhang}, Jun},
        title = "{A Secondary Fan-spine Magnetic Structure in Active Region 11897}",
      journal = {\apj},
         year = 2019,
        month = jan,
       volume = {871},
       number = {1},
          eid = {4},
        pages = {4},
          doi = {10.3847/1538-4357/aaf4f4},
archivePrefix = {arXiv},
       eprint = {1811.11936},
 primaryClass = {astro-ph.SR},
       adsurl = {https://ui.adsabs.harvard.edu/abs/2019ApJ...871....4H}
}

@ARTICLE{HouYJ2017,
       author = {{Hou}, Yijun and {Zhang}, Jun and {Li}, Ting and {Yang}, Shuhong and {Li}, Xiaohong},
        title = "{Simultaneous Observations of p-mode Light Walls and Magnetic Reconnection Ejections above Sunspot Light Bridges}",
      journal = {\apjl},
         year = 2017,
        month = oct,
       volume = {848},
       number = {1},
          eid = {L9},
        pages = {L9},
          doi = {10.3847/2041-8213/aa8edd},
       adsurl = {https://ui.adsabs.harvard.edu/abs/2017ApJ...848L...9H}
}

@ARTICLE{HouYJ2020,
       author = {{Hou}, Y.~J. and {Li}, T. and {Zhong}, S.~H. and {Yang}, S.~H. and {Guo}, Y.~L. and {Li}, X.~H. and {Zhang}, J. and {Xiang}, Y.~Y.},
        title = "{Sunspot penumbral filaments intruding into a light bridge and the resultant reconnection jets}",
      journal = {\aap},
         year = 2020,
        month = oct,
       volume = {642},
          eid = {A44},
        pages = {A44},
          doi = {10.1051/0004-6361/202038668},
archivePrefix = {arXiv},
       eprint = {2008.01284},
 primaryClass = {astro-ph.SR},
       adsurl = {https://ui.adsabs.harvard.edu/abs/2020A&A...642A..44H}
}

@ARTICLE{HouYJ2022,
       author = {{Hou}, Yijun and {Li}, Ting and {Yang}, Shuhong and {Toriumi}, Shin and {Guo}, Yilin and {Zhang}, Jun},
        title = "{Various Activities above Sunspot Light Bridges in IRIS Observations: Classification and Comparison}",
      journal = {\apj},
         year = 2022,
        month = apr,
       volume = {929},
       number = {1},
          eid = {12},
        pages = {12},
          doi = {10.3847/1538-4357/ac5912},
archivePrefix = {arXiv},
       eprint = {2202.10159},
 primaryClass = {astro-ph.SR},
       adsurl = {https://ui.adsabs.harvard.edu/abs/2022ApJ...929...12H}
}

@ARTICLE{HouZY2021,
       author = {{Hou}, Zhenyong and {Tian}, Hui and {Berghmans}, David and {Chen}, Hechao and {Teriaca}, Luca and {Sch{\"u}hle}, Udo and {Gao}, Yuhang and {Chen}, Yajie and {He}, Jiansen and {Wang}, Linghua and {Bai}, Xianyong},
        title = "{Coronal Microjets in Quiet-Sun Regions Observed with the Extreme Ultraviolet Imager on Board the Solar Orbiter}",
      journal = {\apjl},
         year = 2021,
        month = sep,
       volume = {918},
       number = {1},
          eid = {L20},
        pages = {L20},
          doi = {10.3847/2041-8213/ac1f30},
archivePrefix = {arXiv},
       eprint = {2108.08718},
 primaryClass = {astro-ph.SR},
       adsurl = {https://ui.adsabs.harvard.edu/abs/2021ApJ...918L..20H}
}

@ARTICLE{Kleczek1960,
       author = {{Kleczek}, J. and {K{\v{r}}ivsk{\'y}}, L.},
        title = "{Solar Limb Surges accompanied by X-Ray Emission}",
      journal = {\nat},
         year = 1960,
        month = jun,
       volume = {186},
       number = {4730},
        pages = {1035-1036},
          doi = {10.1038/1861035b0},
       adsurl = {https://ui.adsabs.harvard.edu/abs/1960Natur.186.1035K}
}

@ARTICLE{Kumar2018,
       author = {{Kumar}, Pankaj and {Karpen}, Judith T. and {Antiochos}, Spiro K. and {Wyper}, Peter F. and {DeVore}, C. Richard and {DeForest}, Craig E.},
        title = "{Evidence for the Magnetic Breakout Model in an Equatorial Coronal-hole Jet}",
      journal = {\apj},
         year = 2018,
        month = feb,
       volume = {854},
       number = {2},
          eid = {155},
        pages = {155},
          doi = {10.3847/1538-4357/aaab4f},
archivePrefix = {arXiv},
       eprint = {1801.08582},
 primaryClass = {astro-ph.SR},
       adsurl = {https://ui.adsabs.harvard.edu/abs/2018ApJ...854..155K}
}

@ARTICLE{LiFY2021,
       author = {{Li}, Fuyu and {Chen}, Yajie and {Hou}, Yijun and {Tian}, Hui and {Bai}, Xianyong and {Song}, Yongliang},
        title = "{Small-scale Bright Blobs Ejected from a Sunspot Light Bridge}",
      journal = {\apj},
         year = 2021,
        month = feb,
       volume = {908},
       number = {2},
          eid = {201},
        pages = {201},
          doi = {10.3847/1538-4357/abd322},
archivePrefix = {arXiv},
       eprint = {2102.13545},
 primaryClass = {astro-ph.SR},
       adsurl = {https://ui.adsabs.harvard.edu/abs/2021ApJ...908..201L}
}

@ARTICLE{LiT2018,
       author = {{Li}, Ting and {Yang}, Shuhong and {Zhang}, Qingmin and {Hou}, Yijun and {Zhang}, Jun},
        title = "{Two Episodes of Magnetic Reconnections during a Confined Circular-ribbon Flare}",
      journal = {\apj},
         year = 2018,
        month = jun,
       volume = {859},
       number = {2},
          eid = {122},
        pages = {122},
          doi = {10.3847/1538-4357/aabe84},
archivePrefix = {arXiv},
       eprint = {1804.05458},
 primaryClass = {astro-ph.SR},
       adsurl = {https://ui.adsabs.harvard.edu/abs/2018ApJ...859..122L}
}

@ARTICLE{LiT2020,
       author = {{Li}, Ting and {Hou}, Yijun and {Zhang}, Jun and {Xiang}, Yongyuan},
        title = "{NVST observations of collision-induced apparent fan-shaped jets}",
      journal = {\mnras},
         year = 2020,
        month = feb,
       volume = {492},
       number = {2},
        pages = {2510-2516},
          doi = {10.1093/mnras/stz3630},
       adsurl = {https://ui.adsabs.harvard.edu/abs/2020MNRAS.492.2510L}
}

@ARTICLE{LiXH2018,
       author = {{Li}, Xiaohong and {Zhang}, Jun and {Yang}, Shuhong and {Hou}, Yijun and {Erd{\'e}lyi}, Robert},
        title = "{Observing Kelvin-Helmholtz instability in solar blowout jet}",
      journal = {Scientific Reports},
         year = 2018,
        month = may,
       volume = {8},
          eid = {8136},
        pages = {8136},
          doi = {10.1038/s41598-018-26581-4},
archivePrefix = {arXiv},
       eprint = {1806.05867},
 primaryClass = {astro-ph.SR},
       adsurl = {https://ui.adsabs.harvard.edu/abs/2018NatSR...8.8136L}
}

@ARTICLE{Masson2009,
       author = {{Masson}, S. and {Pariat}, E. and {Aulanier}, G. and {Schrijver}, C.~J.},
        title = "{The Nature of Flare Ribbons in Coronal Null-Point Topology}",
      journal = {\apj},
         year = 2009,
        month = jul,
       volume = {700},
       number = {1},
        pages = {559-578},
          doi = {10.1088/0004-637X/700/1/559},
       adsurl = {https://ui.adsabs.harvard.edu/abs/2009ApJ...700..559M}
}

@ARTICLE{Moore2010,
       author = {{Moore}, Ronald L. and {Cirtain}, Jonathan W. and {Sterling}, Alphonse C. and {Falconer}, David A.},
        title = "{Dichotomy of Solar Coronal Jets: Standard Jets and Blowout Jets}",
      journal = {\apj},
         year = 2010,
        month = sep,
       volume = {720},
       number = {1},
        pages = {757-770},
          doi = {10.1088/0004-637X/720/1/757},
       adsurl = {https://ui.adsabs.harvard.edu/abs/2010ApJ...720..757M}
}

@ARTICLE{Moore2013,
       author = {{Moore}, Ronald L. and {Sterling}, Alphonse C. and {Falconer}, David A. and {Robe}, Dominic},
        title = "{The Cool Component and the Dichotomy, Lateral Expansion, and Axial Rotation of Solar X-Ray Jets}",
      journal = {\apj},
         year = 2013,
        month = jun,
       volume = {769},
       number = {2},
          eid = {134},
        pages = {134},
          doi = {10.1088/0004-637X/769/2/134},
       adsurl = {https://ui.adsabs.harvard.edu/abs/2013ApJ...769..134M}
}

@ARTICLE{Morton2012,
       author = {{Morton}, R.~J. and {Verth}, G. and {McLaughlin}, J.~A. and {Erd{\'e}lyi}, R.},
        title = "{Determination of Sub-resolution Structure of a Jet by Solar Magnetoseismology}",
      journal = {\apj},
         year = 2012,
        month = jan,
       volume = {744},
       number = {1},
          eid = {5},
        pages = {5},
          doi = {10.1088/0004-637X/744/1/5},
archivePrefix = {arXiv},
       eprint = {1109.4851},
 primaryClass = {astro-ph.SR},
       adsurl = {https://ui.adsabs.harvard.edu/abs/2012ApJ...744....5M}
}

@ARTICLE{NiL2017,
       author = {{Ni}, Lei and {Zhang}, Qing-Min and {Murphy}, Nicholas A. and {Lin}, Jun},
        title = "{Blob Formation and Ejection in Coronal Jets due to the Plasmoid and Kelvin-Helmholtz Instabilities}",
      journal = {\apj},
         year = 2017,
        month = may,
       volume = {841},
       number = {1},
          eid = {27},
        pages = {27},
          doi = {10.3847/1538-4357/aa6ffe},
archivePrefix = {arXiv},
       eprint = {1705.00180},
 primaryClass = {astro-ph.SR},
       adsurl = {https://ui.adsabs.harvard.edu/abs/2017ApJ...841...27N}
}

@ARTICLE{Pesnell2012,
       author = {{Pesnell}, W. Dean and {Thompson}, B.~J. and {Chamberlin}, P.~C.},
        title = "{The Solar Dynamics Observatory (SDO)}",
      journal = {\solphys},
         year = 2012,
        month = jan,
       volume = {275},
       number = {1-2},
        pages = {3-15},
          doi = {10.1007/s11207-011-9841-3},
       adsurl = {https://ui.adsabs.harvard.edu/abs/2012SoPh..275....3P}
}

@ARTICLE{Raouafi2016,
       author = {{Raouafi}, N.~E. and {Patsourakos}, S. and {Pariat}, E. and {Young}, P.~R. and {Sterling}, A.~C. and {Savcheva}, A. and {Shimojo}, M. and {Moreno-Insertis}, F. and {DeVore}, C.~R. and {Archontis}, V. and {T{\"o}r{\"o}k}, T. and {Mason}, H. and {Curdt}, W. and {Meyer}, K. and {Dalmasse}, K. and {Matsui}, Y.},
        title = "{Solar Coronal Jets: Observations, Theory, and Modeling}",
      journal = {\ssr},
         year = 2016,
        month = nov,
       volume = {201},
       number = {1-4},
        pages = {1-53},
          doi = {10.1007/s11214-016-0260-5},
archivePrefix = {arXiv},
       eprint = {1607.02108},
 primaryClass = {astro-ph.SR},
       adsurl = {https://ui.adsabs.harvard.edu/abs/2016SSRv..201....1R}
}

@ARTICLE{Samanta2019,
       author = {{Samanta}, Tanmoy and {Tian}, Hui and {Yurchyshyn}, Vasyl and {Peter}, Hardi and {Cao}, Wenda and {Sterling}, Alphonse and {Erd{\'e}lyi}, Robertus and {Ahn}, Kwangsu and {Feng}, Song and {Utz}, Dominik and {Banerjee}, Dipankar and {Chen}, Yajie},
        title = "{Generation of solar spicules and subsequent atmospheric heating}",
      journal = {Science},
         year = 2019,
        month = nov,
       volume = {366},
       number = {6467},
        pages = {890-894},
          doi = {10.1126/science.aaw2796},
archivePrefix = {arXiv},
       eprint = {2006.02571},
 primaryClass = {astro-ph.SR},
       adsurl = {https://ui.adsabs.harvard.edu/abs/2019Sci...366..890S}
}

@ARTICLE{Savcheva2007,
       author = {{Savcheva}, Antonia and {Cirtain}, Jonathan and {Deluca}, Edward E. and {Lundquist}, Loraine L. and {Golub}, Leon and {Weber}, Mark and {Shimojo}, Masumi and {Shibasaki}, Kiyoto and {Sakao}, Taro and {Narukage}, Noriyuki and {Tsuneta}, Saku and {Kano}, Ryouhei},
        title = "{A Study of Polar Jet Parameters Based on Hinode XRT Observations}",
      journal = {\pasj},
         year = 2007,
        month = nov,
       volume = {59},
        pages = {S771},
          doi = {10.1093/pasj/59.sp3.S771},
       adsurl = {https://ui.adsabs.harvard.edu/abs/2007PASJ...59S.771S}
}

@ARTICLE{ShenYD2012,
       author = {{Shen}, Yuandeng and {Liu}, Yu and {Su}, Jiangtao and {Deng}, Yuanyong},
        title = "{On a Coronal Blowout Jet: The First Observation of a Simultaneously Produced Bubble-like CME and a Jet-like CME in a Solar Event}",
      journal = {\apj},
         year = 2012,
        month = feb,
       volume = {745},
       number = {2},
          eid = {164},
        pages = {164},
          doi = {10.1088/0004-637X/745/2/164},
archivePrefix = {arXiv},
       eprint = {1110.5243},
 primaryClass = {astro-ph.SR},
       adsurl = {https://ui.adsabs.harvard.edu/abs/2012ApJ...745..164S}
}

@ARTICLE{ShenYD2019,
       author = {{Shen}, Yuandeng and {Qu}, Zhining and {Yuan}, Ding and {Chen}, Huadong and {Duan}, Yadan and {Zhou}, Chengrui and {Tang}, Zehao and {Huang}, Jin and {Liu}, Yu},
        title = "{Stereoscopic Observations of an Erupting Mini-filament-driven Two-sided-loop Jet and the Applications for Diagnosing a Filament Magnetic Field}",
      journal = {\apj},
         year = 2019,
        month = sep,
       volume = {883},
       number = {1},
          eid = {104},
        pages = {104},
          doi = {10.3847/1538-4357/ab3a4d},
archivePrefix = {arXiv},
       eprint = {1908.03660},
 primaryClass = {astro-ph.SR},
       adsurl = {https://ui.adsabs.harvard.edu/abs/2019ApJ...883..104S}
}

@ARTICLE{ShenYD2021,
       author = {{Shen}, Yuandeng},
        title = "{Observation and modelling of solar jets}",
      journal = {Proceedings of the Royal Society of London Series A},
         year = 2021,
        month = feb,
       volume = {477},
       number = {2246},
        pages = {217},
          doi = {10.1098/rspa.2020.0217},
archivePrefix = {arXiv},
       eprint = {2101.04846},
 primaryClass = {astro-ph.SR},
       adsurl = {https://ui.adsabs.harvard.edu/abs/2021RSPSA.47700217S}
}

@ARTICLE{Shibata1992,
       author = {{Shibata}, Kazunari and {Ishido}, Yoshinori and {Acton}, Loren W. and {Strong}, Keith T. and {Hirayama}, Tadashi and {Uchida}, Yutaka and {McAllister}, Alan H. and {Matsumoto}, Ryoji and {Tsuneta}, Saku and {Shimizu}, Toshifumi and {Hara}, Hirohisa and {Sakurai}, Takashi and {Ichimoto}, Kiyoshi and {Nishino}, Yohei and {Ogawara}, Yoshiaki},
        title = "{Observations of X-Ray Jets with the Yohkoh Soft X-Ray Telescope}",
      journal = {\pasj},
         year = 1992,
        month = nov,
       volume = {44},
       number = {5},
        pages = {L173-L179},
          doi = {10.1093/pasj/44.5.L173},
       adsurl = {https://ui.adsabs.harvard.edu/abs/1992PASJ...44L.173S}
}

@ARTICLE{Shibata2007,
       author = {{Shibata}, Kazunari and {Nakamura}, Tahei and {Matsumoto}, Takuma and {Otsuji}, Kenichi and {Okamoto}, Takenori J. and {Nishizuka}, Naoto and {Kawate}, Tomoko and {Watanabe}, Hiroko and {Nagata}, Shin'ichi and {UeNo}, Satoru and {Kitai}, Reizaburo and {Nozawa}, Satoshi and {Tsuneta}, Saku and {Suematsu}, Yoshinori and {Ichimoto}, Kiyoshi and {Shimizu}, Toshifumi and {Katsukawa}, Yukio and {Tarbell}, Theodore D. and {Berger}, Thomas E. and {Lites}, Bruce W. and {Shine}, Richard A. and {Title}, Alan M.},
        title = "{Chromospheric Anemone Jets as Evidence of Ubiquitous Reconnection}",
      journal = {Science},
         year = 2007,
        month = dec,
       volume = {318},
       number = {5856},
        pages = {1591},
          doi = {10.1126/science.1146708},
archivePrefix = {arXiv},
       eprint = {0810.3974},
 primaryClass = {astro-ph},
       adsurl = {https://ui.adsabs.harvard.edu/abs/2007Sci...318.1591S}
}

@ARTICLE{Shimojo2000,
       author = {{Shimojo}, Masumi and {Shibata}, Kazunari},
        title = "{Physical Parameters of Solar X-Ray Jets}",
      journal = {\apj},
         year = 2000,
        month = oct,
       volume = {542},
       number = {2},
        pages = {1100-1108},
          doi = {10.1086/317024},
       adsurl = {https://ui.adsabs.harvard.edu/abs/2000ApJ...542.1100S}
}

@ARTICLE{Solanki2003,
       author = {{Solanki}, Sami K.},
        title = "{Sunspots: An overview}",
      journal = {\aapr},
         year = 2003,
        month = jan,
       volume = {11},
       number = {2-3},
        pages = {153-286},
          doi = {10.1007/s00159-003-0018-4},
       adsurl = {https://ui.adsabs.harvard.edu/abs/2003A&ARv..11..153S}
}

@ARTICLE{Sterling2015,
       author = {{Sterling}, Alphonse C. and {Moore}, Ronald L. and {Falconer}, David A. and {Adams}, Mitzi},
        title = "{Small-scale filament eruptions as the driver of X-ray jets in solar coronal holes}",
      journal = {\nat},
         year = 2015,
        month = jul,
       volume = {523},
       number = {7561},
        pages = {437-440},
          doi = {10.1038/nature14556},
archivePrefix = {arXiv},
       eprint = {1705.03373},
 primaryClass = {astro-ph.SR},
       adsurl = {https://ui.adsabs.harvard.edu/abs/2015Natur.523..437S}
}

@ARTICLE{TanS2023,
       author = {{Tan}, Song and {Shen}, Yuandeng and {Zhou}, Xinping and {Tang}, Zehao and {Zhou}, Chengrui and {Duan}, Yadan and {Yao}, Surui},
        title = "{Stereoscopic observation of simultaneous longitudinal and transverse oscillations in a single filament driven by two-sided-loop jet}",
      journal = {\mnras},
         year = 2023,
        month = apr,
       volume = {520},
       number = {2},
        pages = {3080-3088},
          doi = {10.1093/mnras/stad295},
archivePrefix = {arXiv},
       eprint = {2301.07989},
 primaryClass = {astro-ph.SR},
       adsurl = {https://ui.adsabs.harvard.edu/abs/2023MNRAS.520.3080T}
}

@ARTICLE{TianH2018,
       author = {{Tian}, Hui and {Yurchyshyn}, Vasyl and {Peter}, Hardi and {Solanki}, Sami K. and {Young}, Peter R. and {Ni}, Lei and {Cao}, Wenda and {Ji}, Kaifan and {Zhu}, Yingjie and {Zhang}, Jingwen and {Samanta}, Tanmoy and {Song}, Yongliang and {He}, Jiansen and {Wang}, Linghua and {Chen}, Yajie},
        title = "{Frequently Occurring Reconnection Jets from Sunspot Light Bridges}",
      journal = {\apj},
         year = 2018,
        month = feb,
       volume = {854},
       number = {2},
          eid = {92},
        pages = {92},
          doi = {10.3847/1538-4357/aaa89d},
archivePrefix = {arXiv},
       eprint = {1801.06802},
 primaryClass = {astro-ph.SR},
       adsurl = {https://ui.adsabs.harvard.edu/abs/2018ApJ...854...92T}
}

@ARTICLE{Toriumi2015,
       author = {{Toriumi}, Shin and {Katsukawa}, Yukio and {Cheung}, Mark C.~M.},
        title = "{Light Bridge in a Developing Active Region. I. Observation of Light Bridge and its Dynamic Activity Phenomena}",
      journal = {\apj},
         year = 2015,
        month = oct,
       volume = {811},
       number = {2},
          eid = {137},
        pages = {137},
          doi = {10.1088/0004-637X/811/2/137},
archivePrefix = {arXiv},
       eprint = {1509.00183},
 primaryClass = {astro-ph.SR},
       adsurl = {https://ui.adsabs.harvard.edu/abs/2015ApJ...811..137T}
}

@ARTICLE{Thomas2002,
       author = {{Thomas}, John H. and {Weiss}, Nigel O. and {Tobias}, Steven M. and {Brummell}, Nicholas H.},
        title = "{Downward pumping of magnetic flux as the cause of filamentary structures in sunspot penumbrae}",
      journal = {\nat},
         year = 2002,
        month = nov,
       volume = {420},
       number = {6914},
        pages = {390-393},
          doi = {10.1038/nature01174},
       adsurl = {https://ui.adsabs.harvard.edu/abs/2002Natur.420..390T}
}

@ARTICLE{WeiHY2020,
       author = {{Wei}, H. and {Huang}, Z. and {Hou}, Z. and {Qi}, Y. and {Fu}, H. and {Li}, B. and {Xia}, L.},
        title = "{How eruptions of a small filament feed materials to a nearby larger-scaled filament}",
      journal = {\mnras},
         year = 2020,
        month = nov,
       volume = {498},
       number = {1},
        pages = {L104-L108},
          doi = {10.1093/mnrasl/slaa134},
archivePrefix = {arXiv},
       eprint = {2007.12301},
 primaryClass = {astro-ph.SR},
       adsurl = {https://ui.adsabs.harvard.edu/abs/2020MNRAS.498L.104W}
}

@ARTICLE{Wyper2017,
       author = {{Wyper}, Peter F. and {Antiochos}, Spiro K. and {DeVore}, C. Richard},
        title = "{A universal model for solar eruptions}",
      journal = {\nat},
         year = 2017,
        month = apr,
       volume = {544},
       number = {7651},
        pages = {452-455},
          doi = {10.1038/nature22050},
       adsurl = {https://ui.adsabs.harvard.edu/abs/2017Natur.544..452W}
}

@ARTICLE{Wyper2018,
       author = {{Wyper}, P.~F. and {DeVore}, C.~R. and {Antiochos}, S.~K.},
        title = "{A Breakout Model for Solar Coronal Jets with Filaments}",
      journal = {\apj},
         year = 2018,
        month = jan,
       volume = {852},
       number = {2},
          eid = {98},
        pages = {98},
          doi = {10.3847/1538-4357/aa9ffc},
archivePrefix = {arXiv},
       eprint = {1712.00134},
 primaryClass = {astro-ph.SR},
       adsurl = {https://ui.adsabs.harvard.edu/abs/2018ApJ...852...98W}
}

@ARTICLE{YanXL2021,
       author = {{Yan}, Xiaoli and {Wang}, Jincheng and {Guo}, Qiaoling and {Xue}, Zhike and {Yang}, Liheng and {Tan}, Baolin},
        title = "{The Formation Process of the First Halo Coronal Mass Ejection in Solar Cycle 25: Magnetic Cancellation, Bidirectional Jet, and Hot Channel}",
      journal = {\apj},
         year = 2021,
        month = sep,
       volume = {919},
       number = {1},
          eid = {34},
        pages = {34},
          doi = {10.3847/1538-4357/ac116d},
       adsurl = {https://ui.adsabs.harvard.edu/abs/2021ApJ...919...34Y}
}

@ARTICLE{YangLH2025,
       author = {{Yang}, Liheng and {Yan}, Xiaoli and {Zhang}, Jun and {Xue}, Zhike and {Xu}, Zhe and {Wang}, Jincheng and {Hou}, Yijun and {Zhou}, Yian and {Kong}, Defang and {Umar}, Roslan and {Zhang}, Xinsheng and {Li}, Qiaoling and {Yang}, Liping},
        title = "{Recurrent Jetlets Associated with the Disappearance of a Satellite Spot}",
      journal = {\apj},
         year = 2025,
        month = jul,
       volume = {987},
       number = {2},
          eid = {193},
        pages = {193},
          doi = {10.3847/1538-4357/addac1},
archivePrefix = {arXiv},
       eprint = {2505.10995},
 primaryClass = {astro-ph.SR},
       adsurl = {https://ui.adsabs.harvard.edu/abs/2025ApJ...987..193Y}
}

@ARTICLE{YangSH2015,
       author = {{Yang}, Shuhong and {Zhang}, Jun and {Jiang}, Fayu and {Xiang}, Yongyuan},
        title = "{Oscillating Light Wall Above a Sunspot Light Bridge}",
      journal = {\apjl},
         year = 2015,
        month = may,
       volume = {804},
       number = {2},
          eid = {L27},
        pages = {L27},
          doi = {10.1088/2041-8205/804/2/L27},
archivePrefix = {arXiv},
       eprint = {1504.03563},
 primaryClass = {astro-ph.SR},
       adsurl = {https://ui.adsabs.harvard.edu/abs/2015ApJ...804L..27Y}
}

@ARTICLE{Yokoyama1995,
       author = {{Yokoyama}, Takaaki and {Shibata}, Kazunari},
        title = "{Magnetic reconnection as the origin of X-ray jets and H{\ensuremath{\alpha}} surges on the Sun}",
      journal = {\nat},
         year = 1995,
        month = may,
       volume = {375},
       number = {6526},
        pages = {42-44},
          doi = {10.1038/375042a0},
       adsurl = {https://ui.adsabs.harvard.edu/abs/1995Natur.375...42Y}
}

@ARTICLE{Yokoyama1996,
       author = {{Yokoyama}, Takaaki and {Shibata}, Kazunari},
        title = "{Numerical Simulation of Solar Coronal X-Ray Jets Based on the Magnetic Reconnection Model}",
      journal = {\pasj},
         year = 1996,
        month = apr,
       volume = {48},
        pages = {353-376},
          doi = {10.1093/pasj/48.2.353},
       adsurl = {https://ui.adsabs.harvard.edu/abs/1996PASJ...48..353Y}
}

@ARTICLE{ZhangJW2017,
       author = {{Zhang}, Jingwen and {Tian}, Hui and {He}, Jiansen and {Wang}, Linghua},
        title = "{Surge-like Oscillations above Sunspot Light Bridges Driven by Magnetoacoustic Shocks}",
      journal = {\apj},
         year = 2017,
        month = mar,
       volume = {838},
       number = {1},
          eid = {2},
        pages = {2},
          doi = {10.3847/1538-4357/aa63e8},
archivePrefix = {arXiv},
       eprint = {1702.08585},
 primaryClass = {astro-ph.SR},
       adsurl = {https://ui.adsabs.harvard.edu/abs/2017ApJ...838....2Z}
}

@ARTICLE{Nei_1989,
       author = {{Neidig}, Donald F.},
        title = "{The Importance of Solar White-Light Flares}",
      journal = {\solphys},
         year = 1989,
        month = mar,
       volume = {121},
       number = {1-2},
        pages = {261-269},
          doi = {10.1007/BF00161699},
       adsurl = {https://ui.adsabs.harvard.edu/abs/1989SoPh..121..261N}
}

\end{document}